\documentclass[3p,sort&compress]{elsarticle}
\usepackage{graphicx}
\usepackage{bm}
\usepackage{color}
\usepackage{xcolor}
\usepackage{MnSymbol}
\usepackage{enumerate}
\usepackage{bbold}
\usepackage{lipsum}
\usepackage{array}
\usepackage{hyperref}
\usepackage{textgreek}
\usepackage{soul}
\usepackage{multirow}
\usepackage{accents}
\usepackage{trimclip}
\usepackage{tikz}
\usepackage{slashbox}
\usepackage{makecell}

\usepackage{array,booktabs}

\DeclareMathAlphabet\mathbfcal{OMS}{cmsy}{b}{n}

\journal{Physics Report}

\def\la{\langle}
\def\ra{\rangle}

\def\be{\begin{equation}}
\def\ee{\end{equation}}

\newcommand{\beq}{\begin{equation}}
\newcommand{\eeq}{\end{equation}}
\newcommand{\nn}{\nonumber}

\newcommand{\erf}[1]{Eq.~(\ref{#1})}

\newcommand{\dg}{^\dagger}

\newcommand{\bra}[1]{\langle{#1}|}
\newcommand{\ket}[1]{|{#1}\rangle}

\newcommand{\op}[1]{\hat{ #1}}                % Operator

\newcommand{\tp}{^{\top}}

\renewcommand{\c}{_{\text{D}}}
\newcommand{\etal}{\textit{et al.}}
\newcommand{\ie}{\textit{i.e.}}
\newcommand{\eg}{\textit{e.g.}}

\newcommand{\accentfuturearrowo}[1]{%
\begin{tikzpicture}[#1]%
\draw[line width = 0.2mm] (-0.6mm,0mm) -- (1.2mm,0mm);
\draw (2mm,0) -- (1.2mm,0.3mm) -- (1.2mm,-0.3mm) -- cycle;
\end{tikzpicture}%
}

\newcommand{\tempfuto}{\accentfuturearrowo{}}
\newcommand{\fut}[1]{\accentset{\tempfuto}{#1}}

\newcommand{\accentfuturearrowc}[1]{%
\begin{tikzpicture}[#1]%
\fill (2mm,0) -- (1.2mm,0.3mm) -- (1.2mm,-0.3mm);
\draw[line width = 0.2mm] (-0.6mm,0mm) -- (1.2mm,0mm);
\draw (2mm,0) -- (1.2mm,0.3mm) -- (1.2mm,-0.3mm) -- cycle;
\end{tikzpicture}%
}

\newcommand{\tempfutc}{\accentfuturearrowc{}}
\newcommand{\futp}[1]{\accentset{\tempfutc}{#1}}
\newcommand{\tempsfutc}{\scalebox{0.8}[0.8]{\accentfuturearrowc{}}}
\newcommand{\futps}[1]{\accentset{\tempsfutc}{#1}}
\newcommand{\tempssfutc}{\scalebox{0.6}[0.6]{\accentfuturearrowc{}}}
\newcommand{\futpss}[1]{\accentset{\tempssfutc}{#1}}

\newcommand{\temppast}{\scalebox{-1}[1]{\accentfuturearrowc{}}}
\newcommand{\past}[1]{\accentset{\temppast}{#1}}
\newcommand{\tempspast}{\scalebox{-0.8}[0.8]{\accentfuturearrowc{}}}
\newcommand{\pasts}[1]{\accentset{\tempspast}{#1}}
\newcommand{\temppastss}{\scalebox{-0.6}[0.6]{\accentfuturearrowc{}}}
\newcommand{\pastss}[1]{\accentset{\temppastss}{#1}}

\newcommand{\accentbotharrowc}[1]{%
\begin{tikzpicture}[#1]%
\fill (-0.6mm,0) -- (0.2mm,0.3mm) -- (0.2mm,-0.3mm);
\fill (2mm,0) -- (1.2mm,0.3mm) -- (1.2mm,-0.3mm);
\draw[line width = 0.2mm] (-0.1mm,0mm) -- (1.2mm,0mm);
\draw (2mm,0) -- (1.2mm,0.3mm) -- (1.2mm,-0.3mm) -- cycle;
\draw (-0.6mm,0) -- (0.2mm,0.3mm) -- (0.2mm,-0.3mm) -- cycle;
\end{tikzpicture}%
}

\newcommand{\tempbothc}{\accentbotharrowc{}}
\newcommand{\bothp}[1]{\accentset{\tempbothc}{#1}}
\newcommand{\tempsbothc}{\scalebox{0.8}[0.8]{\accentbotharrowc{}}}
\newcommand{\bothps}[1]{\accentset{\tempsbothc}{#1}}

\newcommand{\accentbotharrowo}[1]{%
\begin{tikzpicture}[#1]%
\fill (-0.6mm,0) -- (0.2mm,0.3mm) -- (0.2mm,-0.3mm);
\draw[line width = 0.2mm] (-0.1mm,0mm) -- (1.2mm,0mm);
\draw (2mm,0) -- (1.2mm,0.3mm) -- (1.2mm,-0.3mm) -- cycle;
\draw (-0.6mm,0) -- (0.2mm,0.3mm) -- (0.2mm,-0.3mm) -- cycle;
\end{tikzpicture}%
}

\newcommand{\tempbotho}{\accentbotharrowo{}}
\newcommand{\both}[1]{\accentset{\tempbotho}{#1}}
\newcommand{\tempsbotho}{\scalebox{0.8}[0.8]{\accentbotharrowo{}}}
\newcommand{\boths}[1]{\accentset{\tempsbotho}{#1}}

\newcommand{\fil}{_{\text F}}
\newcommand{\sm}{_{\text S}}
\newcommand{\god}{_{\text T}}%{^{\rm true}}%{_{\text G}}

\newcommand{\suba}{_{\text A}}
\newcommand{\subb}{_{\text B}}

\newcommand{\bx}{{ x}}

\newcommand{\cx}{{\bf x}}

\newcommand{\trho}{{\tilde \rho}}
\newcommand{\mrho}{\rho}
\newcommand{\true}{{\rm true}}
\newcommand{\ik}{_{t}}
\newcommand{\ost}{_{\rm ost}}

\newcommand{\qp}{\lambda}
\newcommand{\twp}{\tilde \wp}

\newcommand{\tin}{0}

\newcommand{\est}[1]{{#1}^\star}
\newcommand{\rhoOU}{\rho_{\pasts{\bo}, \pasts{\bu}}}
\newcommand{\rhoU}{\rho_{\pasts{\bu}}}
\newcommand{\psiOU}{\hat\psi_{\pasts{\bo}, \pasts{\bu}}}
\newcommand{\wpOU}{\wp_{\pasts{\bo}, \pasts{\bu}}}
\newcommand{\wpU}{\wp_{ \pasts{\bu}}}
\DeclareMathOperator*{\argmax}{arg\,max}
\DeclareMathOperator*{\argmin}{arg\,min}
\newcommand{\Tr}[1]{{\rm Tr}\left(#1\right)}
\newcommand{\wvs}{_{\rm SWV}}
\newcommand{\Gset}{{\mathfrak G}({\mathbb H})}
\newcommand{\Hset}{\mathbb H}

\definecolor{nblue}{rgb}{0.06,0.3,0.73}%229 11R, 61G, 145B
\definecolor{nblack}{rgb}{0,0,0}
\definecolor{nred}{rgb}{0.9,0.1,0.1}
\definecolor{nmagenta}{rgb}{0.7,0.0,0.3}
\definecolor{npurple}{rgb}{0.52,0,0.52}
\definecolor{neditcolor}{rgb}{0.3,0.3,0.9}
\definecolor{colq1}{HTML}{FC0D1B}
\definecolor{colq2}{HTML}{1FBAE8}
\definecolor{colq3}{HTML}{FC28FC}
\definecolor{colq4}{HTML}{FD7F23}
\definecolor{colq5}{HTML}{986638}
\definecolor{colq67}{HTML}{1FCA53}
\definecolor{colq8}{HTML}{239625}

\newcommand{\bo}{O}
\newcommand{\bu}{ U}
\newcommand{\dd}{{\rm d}}
\newcommand{\dt}{\dd t}
\newcommand{\ddt}{\delta t}

\newcommand{\qq}{{\bm q}}
\newcommand{\pp}{{\bm p}}

\begin{document}

\begin{frontmatter}



\title{Unifying theory of quantum state estimation \\ using past and future information}


\author{Areeya Chantasri\corref{cor1}}
\ead{areeya.chn@mahidol.ac.th}
\address{Centre for Quantum Computation and Communication Technology (Australian Research Council), \\ Centre for Quantum Dynamics, Griffith University, Nathan, Queensland 4111, Australia}
\address{Optical and Quantum Physics Laboratory, Department of Physics, \\Faculty of Science, Mahidol University, Bangkok, 10400, Thailand}%
\author{Ivonne Guevara\corref{}}
%\ead{i.guevaraprieto@griffith.edu.au}
\author{Kiarn T. Laverick\corref{}}
%\ead{kiarn.laverick@griffithuni.edu.au }
\author{Howard M. Wiseman\corref{cor1}}
\ead{h.wiseman@griffith.edu.au}
\address{Centre for Quantum Computation and Communication Technology (Australian Research Council), \\ Centre for Quantum Dynamics, Griffith University, Nathan, Queensland 4111, Australia}
\cortext[cor1]{Corresponding authors}

\begin{abstract}
Quantum state estimation for continuously monitored dynamical systems involves assigning a quantum state to an individual system at some time, conditioned on the results of continuous observations. The quality of the estimation depends on how much observed information is used and on how optimality is defined for the estimate. In this work, we consider problems of quantum state estimation where some of the measurement records are not available, but where the available records come from both before (past) and after (future) the estimation time, enabling better estimates than is possible using the past information alone. Past-future information for quantum systems has been used in various ways in the literature, in particular, the quantum state smoothing, the most-likely path, and the two-state vector and related formalisms. To unify these seemingly unrelated approaches, we propose a framework for partially observed quantum systems with continuous monitoring, wherein the first two existing formalisms can be accommodated, with some generalization. The unifying framework is based on state estimation with expected cost minimization, where the cost can be defined either in the space of the unknown record or in the space of the unknown true state. Moreover, we connect all three existing approaches conceptually by defining five new cost functions, and thus new types of estimators, which bridge the gaps between them. We illustrate the applicability of our method by calculating all seven estimators we consider for the example of a driven two-level system dissipatively coupled to bosonic baths. Our theory also allows connections to classical state estimation, which create further conceptual links between our quantum state estimators. 
\end{abstract}
%\maketitle
\date{today}


%item Research highlight 1

\begin{keyword}

Quantum state estimation \sep Filtering \sep Smoothing \sep Continuous quantum measurement \sep Quantum trajectory
\sep Cost functions \sep Stochastic processes



\end{keyword}

\end{frontmatter}

\tableofcontents



\section{Introduction}

%For Mankei, he said that he can show the connection between the smoothing in quantum case with the classical information theoretic approach, and show in general case. But the case he shows is only when is to see the essense of the operators in the mean square form of operators that are not neceesary commute, or even in the different Hilbert space. especially that the hidden or unknown variable are in one hilbert space and the estimator lives in another hilbert space. For example, the weak value, the hidden variable is the unknown state of an observable X, but the estimator is an operator in a different Hilbert space (but with the connection through how the filtered state connect with the POVM)

Estimation problems deal with assigning appropriate values to quantities that are unknown because they are not completely accessible via observation. %Such quantities can be physical parameters or some unknown functions, and they can be constant or time-dependent. 
Given available information, such as observed measurement records and the relationships between the records and the quantities of interest, one can devise an optimal strategy to extract appropriate estimators of the unknowns. We are interested in dynamical systems, where quantities of interest at time $t$ can be estimated conditioned on the observation at other times. Examples are classical \emph{filtering} and \emph{smoothing} techniques which give optimal estimators, typically minimizing the mean square errors of the estimation~\cite{BookJazwinski,KalBuc1961,Kushner1964,FraPot1969,BookWienerSmt}. The filtering uses only records before time $t$ (\emph{past} records) for the estimation, while the smoothing uses records both before and after $t$ (\emph{past-future} records). It has been shown that the smoothed estimate, using both past and future records, is statistically closer to the \emph{true} values of the estimated quantity, because it uses more information from the observation than the filtered one, and thus is preferable provided that real-time estimation is not required~\cite{Wheatley2010,Ivonne2015,Huang2018,Laverick2018,LavCha2019}.

Estimation techniques from classical cases have been adopted for quantum systems. However, the implementation is not always straightforward. For problems where the quantities to be estimated can be treated as ``classical", such as parameters that affect quantum systems, techniques in classical parameter estimation (including classical filtering and smoothing) can be applied relatively straightforwardly~\cite{BookHelstrom,Tsangsmt2009,Wheatley2010}. However, for estimating quantities that are inherently ``quantum'', such as quantum states or observables, naively applying classical techniques can lead to strange results if operators representing the quantities of interest and the observed operators do not commute. This occurs when we include both past and future information in the estimation~\cite{Wise2002,Ivonne2015,Budini2017,LuisPsmooth2017,Budini2018b}, as a quantum system's observables at time $t$ do not commute with operators representing measurement results of the system at later time~\cite{BookWiseman}. 
%Using the filtering and smoothing as examples, where quantum states at any time $t$ commute with the past, but not the future, record observables; the classical filtering has a direct analogue as the quantum trajectory approach, while the classical smoothing on the other hand leads to distinctive approaches that seem unrelated, depending on whether the estimated quantities are quantum observables or states.
%which naturally include techniques for quantum observables and ones conditioned on past-only and future-only information. We then aim to make connections among 
It is precisely this class of estimation problems, for \emph{quantum states and observables}, involving the use of past-future information that we are concerned with in this paper. 

The uses of past-future information for quantum systems have been investigated since the 1950's~\cite{Watanabe1955} and is still a topic that creates debate~\cite{Vaidman2017,Tsang2019}. In this work, we identify three types of approaches in the literature, based on their mathematical similarity and underlying concepts. The main result of this paper is to unify these approaches, after first generalizing them by using the idea of optimal estimators defined with distinct \emph{cost functions}. The existing approaches are as follows, in chronological order.

The first category (including most of the work in the literature), we name the \emph{two-state formalism}.  We chose this name to encompass a number of closely related formalisms~\cite{Watanabe1955,ABL1964,AV2002,Gammelmark2013} that use two states (one forward-in-time and one backward-in-time) to utilise the past-future information. %The formalism was proposed to use both forward (from the past) and backward (from the future) evolving quantum states to completely describe a quantum system at an intermediate time. 
This idea also led to the formulation of the theory of weak values~\cite{AAV1988}. 
The real part of the weak value has a simple interpretation as the ensemble-average of weak measurement results 
with appropriate preselection (forward-in-time state) 
and postselection (backward-in-time state). %, describing average measurement results of an infinitely weak measurement conditioned on the pre and postselected states of the system of interest. 
Here, a weak measurement is one whose result contains an arbitrarily small amount of information about a quantity of interest and which is thereby able to have an arbitrarily small measurement back-action upon the system. The two-state formalism is also useful 
within the context of classical smoothing theory, for estimating  the probabilities of outcomes of measurements 
that are not weak, at some time between the 
preselection and postselection~\cite{Tsangsmt2009-2,Gammelmark2013}. 
%Under this category, we also include the application of classical smoothing for estimating unknown measurement results of quantum observables 
%at some intermediate times
%, as such an estimate reduces to the real part of the weak value, in the limit of a weak measurement. 
Finally, a smoothing technique for quantum observables has been formulated based on the least-mean-squared errors~\cite{Ohki15,Ohki19,Tsang2019}, analogous to standard classical smoothing, and shown to equal to the real part of the weak value.

The second category is the \emph{most-likely path formalism}, which considers quantum state trajectories from all possible measurement records associated with initial and final boundary conditions. These trajectories result from continuous measurements, which can be thought of as infinitely many consecutive weak measurements, a time $\dt$ apart, where the average measurement back-action scales as $\dt$. The most-likely path is defined as the one with the maximum-likelihood measurement record.  The formalism resembles the most-probable paths (also known by many closely related names) considered for classical stochastic processes~\cite{ZeiDem1987,DurBac1978,DAD2014}, which have been useful in computing transition rates between initial and final states~\cite{Dykman1994}. Similar techniques were developed for quantum systems, in particular, for diffusive-type continuous measurement in Refs.~\cite{Chantasri2013,chantasri2015stochastic}, where the boundary conditions are defined as initial and final quantum states. This use of past-future information  is compatible with formulating the estimation problem in the space of all possible records because there is a one-to-one relation between a quantum-state path and a scalar measurement record.  %, at least for the case of a single type of measurement. 

%In the language of estimation, one can treat the measurement record as \red the unknown to be estimated, since it is  with a  path, and 

% and the most-likely path as an estimator (since record),  where the is implemented in the form of initial and final boundary conditions for the quantum state.

The last category is the \emph{quantum state smoothing formalism}. This is similar to the most-likely path formalism in that it yields a quantum state following a trajectory, but is also related to the two-state formalism in that it is a quantum analogue of the classical state smoothing used in some of the approaches mentioned in the paragraph about that formalism above. Quantum state smoothing considers estimating unknown quantum states in a partially-observed quantum system scenario, where an observer has access to only some measurement records (observed records) and not to the rest (unobserved records). The smoothed quantum state is defined as an observer's estimate conditioned on the past-future observed information, which is a better estimate of unknown true states than the usual past-conditioned quantum trajectory~\cite{Ivonne2015,Budini2017,Budini2018b,chantasri2019,GueWis20}. 
%and is considered an optimal state estimator for specific cost functions~\cite{LavGue2020}. 
The formalism is a quantum analogue of classical state smoothing, in the sense that the smoothed quantum state reduces to its classical counterpart when initial conditions and dynamics of the system can be described probabilistically in a fixed basis~\cite{LavCha2020a}.

\begin{figure}[t]
\centering
\includegraphics[width=12cm]{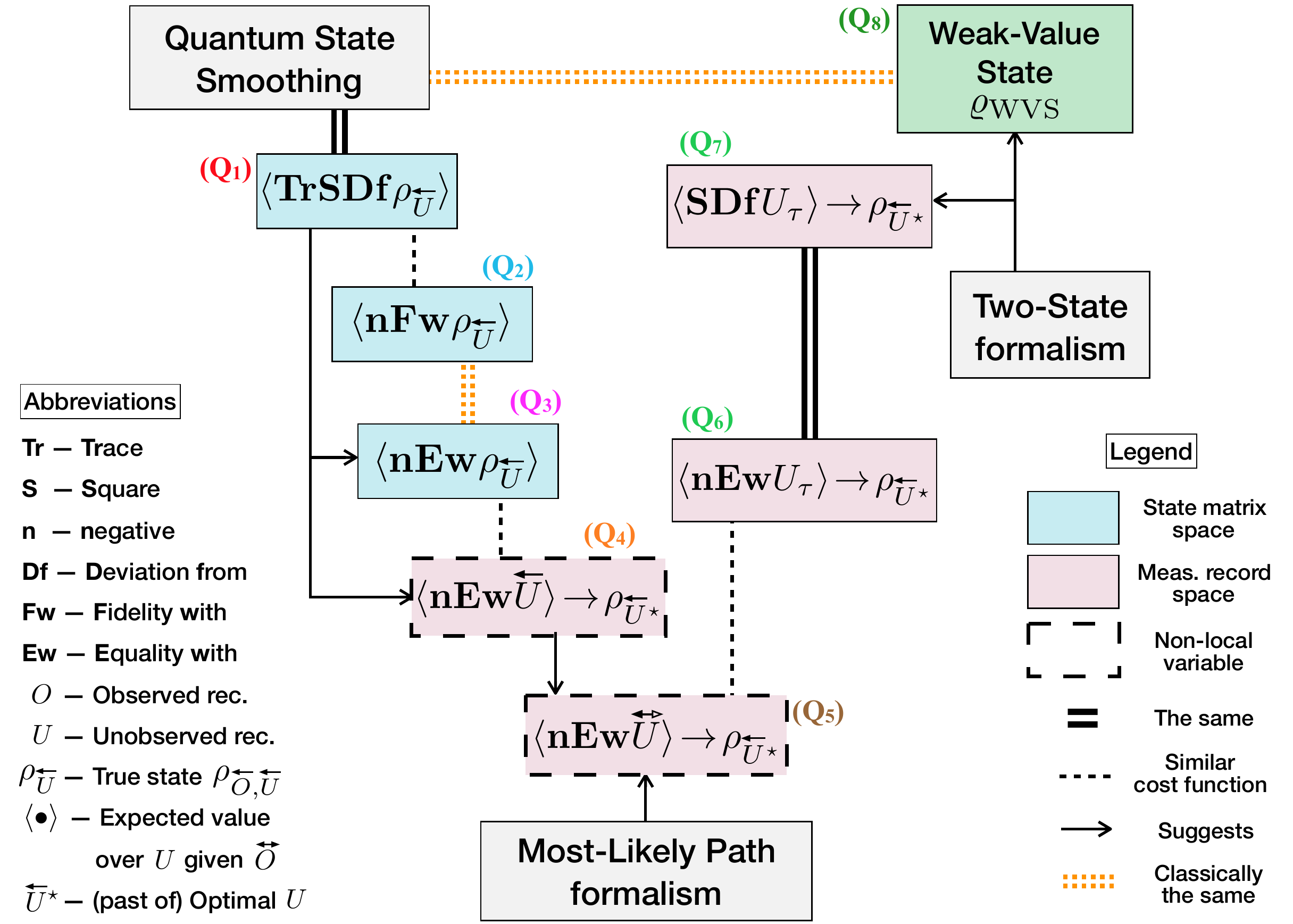}
\caption{Diagrams showing eight different expected costs, for $\textbf{Q}_1$--$\textbf{Q}_8$, which gives eight optimal estimators, connecting the existing formalisms (grey boxes): quantum state smoothing, the CDJ most-likely path formalism, and the two-state  formalism. Expected costs in blue and pink boxes represent the costs defined in the quantum state space and the unknown record space, respectively. The connections are described in different lines and colors, following the discussion in the text in Section~\ref{sec-sevenQSE}. Since the observed record is fixed in the optimization, we omit the $\bo$-dependence in the definition of true states. The labels $(\textbf{Q}_1)$--$(\textbf{Q}_8)$ are colour-coded consistent with the colours used for estimators in the later plots in Section~\ref{sec-example} and Table~\ref{tab-avecost}.}
\label{fig-diagramQ}
\end{figure}

Given these three very different existing formalisms, it is natural to ask whether they can somehow be related. For that purpose, here we introduce a unifying framework for quantum state estimation. The framework is built upon the scenario of an open quantum system with observed (O)  and unobserved (U)  records, with different optimal state estimators corresponding to different cost functions (see Figure~\ref{fig-diagramQ}). Of the above three categorized formalisms, the last two --- most-likely path formalism and quantum state smoothing ---  which directly involve quantum states, can be generalized and rederived by miminizing expected cost functions.  The expected costs are conditioned on the past-future information of the observed record and are  defined in the record space (the most-likely path formalism)  and in the state space (the quantum state smoothing). For the first category (the two-state formalism), even though it does not seek to estimate a quantum state in a conventional sense, we exploit a state-like quantity as originally proposed by Tsang~\cite{Tsangsmt2009}, which we call the smoothed weak-value state~\cite{LavCha2019}. This is defined in our theory by using a cost that relates to estimating weak measurement results. Having established this unifying framework, we also introduce other costs and their associated estimators, showing conceptual connections between all three formalisms, shown as $\textbf{Q}_1$--$\textbf{Q}_8$ in Figure~\ref{fig-diagramQ}. Such a framework can also be applied to classical estimators, which helps in solidifying the structure of the quantum estimators we construct, as shown in Figure~\ref{fig-diagramQ}. 

We start by explaining the required types of estimation problems, \ie, configuration estimation, classical state estimation, quantum state estimation, and record estimation, in Section~\ref{sec-estimation}. We then review the three categories of formalisms in more details, in a way that allows us concurrently introduce the partially-observed quantum system with observed, unobserved, and both records, in Section~\ref{sec-existing}. In Section~\ref{sec-sevenQSE} we describe all eight types of estimators which form the unifying theory, show various equivalences for classical state estimators, and investigate how to calculate the expected costs for an arbitrary (not necessarily optimal) estimator. We show how all eight estimators, and many costs, all can be calculated explicitly using the example of a single qubit coupled to bosonic bath in Section~\ref{sec-example}.  We conclude with a discussion of future work and open questions in Section~\ref{sec-discussion}.  Details of the calculations, numerical methods, and glossary of abbreviations and acronyms are presented in Appendices \ref{sec-app-pdf}--\ref{Glossary}.

\section{Estimation and cost functions}\label{sec-estimation}
In estimation problems, it is usually assumed that definite \emph{true} values of unknown quantities exist, but are unknown to an observer. Given the data from observation, one can devise a systematic approach to best guess the hidden quantities, based on some measure of how far the observer's guesses are from the true quantities. In this section, we develop this concept as cost-minimization estimation. We start with the conventional case of estimating classical variables, which we refer to as \emph{configurations}, and then generalize to the estimation of probability distributions of such configurations, which we refer to as classical \emph{states}. This enables us to naturally introduce the optimal quantum state estimation and the unknown record estimation, which will be used intensively in the rest of the paper.

%As an example, let us consider an unknown vector of parameters $\cx$, which can take one of many values in a set ${\bf X}$. Using the data from observation, the observer can construct his or her knowledge (or believe) of such parameters, in the form of a probability distribution of true vectors, $\wp(\cx^\true)$, which can then be used to define an optimal estimator
%\begin{align}\label{eq-intro}
%\cx^* = \argmin_{\cx \in \cal \bf X} \left\langle c\left(\cx,\cx^\true \right) \right\rangle,
%\end{align}
%that is, a value of $\cx$ in ${\cal \bf X}$ that minimizes an expected cost, where the cost is denoted by $c\left(\cx,\cx^\true \right)$. This cost can be any function that determines a distance measure between a possible estimator $\cx$ and a possible true vector $\cx^\true$. The expected value $\la \cdots \ra$ is calculated as an average weighted by the distribution $\wp(\cx^\true)$. The form of the optimal estimator Eq.~\eqref{eq-intro} is applicable for the parameters both with continuous and discrete values.

\subsection{Configuration estimation}\label{sec-CFE}
The most common type of estimation problems is the configuration estimation. Consider an example of estimating a vector $\cx^\true$ of $d$ parameters, representing the system's true configuration that is unknown. The true configuration can take any value $\cx$ in a set denoted by ${\mathbb X}$, which could be a set of continuous or discrete parameters. Using data from observation and prior knowledge, an observer, named Alice, refines her knowledge of such parameters, in the form of a probability distribution of true configurations, $\wp\c(\cx)$, where the subscript `D' refers to conditioning on the relevant data (either observations or constraints). Given this posterior knowledge, an \emph{optimal estimate of the configuration} is defined as an estimator $\est\cx$ that minimizes an expected cost,
\begin{align}\label{eq-gencostx}
\est\cx = \argmin_{\cx \in \mathbb X} \left\langle c\left(\cx,\cx^\true \right) \right\rangle_{\cx^\true|{\rm D}},
\end{align}
where the expected cost is a cost function $c\left(\cx,\cx^\true \right)$ averaged over possible true configurations $\cx^\true$ conditioned on `D', and can be written explicitly as
\begin{align}\label{eq-confest-int}
 \left\langle c\left(\cx,\cx^\true \right) \right\rangle_{\cx^\true|{\rm D}} \equiv \int\!\!\dd\cx' \,\wp\c(\cx') c\left(\cx,\cx' \right),
\end{align}
using $\cx'$ as a dummy variable. This integral can be replaced by a sum (with associated probability weights), if the possible values of $\cx^\true$ are discrete. Note that, for the rest of the paper, we will use `$\star$' to indicate estimators, and use unadorned variables for dummy variables or arguments of $\argmax$ and $\argmin$ functions.

In this work, we are most interested in two common costs for configuration estimates: Square Deviation (SD) and negative Equality (nE). The estimators that minimize these expected costs are the well-known Bayesian mean estimator (BME) and the maximum likelihood estimator (MLE), respectively. That is, the BME minimizes the expected SD cost (also referred to as the mean square error),
\begin{align}\label{eq-confestBME1}
 \est\cx_{\rm SD} = & \argmin_{\cx \in \mathbb X} \left\la \left(\cx - \cx^\true\right)\tp \left(\cx - \cx^\true\right) \right\ra_{\cx^\true|{\rm D}} \nonumber \\
 = & \left\langle \cx^\true \right\rangle_{\cx^\true|{\rm D}} \equiv \est\cx_{\rm BME}.
\end{align}
This result can be obtained by taking the derivative of the expected cost in the first line of Eq.~\eqref{eq-confestBME1} over $\cx$ and setting it to zero. Similarly, the MLE minimizes an expected nE cost,
\begin{align}\label{eq-mleprobmax}
 \est\cx_{\rm nE}  = & \argmin_{\cx \in \mathbb X} \left\la -\delta^{(d)}\!\left(\cx - \cx^\true\right) \right\ra_{\cx^\true|{\rm D}} \nonumber \\
=&  \argmin_{\cx \in {\mathbb X}}\,[ - \wp\c(\cx) ], \nonumber \\
=&  \argmax_{\cx \in {\mathbb X}}\,\wp\c(\cx) \equiv \est\cx_{\rm MLE},
\end{align}
where we have used $\delta^{(d)}(\cdots)$ as a $d$-dimensional Dirac $\delta$-function, which can be replaced by the Kronecker delta $\delta_{\cx,\cx^\true}$ for a discrete-variable binary cost function. To obtain the last line of Eq.~\eqref{eq-mleprobmax}, we use the definition of an expected cost in Eq.~\eqref{eq-confest-int}, where minimizing $-\wp\c(\cx)$ is equivalent to maximizing $\wp\c(\cx)$. %We will drop the explicit dimension of the $\delta$-function whenever appropriate. 

\subsection{Classical state estimation}\label{sec-CSE}
The optimality concept in configuration estimation can be applied to the estimation of classical states, which are probability density functions (PDFs) of unknown configurations. In the configuration estimation, we assumed that the true configuration of the system does exist as $\cx^\true$. Therefore, if any observer has a perfect knowledge about the true configuration, the observer's classical state of knowledge should be a \emph{true classical state}, which is either the Dirac $\delta$-function $\wp^{\rm true}(\cx) = \delta^{(d)}\!\left(\cx - \cx^{\rm true}\right)$ for continuous variables, or the Kronecker $\delta$-function $\wp^{\rm true}(\cx) = \delta_{\cx, \cx^{\rm true}}$ for discrete variables. Classical state estimation is relevant to the question: What is an optimal posterior distribution $\wp(\cx)$ given that a possible true state is $\wp^{\rm true}(\cx)$ and the observer's belief of true configurations conditioned on observation is given by $\wp\c(\cx)$? We will see below that classical state estimation might seem redundant given configuration estimation; however, for us, it is an important step towards introducing the concept of quantum state estimation in the next subsection.

In order to make a smooth transition to quantum state estimation, we will simplify the analysis in this subsection by only considering a discrete-type configuration. That is, let $\bx$, the discrete variable of interest, take any value in a countable set ${\mathbb X}$, and define a set ${\mathbb P} = \{ \wp(\bx) : \bx \in {\mathbb X}\}$ of all possible normalized PDFs of $\bx$. We can then define an \emph{optimal estimate of the classical state} as that which minimizes an expected cost function,
\begin{align}
\est\wp(\bx) = \argmin_{\wp \in {\mathbb P}} \left\la c\left[\wp, \wp^{\rm true}\right] \right\ra_{\wp^\true|{\rm D}}\!\!(\bx).
\end{align}
Recall that the true state is the Kronecker $\delta$-function $\wp^\true(\bx) = \delta_{\bx,\bx^\true}$. The expected value $\la \cdots \ra_{\wp^\true|{\rm D}}$ is computed by integrating over all possible true states, which in this case is equivalent to summing over all possible true configurations, with weights conditioned on the observation data:
\begin{align}\label{eq-confest-int2}
 \left\langle c\left[\wp,\wp^\true \right] \right\rangle_{\wp^\true|{\rm D}} \equiv & \left\langle c\left[\wp(\bx),\delta_{\bx,\bx^\true} \right] \right\rangle_{\bx^\true|{\rm D}} \nonumber \\
 = & \sum_{\bx' \in {\mathbb X}} \wp\c(\bx') \, c\left[\wp(\bx),\delta_{\bx,\bx'} \right],
\end{align}
where the probability distribution $\wp\c(\bx)$ represents the observer's knowledge of the true configuration conditioned on the observed data. We now consider the two types of cost functions presented in Section~\ref{sec-CFE}.

%However, since the true state is a $\delta$-function of true configuration, an average with the weights imposed on $\delta$-functions is equivalent to an average with the same weights imposed on the corresponding true configurations.

First, the state estimator that minimizes a sum Square Deviation (\textSigma SD) cost for classical states is,
\begin{align}\label{eq-cstateBME1}
\est\wp_{ \text{ \textSigma SD}} (\bx)= & \argmin_{\wp \in {\mathbb P}} \left\la \sum_{\bx' \in {\mathbb X}} \left\{\wp(\bx') - \wp^{\rm true}(\bx')\right\}^2 \right\ra_{\wp^\true|{\rm D}}\!\!\!\!\!\!\!\!\!\!\!\!\!\!\!(\bx)\nonumber \\
 = & \left\langle \delta_{\bx,\bx^\true} \right\rangle_{\bx^\true|{\rm D}} =  \wp\c(\bx),
\end{align}
where, comparing with Eq.~\eqref{eq-confestBME1}, the square deviation becomes an inner product of two identical functional deviations (between $\wp(\bx)$ and $\wp^\true(\bx)$), where $\wp^\true(\bx) = \delta_{\bx,\bx^\true}$. In the second line of Eq.~\eqref{eq-cstateBME1}, it turns out that the optimal state estimator for the \textSigma SD cost is exactly the observer's original state of knowledge of the system's true configuration $\wp\c(\bx)$. However, this coincidence happens only when the true state is assumed to be a $\delta$-function, not necessarily for general cases where true states are of other forms. 

Second, for the negative equality cost, we need to define a $\delta$-functional for an equality between any two functions, $\delta\!\left[\wp, \wp^\true\right]$, which gives a positive value when the arguments are exactly the same and zero otherwise. The optimal state estimator that minimizes this expected negative cost is given by
\begin{align}\label{eq-cstate-ne}
\est\wp_{\rm nE}(\bx) = & \argmin_{\wp \in {\mathbb P}} \left\la -\delta\!\left[\wp, \wp^\true\right] \right\ra_{\wp^\true|{\rm D}}\!\!(\bx) \nonumber \\
=&\, \delta_{\bx,\est\bx_{\rm MLE}} ,
\end{align}
where $\est\bx_{\rm MLE} = \argmax_{\bx} \wp\c(\bx)$ from Eq.~\eqref{eq-mleprobmax} in the previous subsection. To understand the result in the second line above, we use the definition of the expected cost in Eq.~\eqref{eq-confest-int2} to obtain the expected cost,
\begin{align}\label{eq-cstate-necost}
\left\langle -\delta\!\left[\wp, \wp^\true\right] \right\rangle_{\wp^\true|{\rm D}} = &\, -\!\!\!\sum_{\bx' \in \mathbb X} \wp\c(\bx') \, \delta\!\left[\wp, \delta_{\bx,\bx'}\right],
\end{align}
where $\delta\!\left[\wp, \delta_{\bx,\bx'}\right]$ is zero unless the argument $\wp(\bx)$ is also a pure ($\delta$-function) state of the same $\bx'$. Since there is also the weight $\wp\c(\bx)$ in Eq.~\eqref{eq-cstate-necost}, we can conclude that the optimal state that minimizes this expected cost is a $\delta$-function state with a true configuration that maximizes $\wp\c(\bx)$, as stated in the second line of Eq.~\eqref{eq-cstate-ne}.

\subsection{Quantum state estimation}\label{sec-QSE}
Following classical state estimation for a discrete configuration $\bx$, let us consider a quantum system described similarly by a discrete set ${\mathbb X}$ of basis states. The observer's knowledge of the system is represented by a state matrix (also called a density matrix) $\rho$, which we also refer to as a \emph{quantum state}. Diagonal elements of the quantum state equal the probabilities that the system can be found in the basis states, for example when projective measurements are performed on the system. If the quantum state is diagonal, it can be written as a sum of basis projectors with corresponding probabilities, $\rho_{\rm diag} = \sum_{\bx} \wp(\bx) |\bx\ra \la \bx|$. %where $\wp(\bx)$ is the classical state for the discrete configuration $\bx$.

Maximum knowledge of the quantum system is represented by a quantum state with unit purity. Using the notation $\hat\psi \equiv |\psi\ra\la \psi|$ for the projector onto a pure quantum state $|\psi\ra$ in a Hilbert space ${\mathbb H}$, let us denote a pure \emph{true quantum state} by $\hat\psi^\true$. This may be unknown analogously to an unknown true pure classical state $\wp^{\rm true}(\cx) = \delta_{\cx, \cx^{\rm true}}$. The quantum state estimation can therefore be formulated as to find an estimator from the set of possible quantum states. We can define an \emph{optimal estimate of the quantum state} that minimizes an expected cost with the true state,
\begin{align}
\est\rho = \argmin_{\rho \in \Gset} \left\la c\left[\rho, \hat\psi^{\rm true}\right] \right\ra_{\hat\psi^\true | {\rm D}}.
\end{align}
Here we have defined the set of valid state  matrices as $\Gset$, which can be represented as a convex subset of ${\mathbb R}^{2d-1}$~\cite{holevo2001statistical}. For example,  for a qubit, the set is the Bloch ball. The expected cost is defined on the Hilbert space of pure quantum states,
\begin{align}
\left\la c\left[\rho, \hat\psi^\true \right] \right\ra_{\hat\psi^\true | {\rm D}} \!\equiv \!\! \int\!\! \dd \mu_{\rm H}(\hat\psi)\, \wp\c(\hat\psi) \, c\left[\rho, \hat\psi\right],
\end{align}
using the Haar measure $\dd\mu_{\rm H}(\hat\psi)$ as the measure of the integral, with a conditioned PDF of pure states $\wp\c(\hat\psi)$. %It is also important to note that, this expected average can be viewed as either an average over a functional space of pure quantum states, or an average over a set of ``points" on the continuous-value Hilbert space, where these points have one-to-one mapping to distinct pure quantum states of the system. 

As before, we consider two examples for the cost functions. The quantum state estimator that minimizes an expected Trace Square Deviation (TrSD) cost is given by
\begin{align}\label{eq-q-sd}
\est\rho_{\rm TrSD} = & \argmin_{\rho \in {\Gset}} \left\la {\rm Tr} \left [ \left(\rho - \hat\psi^\true \right)^2 \right] \right\ra_{\hat\psi^\true | {\rm D}} \nonumber \\
 = & \left\langle \hat\psi^\true \right\rangle_{\hat\psi^\true | {\rm D}},
\end{align}
where, comparing with Eq.~\eqref{eq-cstateBME1}, the inner product becomes a trace of a square of difference between two quantum states, and the estimator is a conditioned average of all possible true states. For the negative equality cost, we define a $\delta$-functional for pure quantum states using the Haar measure, $\delta_{\rm H}\!\left[\rho, \hat\psi^\true \right]$, giving an infinite value when the arguments are the same and zero otherwise. The quantum state estimator that minimizes the expected negative equality cost is,  
\begin{align}\label{eq-q-ne}
 \est\rho_{\rm nE} = & \argmin_{\rho \in {\Gset}} \left\la -\delta_{\rm H}\!\left[\rho, \hat\psi^\true \right] \right\ra_{\hat\psi^\true | {\rm D}} \nonumber\\
=& \argmin_{\hat\psi\in {\Hset}}\, [- \wp\c(\hat\psi)], \nonumber \\
=& \argmax_{\hat\psi\in {\Hset}}\, \wp\c(\hat\psi),
\end{align}
which is a conditioned most-likely (pure) state. The proof of the second and third lines is similar to the classical case, where the $\delta$-functional picks out a pure state and the argmin chooses a pure state that maximizes the probability distribution $\wp\c(\hat\psi)$.

\subsection{Hidden record estimation}\label{sec-RE}
Following the previous two subsections, if the classical and quantum states to be estimated live in large-dimension functional spaces, with specific constraints such as the Hermitian and positivity constraints, the averages over all possible true states can become overly complex. Therefore, in some cases, it is preferable to devise a simpler set of book-keeping variables which can transform the measure of the integral of the state variables to that of simpler parameters, such as unknown measurement records that determine the system's true states. However, this is possible only if there exist the records that can be unambiguously mapped to all possible true states of interest.

In this work, we consider the case for continuously monitored systems, where there exists an unknown \emph{true record} $R^\true$ that can unambiguously determine, at any time, a true classical configuration $\bx^\true = \bx_{R^\true}$ and its corresponding true classical state $\wp^\true(\bx) = \delta_{\bx, \bx_{R^\true}}$, or a true quantum state $\hat\psi^\true = |\psi_{R^\true}\ra \la \psi_{R^\true}|$. Let us discretize the total time of measurement into $d_r$ timesteps and describe the measurement record as a string of $d_r$ real-continuous variables, where its realization is denoted by $R = \{ R_1, R_2, ... , R_{d_r} \}\tp$, that is, $R \in {\mathbb R}^{d_r}$. Consequently, we can replace the average over all possible true states by an average over all possible hidden measurement records with appropriate probability measures. For example, an average of a quantum state function ${\cal A}[\hat\psi]$ can be written in two ways,
\begin{subequations}\label{eq-r-ave}
\begin{align}
\left\la {\cal A} \right\ra_{\hat\psi^\true | {\rm D}} \equiv & \int \!\! \dd \mu_{\rm H}(\hat\psi)\, \wp\c(\hat\psi) \, {\cal A}\!\left[\hat\psi \right]\\
 = & \int \!\! \dd\mu( R)\, \,\wp\c(R) \, {\cal A} \!\left[\hat\psi_{R} \right] \equiv  \left\la {\cal A} \right\ra_{R^\true | {\rm D}},
\end{align}
\end{subequations}
where $\hat\psi_R$ is a true state corresponding to a possible hidden record $R$, and, in the second line, the usual multi-variable measure of the integral is defined as $\int\! \dd\mu(R) \equiv \int \prod_{k =1}^{d_r} \dd R_k$. The first line is the average over all pure quantum states $\hat\psi$ with the specific Haar measure, and is equivalent to the average over all possible records that determine the pure states in the second line. %given that $\dd \mu_{\rm H}(\hat\psi')\, \wp\c(\hat\psi') =\prod_{k =1}^{d_r} \dd R'_k\, \,\wp\c(R')$ is satisfied.

In addition to the above advantage in transforming the measures, we can also formulate another type of optimal estimation based on the costs defined in hidden record variables. That is, we define an \emph{optimal estimate of the measurement record} as an estimator that minimizes an expected cost function,
\begin{align}\label{eq-gencostx}
\est R = \argmin_{R \in {\mathbb R}^{d_r}} \left\langle c\left(R,R^\true \right) \right\rangle_{R^\true|{\rm D}},
\end{align}
where the expected cost is
\begin{align}
\left\la c(R, R^\true) \right\ra_{R^\true | {\rm D}} \equiv \!\int\!\! \dd\mu(R') \, \wp\c(R')\, c(R, R').
\end{align}
This is similar to the estimation of unknown configurations because the unknown measurement records are simply classical variables.

Therefore, an optimal estimator for the hidden record that minimizes an expected square deviation cost can be defined as
\begin{align}
\est R_{\rm SD} = & \argmin_{R \in {\mathbb R}^{d_r}} \left\la \left(R - R^\true \right)\tp \left(R - R^\true \right) \right\ra_{R^\true | {\rm D}} \nonumber \\
 = & \left\langle R^\true \right\rangle_{R^\true | {\rm D}}.
\end{align}
The hidden record estimator that minimizes an expected negative equality cost is  
\begin{align}
\est R_{\rm nE}  = & \argmin_{R\in {\mathbb R}^{d_r}} \left\la -\delta^{(d_r)}\!\left(R - R^\true \right) \right\ra_{R^\true | {\rm D}} \nonumber \\
=& \argmin_{R\in {\mathbb R}^{d_r}}\,[ -\wp\c(R)], \nonumber \\
=& \argmax_{R\in {\mathbb R}^{d_r}}\, \wp\c(R),
\end{align} 
which is the maximum likelihood record given the conditioned PDF $\wp\c(R)$.

It is important to note here that, since we have assumed that the hidden records can determine true configurations and true states, we can use the record estimators $\est R$ derived above to compute estimators for: configurations $\bx_{\est R}$ and classical states $\delta_{\bx,\bx_{\est R}}$, or quantum states $|\psi_{\est R} \ra \la \psi_{\est R} |$. However, one has to keep in mind that these estimators are not ``optimal'' in the configuration or state spaces; the optimality is defined in the record space.

For the rest of the paper, we consider a dynamical system under continuous measurements, where we define each measurement record as a string of $n = T/\dt$ measurement results, from an initial time $t_0 =0$ to a final time $T$ with an infinitesimal time resolution $\dt$.  Given complete information of the system at the initial time, a true configuration/classical state, or a true quantum state, at any intermediate time $\tau \in (0, T]$, can be determined from a complete knowledge of true records up to that time. However, in the scenario that the complete knowledge of measurement records are not possible, for example there are observed and unobserved records, we will then use our concepts of estimation with defined cost functions to assign optimal estimates of those quantities. 

Throughout this paper, we follow the notations for measurement records with overhead arrows introduced in Ref.~\cite{Ivonne2015}, but with a slight modification. Denoting $R_t$ as a measurement result acquired between time $t$ and $t+\dt$, we then define: (a) the \emph{past} record $\past R_\tau = \{ R_t : t \in [\tin, \tau)\}\tp$ for a string of measurement results from the initial time to an estimation time $\tau$, (b) the \emph{future} record $\futp{R}_\tau = \{ R_t : t \in [\tau, T]\}\tp$ for a string of results from the estimation time $\tau$ to the final time (including a result at the final time), and (c) the \emph{past-future} record $\bothp{R} = \{ R_t : t \in [\tin,T] \}\tp$ for a string of results of all times. For convenience, we also define a future and past-future record excluding a measurement at the final time $T$. We denote such a record with an open-faced arrow \ie, $\fut R_\tau = \{ R_t : t \in [\tau, T)\}\tp$ and $\both R = \{ R_t : t \in [\tin, T) \}\tp$, respectively. The subscript $\tau$ on these records is omitted unless needed for clarity. Note, these superscript arrows are defined in a slightly different way from those in Refs.~\cite{Ivonne2015,LavCha2019,LavCha2020a,GueWis20,LavCha2020b}. In those papers, generic arrowheads, such as $\overleftrightarrow{R}$, were used as there was no need to distinguish between a record including a measurement at the final time $T$ and one without. All other conventions remain the same.

%%%%%%%%%%%%%%%%%%%
% Section III
%%%%%%%%%%%%%%%%%%%
\section{Existing formalisms and their roles in partially observed systems}\label{sec-existing}
%The idea of using past and future information in quantum mechanics has been proposed as a time-symmetric interpretation of quantum mechanics. However, it has been developed mostly to solve the problem that quantum mechanics is incomplete. As mentioned in the introduction, the past-future information has already been implemented in classical estimation technique, to maximally use available information in the process of estimating unknown quantities. 

In this section, we briefly overview the three main existing quantum formalisms that utilize past-future 
information. We present the formalisms in a chronological order, which handily allows us to introduce the 
observed record (two-state formalism), the unobserved record (the most-likely path formalism), and both 
records (quantum state smoothing formalism), one at a time. 

\subsection{Two-state formalisms}\label{sec-wv}
\subsubsection{Two-state vector formalism}
%The first work in this category is the two-state vector formalism (TSVF) \cite{AhaRoh91}, 
%which introduced the use of both forward (from the past) and backward (from the future) evolving quantum 
%states to completely describe a quantum system at an intermediate time. 

%While time-symmetric 
%formulations of quantum mechanics have been discussed informally by physicists as early as in the 1920s~\cite{Schottky1921,Eddington1928}, 
%the first formal theory, {\blu referred to as the \emph{double inferential state-vector} formalism}, was provided

The notion of a time-symmetric formulation of quantum mechanics can be traced as far back as 1928, in a remarkable footnote  by Eddington~\cite{Eddington1928} (our italics):
\begin{quote}
``The probability is often stated to be proportional to $\psi^2$ ... The whole interpretation is very obscure, but it seems to depend on whether you are considering the probability after you know what has happened or the probability for the purpose of prediction. The $\psi^2$ is obtained by {\em introducing two symmetrical systems of $\psi$-waves travelling in opposite directions in time}; one of these must presumably corresponds to probable inference from what is known (or is stated) to have been the condition at a later time. Probability necessarily means ``probability in the light of certain given information," so that the probability cannot possibly be represented by the same function in different classes of problems with different initial data''.
\end{quote} 
While one might expect $|\psi|^2$ here, rather than $\psi^2$, it should be born in mind that this was a book for a popular audience, and one published less than two years after Born introduced the $|\psi|^2$ probability rule (for scattering problems), also in a footnote~\cite{Born26}. 
%DETAILS: "I.2".  Max Born, Zur Quantenmechanik der Stoßvorgänge [On the quantum mechanics of collisions]. Zeitschrift für Physik. 37. (!926). In Wheeler, J. A.; Zurek, W. H. (eds.), {\em Quantum Theory and Measurement} Princeton University Press (published 1983). pp. 863–867}. 

This idea of Eddington seems to have been forgotten, but was discovered again and investigated in full detail 
%approach to quantum theory, however, was not investigated until 
in the 1950s %over two decades later, 
by Watanabe, 
as his \emph{double inferential state-vector} formalism~\cite{Watanabe1955,Watanabe1956}.  Watanabe introduced a backward-evolving (from the future) state vector, referred to as the \emph{retrodictive} state, to be used in conjunction with the conventional forward-evolving (from the past) state vector, referred to as the \emph{predictive} state, to completely describe a quantum system at intermediate times.  This theory was subsequently rediscovered as the \emph{two-state vector} formalism (TSVF) by Aharonov, Bergmann and Leibwitz \cite{ABL1964} in the 1960s, receiving considerably more attention and debate
\cite{VAA1987,AhaRoh91,AV91,Aharonov1998,AV2002,Qi2010,Yang2013,DanVaid13,Aharonov2014,CampagneI2014,Hashmi2016,Nowakowski2018} 
than its predecessor.

Most notably, the TSVF led to the formulation 
of weak values~\cite{AAV1988} by Aharonov, Albert, and Vaidman in 1988. They considered a \emph{weak} 
measurement of a quantum system, by coupling the system to another system, called the measuring device or probe,  with preselection (before 
the weak measurement) and postselection (after) conditions on the system's quantum state. Specifically, the 
original work~\cite{AAV1988} considered a weak measurement of an observable described by an 
Hermitian operator $\op{X}$, via a weak von-Neumann-type coupling to a probe with a Gaussian wavefunction at some time $\tau$~\cite{BookVon1932},  
 on a system prepared in state $|\psi_i\ra$ and subsequently subjected to a 
projective measurement onto state $|\psi_f \ra$. The weak value, defined as 
\begin{align}\label{eq-wv-orig}
%{\cal X}_w  = 
\frac{ \la \psi_f | \op{X} | \psi_i \ra }{\la \psi_f | \psi_i \ra },
 \end{align} 
 is a complex quantity which is sufficient to describe the change in the Gaussian wavefunction of the probe conditioned on 
the preselected and postselected states of the system, $\hat\psi_i$ and $\hat\psi_f$, respectively.  This is known as an anomalous 
weak values when it lies outside the eigenvalue range of the quantum observable $\op X$ which is measured. See Refs.~\cite{Kofman2012,Tamir2013,Aharonov2014-2,Dressel2014} for some recent reviews on this topic.
Weak values have been investigated for their use in quantum metrology or tomography~\cite{Hosten2008,Starling2009,Brunner2010,Hofmann2011a,Lundeen11,Kedem2012,Viza2013,Dressel2014,Jordan2014,Knee2014,Salazar2014,Gross2015,VMH15,Zhang2015,Knee2016,Steinberg2017,Ren2020} and for illuminating numerous other phenomena in quantum theory and its interpretations~\cite{RohAha02,Wiseman2007-2,Brunner2004,Mir07,Lundeen09,Yokota09,Dressel2011,KBR2011,Pryde2011,RozSte12,HWP13,Kaneda14,HWP15,MRF2016,YX2017,XW2019,Ramos2020}.
%AND THIS: https://www.nature.com/articles/s41586-020-2490-7
%AND the references therein about earlier (theory and experiment) uses of weak values in tunnelling time. 
The phenomenon of weak values have now been demonstrated in many experiments (some already cited above)~\cite{Hulet1997,Brunner2004,PryWis2005,Mir07,Lundeen09,Yokota09,Dressel2011,Hofmann2011c,KBR2011,Pryde2011,RozSte12,HWP13,Kaneda14,HWP15,SDG2015,MRF2016,Vaidman2017-2,YX2017,XW2019,Ramos2020}.

\subsubsection{Weak measurements and weak-values}
While there have been debates~\cite{Johansen2004,Jozsa2007,Hosoya2010,DreAga2010,Dressel2012,Dressel2015,Hall2016} on the physical meaning of the above complex weak 
value in Eq.~\eqref{eq-wv-orig}, its real part 
has a simple operational meaning as the conditioned average of the weak measurement result. Consider the 
 probe's   state to be Gaussian as in Ref.~\cite{AAV1988} and denote the measurement result and its 
possible value by $X^w$ and $x$, respectively. Then one can write the measurement operator~\cite{nielsen2010quantum,BookWiseman,BookJacobs}, 
also called a Kraus operator~\cite{BookKraus}, describing the 
measurement backaction on the system's state as
\begin{align}\label{eq-wv-kraus}
\op K_{x} =&\, (\epsilon/2\pi)^{1/4}\exp\left\{ - \epsilon(x - \op X)^2/4 \right\} \nonumber \\
\approx & \, \sqrt{\wp_0(x)} \left \{ \hat 1 + \tfrac{1}{2} \epsilon \, x \hat X - \tfrac{1}{4}\epsilon \hat X^2 + \tfrac{1}{8} \epsilon^2 x^2 \hat X^2\right\}.
\end{align}
Here $\wp_0(x) = (\epsilon/2\pi)^{1/2} \exp(-\epsilon x^2/2)$ is a zero-mean Gaussian function.  
The limit of a weak measurement is when $\epsilon 
\rightarrow 0$. This limit is best understood from the second line of Eq.~\eqref{eq-wv-kraus}, 
which is the expansion of the operator parts of $\op{K}_x$ up to first order in $\epsilon$. 
The last term, which on the face of it is ${\cal O}(\epsilon^2)$, is in fact ${\cal O}(\epsilon)$ because 
$x^2$ is typically of order $1/\epsilon$. This follows from the form of $\wp_0(x)$ and the fact that, in this limit, the PDF 
for the result $x$,
$\wp(x|\hat\psi_i) =  ||\hat K_x \ket{\psi_i}||^2$,  %is, to zeroth order in $\epsilon$, just 
approaches $\wp_0(x)$, 
which is independent of the initial system state $\hat\psi_i$.  

The completeness relation $\int \!\! \dd x \hat K_{x}^\dagger \hat K_{x} 
= \hat 1 + {\cal O}(\epsilon^n)$ is satisfied, to all orders of $\epsilon$ ($n = \infty$) for the first line of 
Eq.~\eqref{eq-wv-kraus}, and to first order of $\epsilon$ ($n = 2$) for the second line. Using the Kraus operator, 
we can calculate the conditional PDF of the weak measurement result  with both preselection and postselection as 
\begin{align} \label{preposPDF}
\wp(x | \hat\psi_i, \hat\psi_f)  \propto  \,  
\frac{|\la \psi_f | \hat K_{x} | \psi_i \ra |^2}{\wp(\hat\psi_f|\hat \psi_i)},
\end{align} 
where  $\wp(\hat\psi_f|\hat \psi_i) =  \int \!\! \dd x\,|\la \psi_f | \hat K_{x} | \psi_i \ra |^2$   is independent of $x$. 
 From \erf{preposPDF}, it is not 
difficult to  show that the average  result,  over many trials, conditioned on the preparation and the final postselection, 
 and in the limit $\epsilon\rightarrow 0$, is 
\begin{align}\label{eq-wv-avg}
_{\hat\psi_f}\!\la X^w \ra_{\hat\psi_i} =&  \lim_{\epsilon \rightarrow 0}  \int_{-\infty}^\infty \!\! \dd x \,  x \,\wp(x|
\hat\psi_i,\hat\psi_f) \nonumber \\
= & \, {\rm Re}\, \frac{ \la \psi_f | \op{X} | \psi_i \ra }{ \la \psi_f | \psi_i \ra },
\end{align}
which is the real part of the weak value. We have used $_{\hat\psi_f}\!\la \cdots 
\ra_{\hat\psi_i}$ as an expectation conditioned on the preselected and postselected states,  following the 
notation of Ref.~\cite{Wise2002}, with the $X^w$  
denoting the  result of a weak measurement of $\hat X$, rather than the weak value of $\hat X$ as is common 
in the literature. In this limit, the average back-action from the 
weak measurement is negligible so that $\wp(\hat\psi_f|\hat \psi_i) \approx |\la \psi_f | \psi_i \ra |^2$. Thus we can see that 
averaging the weak value over a complete set of final states with this probability gives the usual quantum expectation value, 
as expected: 
\begin{align} 
\la X^w \ra_{\hat\psi_i} \,&= \, \sum_{\psi_f}  \, \wp(\hat\psi_f|\hat \psi_i) \, \times \,   _{\hat\psi_f}\!\la X^w \ra_{\hat\psi_i} \nn \\
&\approx  \,\sum_{\psi_f} \,  \la \psi_i | \psi_f \ra \la \psi_f | \op{X} | \psi_i \ra  \,=\, \la \psi_i | \op{X} | \psi_i \ra. \label{sumoverfin}
\end{align}  %\, {\rm Re}\,

%We have also used $\wp(A_w|\hat\psi,\hat\phi)$ as the conditioned probability for the detector readout $A_w$. 
 
The weak value formula, Eq.~\eqref{eq-wv-avg}, involving pure states for preselection and postselection, has some unique properties that are evident in experiments with weakish-strength measurements~\cite{Vai14}. Nevertheless, it is very natural to consider generalizing Eq.~\eqref{eq-wv-avg} to weak measurement with mixed states for the preselection and 
postselection, and also for a non-Hermitian measurement coupling~\cite{Wise2002,Tsangsmt2009-2,DreAga2010,Gammelmark2013}. Say the system's measured 
observable can be written as $\op X = \op c + \op c^\dagger$, where $\hat c$ is now the operator describing the 
system-detector interaction. By this we mean that the Kraus operator is generalized from Eq.~\eqref{eq-wv-kraus} to 
\begin{align}\label{eq-wv-kraus-gen}
\hat{K}_x \approx & \, \sqrt{\wp_0(x)}\left \{ \op{1}  - i \epsilon \hat H + \epsilon \, x \op{c}  - \tfrac{1}{2} \epsilon \, \op{c}^\dagger \op{c}  - \tfrac{1}{2} \epsilon\, \op{c}^2 + \tfrac{1}{2} \epsilon^2 x^2 \op{c}^2 \right\},
\end{align}
where $\hat H$ is Hermitian. Again, this satisfies the completeness relation to first order: $\int \!\! \dd x \hat K_{x}^\dagger \hat K_{x} 
= \hat 1 + {\cal O}(\epsilon^2)$. 
 Say  the system's state just before the time $\tau$ of the weak measurement  is $\rho$ (which can be 
mixed), and the final measurement (which can be non-projective) is described by a positive operator (POVM 
element) $\op{E}$~\cite{BookHelstrom,BookKraus,BookWiseman}  at the time just after the weak measurement time $\tau$.  Then the PDF conditioned 
on the preparation $\rho$ and the postselection positive operator $\hat E$ (also known as an {\em effect}~\cite{BookDavies,BookKraus,BookWiseman}) 
%PDF
 is given by
\begin{align}\label{eq-wv-prob-gen}
\wp(x | \rho, \op{E})  =  \frac{\Tr{ {\op E} {\hat K}_{x}\rho \hat K_{x}^\dagger}}{\wp(\hat E|\rho)},
\end{align} 
where  $\wp(\hat E|\rho)=  \int \!\! \dd x\, \Tr{\hat E \hat K_x \rho \hat K_x\dg }$.  
Following a similar calculation to Eq.~\eqref{eq-wv-avg}, the conditioned average of the detector readout is 
found to be
\begin{align}\label{eq-wv-gen}
 _{\op{E}}\la X^w \ra_{\rho} = \,2\,{\rm Re}\,\frac{{\rm Tr}\left( \op{E} \,\op c\, \rho \right) }{{\rm Tr}\left( \op{E}  \rho\right)},
\end{align}
to lowest order in $\epsilon$.  This reduces to Eq.~\eqref{eq-wv-avg}, for the case of pure-state preparation, 
$\rho = |\psi_i \ra \la \psi_i|$, rank-one final measurement $\op{E} \propto |\psi_f \ra \la \psi_f|$, 
and Hermitian weak coupling $\op c = \op c^\dagger \equiv \op X/2$.  
  Note that, for no postselection, we can replace $\op E$ with the identity and 
Eq.~\eqref{eq-wv-gen} reduces to the usual expectation value, $\la X^w \ra_{\rho} = {\rm Tr}\left( \op X \rho 
\right)$, which justifies considering Eq.~\eqref{eq-wv-kraus-gen} to describe a measurement of $\op X$. 
 Similar to the average in \erf{sumoverfin}, this no-postselection expectation value can also be obtained by averaging 
the weak value with the appropriate probability for the final measurement result, using the POVM completeness condition $\sum_{\hat E}\hat E = \hat{1}$. 

\subsubsection{Weak-value states}

 For the case for Hermitian weak coupling,  $\op c = \op c^\dagger \equiv \op X/2$, 
where the Kraus operator, Eq.~\eqref{eq-wv-kraus-gen}, reduces to Eq.~\eqref{eq-wv-kraus}, the 
general expression, Eq.~\eqref{eq-wv-gen}, for the weak value becomes 
\begin{align}\label{eq-wv}
 _{\op{E}}\la X^w \ra_{\rho} = \frac{ {\rm Tr} [  \op{X} (\op{E}  \rho + \rho \op{E}) ] }{{\rm Tr}( \op{E}  \rho + \rho \op{E})}.
\end{align}
From Eq.~\eqref{eq-wv}, we see that the 
conditioned average of the weak (Hermitian) measurement results can be written as an expectation value of the 
observable $\op X$ and an Hermitian matrix as $ _{\op{E} }\la X^w \ra_{\rho}   = \Tr{ \op{X} \varrho\wvs}$, 
where the symmetrized matrix is given by~\cite{Tsangsmt2009-2,Gammelmark2013,LavCha2019}
\begin{align}\label{eq-wvs}
\varrho\wvs = \frac{\op{E} \rho+ \rho\op{E}} {\Tr{\op{E}\rho + \rho \op{E}}}.
\end{align}
This has been referred to as a Smoothed Weak-Value (SWV) state~\cite{LavCha2019,LavCha2020a,LavCha2020b}; the 
relation to smoothing will become clear in our main results in Section~\ref{sec-sevenQSE}. Note that  $
\varrho\wvs \in { {\mathfrak G}'({\mathbb H})}$, { where ${\mathfrak G}'({\mathbb H})$ is a superset of ${\mathfrak G}({\mathbb H})$ made by dropping the positivity requirement}. Thus $\varrho\wvs$ not generally positive-semi definite, and therefore cannot represent a proper quantum state.
This is necessary because, as noted earlier, the weak value can be outside the eigenvalue range of the 
observable $\op X$, which is in contrast to the expectation value, ${\rm Tr}\left( \op X \rho \right)$, for a proper 
positive quantum state $\rho$~\cite{Tsangsmt2009-2}.

\subsubsection{Related two-state formalisms}\label{sec-reltwostate}

In this work, we are interested in the case where, as well as the  measurement at the intermediate time $\tau$,  there 
is an \emph{observed} continuous measurement on the system, before and after $\tau$. Therefore, the 
preselected state is replaced by a \emph{filtered state}, $\rho = \rho_{\pasts{\bo}}$, computed from the past 
observed record up to time $\tau$ using the usual quantum trajectory approach~\cite{BookCarmichael,BookWiseman,BookJacobs}. 
Similarly, the postselected retrodictive effect is replaced by a \emph{retrofiltered effect}, $\op{E} = \op{E}_{\futps{\bo}}$, that 
is a POVM element giving the probability of the future observed record. 
%\begin{align}\label{eq-retroprob}
%{\rm Tr}\left(\op{E}_{\futps{\bo}}\, \rho \right) \red = 
%\wp(\hat E_{\futps{\bo}}|\rho_{\pasts{\bo}})  \equiv \wp\left(\futp{\bo}\,|\red \pasts{\bo}\right)  , 
%\end{align}
%given \red the filtered state $\rho_{\pasts{\bo}}$.  
The retrofiltered effect is a solution of an adjoint equation of the 
quantum trajectory, evolving backward in time given the record $\futp \bo$. Gammelmark, Julsgaard  and M\o lmer have introduced the terminology ``past quantum state'' for the pair $\{ \rho_{\pasts{\bo}}, \op{E}
_{\futps{\bo}}\}$~\cite{Gammelmark2013}. As well as weak values, they considered a measurement of arbitrary 
strength at the intermediate time $\tau$,  and calculated its probability distribution conditioned on the past-future 
observation using the pair of operators as in Eq.~\eqref{eq-wv-prob-gen}. However, 
unlike in the weak-measurement limit of Eq.~\eqref{eq-wv-gen}, one cannot use the approximation 
that  $\wp(\hat E_{\futps{\bo}}|\rho_{\pasts{\bo}})  \approx \Tr{\hat E_{\futps{\bo}} \rho_{\pasts{\bo}}}$. 
%in Eq.~\eqref{eq-wv-prob-gen}. 
%but without the approximation (valid for weak intermediate measurements)  that $\wp(\hat E|\rho)  = \Tr{\hat E\rho}$. 
 This general two-state formalism  
%The past quantum state 
has been 
applied to the special case of linear Gaussian quantum systems~\cite{ZhaMol17,Huang2018}, 
to investigate non-Markovianity in quantum systems~\cite{Budini2018a,Budini2019}, %quantum teleportation~\cite{}, 
 and to many other theoretical 
topics~\cite{GreMol15,GreMol16,GreMol17,Gough2020},  %. Furthermore, the theory of past quantum state has been applied 
as well as  to  experiments in cavity-QED~\cite{RybMol15}, circuit-QED~\cite{TanMol15,TanMol16,FTM16,TanMol17}, and large atomic-ensembles~\cite{BaoMol20a,BaoMol20b}.  % It was also shown in Ref.~\cite{Gammelmark2013} that, in the weak measurement limit, the conditioned average does converges to the weak value of the form in Eq.~\eqref{eq-wv-gen}.

%Originally introduced as a tool to investigate statistics of quantum trajectories for diffusive continuous quantum measurement, the most-likely path formalism proposed by Chantasri, Dressel and Jordan (CDJ) \cite{Chantasri2013,chantasri2015stochastic} presents an optimization approach for a quantum state's path that maximizes a joint PDF of the diffusive records. The formalism is used to solve for quantum paths that correspond to maximum-likelihood records between any two quantum states at two different times. Therefore, this can be considered as a kind of optimal estimation conditioned on the past (initial conditions) and future (final conditions) information. 

\subsection{Most-likely path formalism}\label{sec-cdj}
The most-likely path formalism was proposed by Chantasri, Dressel and Jordan (CDJ) 
\cite{Chantasri2013,chantasri2015stochastic}, as a tool to investigate statistics of quantum trajectories for 
diffusive time-continuous quantum measurement. This measurement can be considered as a series of weak 
measurements with their strength determined by the time resolution for acquiring the measurement record, \ie, 
$\epsilon \propto \dt $.  Given the record, one can compute the corresponding quantum 
state evolution (or quantum trajectory) of the measured system, via a Kraus operator of the form in 
Eq.~\eqref{eq-wv-kraus-gen}~\cite{BookKraus}. The resulting quantum trajectories are called diffusive because they 
realize a diffusive stochastic process in the quantum state space~\cite{BookCarmichael,BookGardiner1,BookWiseman}. The most-likely path formalism has been verified in circuit-QED experiments~\cite{Weber2014,ChaKim2016,JorMur2017} and has been used in studying entanglement generation~\cite{ChaKim2016,Silveri2016}, quantum chaos~\cite{Lewalle2018chaos}, and quantum correlations of non-commuting observables~\cite{JorCha2016,ChaAta2018}, with time-continuous quantum measurements.

\subsubsection{Continuous diffusive measurement}\label{sec-contmeas}

The CDJ most-likely path formalism, based on techniques in classical stochastic processes, yields a 
deterministic optimization problem for a quantum state's path to maximize the PDF for the diffusive records (\ie, 
giving maximum-likelihood records) between any two quantum states at two different times. Therefore, in the 
language adopted in this paper, we can treat the diffusive measurement record as \emph{unknown or 
unobserved}. The most-likely path formalism can be considered as a kind of optimal estimation for the unknown 
record and its corresponding quantum state, conditioned on the past (initial condition) and future (final 
condition) information. 

 Let us discretize the time into steps of size $\ddt$, where the continuous time range $[\tin, T]$ becomes a set 
 of discretized times $\{ \tin, \ddt, 2\ddt,..., T\}$. We then denote the unknown measurement record by $\both 
 \bu = \{ \bu_t : t\in \{ \tin, \ddt, ..., T-\ddt \} \}\tp$, where $\bu_t$ is a measurement result acquired between time 
 $t$ and $t+\ddt$.  The reason we have defined the unobserved record excluding a measurement at the final time will become clear later. Given a possible realization of the unknown record denoted by $\both u = \{ u_t : t\in \{ \tin, 
 \ddt, ..., T-\ddt \}\}\tp$, one can compute the system's state dynamics conditioned on the record from an update 
 equation,
\begin{align}\label{eq-stup}
\rho_{t+\ddt} =& \frac{{\cal M}_{u_t} \rho_t }{ {\rm Tr}( {\cal M}_{u_t} \rho_t  ) },
\end{align}
where the completely positive map ${\cal M}_{u_t}$ is a function of $u_t$. Denoting the initial state of the 
system by $\rho_0$, we can compute the system's state at any time $\tau$,
\begin{align}\label{eq-qtrajmap}
\rho_\tau = \rho_{\,\pasts{u}} = &  \frac{{\cal M}_{u_{\tau-\ddt}} \cdots {\cal M}_{u_0} \rho_0}{{\rm Tr}\left({\cal M}
_{u_{\tau-\ddt}} \cdots {\cal M}_{u_0} \rho_0 \right)}.
\end{align}
In this subsection, we only consider the system's dynamics arising from  a Hamiltonian $\hat H$  plus the 
measurement backaction from conditioning on the unknown record $\both u$. Therefore, the completely 
positive map becomes a purity-preserving measurement operation defined as ${\cal M}_{u_t} \bullet = \hat{M}
_{u_t} \bullet \hat{M}_{u_t}^\dagger$. Here
\begin{align}\label{eq-moput}
\hat{M}_{u_t}\!  =  \op{1}  - i \, \ddt \op H + \ddt \, u_t \op{c} - \tfrac{1}{2} \ddt \, \op{c}^\dagger \op{c}   - \tfrac{1}
{2} \ddt \, \op{c}^2 + \tfrac{1}{2} \ddt^2 u_t^2 \op{c}^2,
\end{align}
is the so-called diffusive measurement operator,  with high-order terms introduced in Refs.~\cite{Amini2011,Rouchon2015}.  Apart from the Hamiltonian term, this operator is simply obtained from 
the general (not necessarily Hermitian) weak-measurement Kraus operator in Eq.~\eqref{eq-wv-kraus-gen} by taking $\epsilon = \ddt$ and defining 
$\hat{K}_{u_t} = \sqrt{\wp_0(u_t)}\, \hat{M}_{u_t}$. Following common terminology~\cite{BookWiseman}, 
we call $\wp\ost(u_t)  \equiv \wp_0(u_t)$ the \emph{ostensible} probability:  
\beq \label{postend}
\wp\ost(u_t) = (\ddt/2\pi)^{1/2}  \exp(- u_t^2\ddt/2). 
\eeq
The completeness relation for this measurement operator is given by $\int \!\! \dd u_t \, \wp\ost(u_t) \hat 
M_{u_t}^\dagger \hat M_{u_t} = \hat 1 + {\cal O}(\ddt^2)$.

%$\{ \qq_t : t \in [\tin, T] \}\tp$, where a vector $\qq\ik$ parametrizes the quantum state $\rho\ik$ at time $t$, \eg, a Bloch vector for a qubit state. The state update an update equation denoted by $\qq\ikd = \bm{{\cal E}}(\qq\ik, u\ik)$,  This so-called measurement operator describes the unitary dynamics from the system's Hamiltonian $\hat{H}$ and the measurement backaction on the system between time $t$ and $t+\dt$.

\subsubsection{Quantum state paths with most-likely unknown records}

The goal of the CDJ formalism is to find the most-likely record estimator, which can be formulated as
\begin{align}\label{eq-originalmlp}
\both\bu_{\rm CDJ}^\star  = \argmax_{\boths{u}} \, \wp\c(\,\both{u}\,),
\end{align}
where the PDF of unknown records $\both u$ is given by \cite{Chantasri2013}
\begin{align}\label{eq-mlpprob}
&\wp\c(\,\both{u}\,)  \dd\mu(\,\both{u}\,)  =  \int \!\! \dd\mu(\,\bothp{\qq}\,) \, {\cal B}\c \! \prod_{t=\tin}^{T-\ddt} \dd u\ik \, \wp(u\ik | \qq\ik) \delta[\qq_{t+\ddt} - 
\bm{{\cal E}}(\qq\ik, u\ik)].
\end{align}
Here we have defined the vector $\qq_t$ that parametrizes the quantum state matrix, such that $\rho_t = {\hat 
S}(\qq\ik)$ for a state-operator function $\hat S$. An example is $\qq_t = \{ x_t , y_t, z_t \}\tp$, which is a Bloch 
vector for a qubit state matrix, for which ${\hat S}(\qq\ik) = \tfrac{1}{2}(\hat 1 +  \qq\ik \cdot \hat {\bm \sigma} $) 
and the vector of Pauli matrices is $\hat {\bm \sigma} = \{ \hat \sigma_x, \hat \sigma_y, \hat \sigma_z \}\tp$. We 
also defined the measure $\dd\mu(\,\bothp{\qq}\,) = \prod_{t =\tin}^{T} \dd \qq_{t}$ for the path integral over the 
state vectors and the measure $\dd\mu(\,\both{u}\,) = \prod_{t=\tin}^{T-\ddt} \dd u_t$ for the unknown records. This 
measure is to be understood in any PDFs of the unknown records in the rest of the paper. 

% via \red
%\begin{align}\label{eq-gellmann}
%\rho_t = \tfrac{1}{d} \hat I + \tfrac{1}{2} \hat{\bm \Lambda}\cdot \qq\ik,
%\end{align}
%using the generalized Gell-Mann matrices denoted by $ \hat{\bm \Lambda} = \{ \op{\Lambda}_1, \op{\Lambda}_2, \cdots, \op{\Lambda}_{d^2-1} \}\tp$~\cite{NLevelBloch}. These matrices satisfy $\hat \Lambda_j = \hat \Lambda_j^\dagger$, ${\rm Tr}\{ \hat \Lambda_j \} = 0$, and ${\rm Tr}\{ \hat \Lambda_j \hat \Lambda_k \} = 2 \delta_{j,k}$, for $j,k \in \{1, 2,..., d^2-1\}$, and can be used to define the Bloch vector for a $d$-level system. For qubit states, the matrices are the Pauli matrices. For the state matrix in Eq.~\eqref{eq-gellmann}, the corresponding state vector is given by $\qq\ik = \{ {\rm Tr}(\op{\Lambda}_1 \rho\ik ), {\rm Tr}(\op{\Lambda}_2 \rho\ik ), ..., {\rm Tr}(\op{\Lambda}_{d^2-1} \rho\ik ) \}^\top$. %$\op{\Lambda}_1=\op{\sigma}_x$, $\op{\Lambda}_2=\op{\sigma}_y$, and $\op{\Lambda}_3=\op{\sigma}_z$. 

The term $\wp(u\ik | \qq\ik)$ in the product of Eq.~\eqref{eq-mlpprob} is the probability of a time-local unknown 
measurement result $u\ik$, conditioned on a state $\qq\ik$ right before its measurement. We have $\wp(u\ik | 
\qq\ik) = \wp\ost(u_t) {\rm Tr}[ {\cal M}_{u_t} \hat S(\qq_t) ]$ for the map in Eq.~\eqref{eq-moput}. The delta-function term $\delta[\qq_{t+\ddt} - \bm{{\cal E}}(\qq\ik, u\ik)]$ is to ensure that the quantum states at all times 
satisfy an update equation denoted by $\qq_{t+\ddt} = \bm{{\cal E}}(\qq\ik, u\ik)$, which describes the evolution 
of the system's state in Eq.~\eqref{eq-stup} in terms of the vector $\qq\ik$. The boundary condition term ${\cal 
B}\c$ takes care of any constraints imposed on the evolution. 

\subsubsection{Calculational techniques and boundary conditions}

In the original work~\cite{Chantasri2013}, the CDJ approach was used for fixed boundary states at the initial 
and final times. The solution is a quantum state path and a measurement record that maximize the probability 
distribution in Eq.~\eqref{eq-mlpprob} under the following constraints: the correct evolution $\qq_{t+\ddt} = 
\bm{{\cal E}}(\qq\ik, u\ik)$ for $t \in \{ \tin, \ddt, ..., T-\ddt \}$; and the boundary conditions, $\qq_0 = \qq_I$ and $
\qq_T = \qq_F$ corresponding to ${\cal B}\c = \delta(\qq_0 -\qq_I)\delta(\qq_T-\qq_F)$. This final condition is the reason the past-future unobserved record is defined on the open interval $[0,T)$. The technique uses the 
Lagrange multiplier method, where in this case the Lagrangian function is
\begin{align}
{\cal S} =&  - \pp_{-\ddt}(\qq_0 - \qq_I) - \pp_{T}(\qq_T - \qq_F) + \sum_{t=\tin}^{T-\ddt} \left\{ - \pp\ik  [ \qq_{t+\ddt} - \bm{{\cal E}}(\qq\ik, u\ik) ]  + \ln \wp(u\ik | \qq\ik) \right\},
\end{align}
given the multipliers $\pp\ik$ for times $t \in \{ -\ddt, \tin, \ddt, ..., T \}$. By extremizing this Lagrangian function 
over its variables $\qq\ik$, $\pp\ik$, and $u\ik$, one arrives at difference equations describing an optimal 
solution, 
\begin{subequations}\label{eq-original-diffeq}
\begin{align}
\qq_{t+\ddt} = & \,\, \bm{{\cal E}}( \qq\ik , u\ik), \\
\pp_{t-\ddt} =&\,\,  \bm{\nabla}_{\qq\ik} \big[ \pp\ik \cdot \bm{{\cal E}}(\qq\ik, u\ik ) + \ln \wp (u\ik | \qq\ik)\big], \\
0 = & \,\, \frac{\partial}{\partial u\ik}   \big[ \pp\ik \cdot \bm{{\cal E}}(\qq\ik, u\ik)+ \ln \wp( u\ik | \qq\ik) \big] ,
\end{align}
\end{subequations}
where $\bm{\nabla}_{\qq\ik}$ denotes the gradient for the $d^2-1$ dimensional vector $\qq_t$ at time $t$. 
Eqs.~\eqref{eq-original-diffeq} can be solved with the boundary condition $\qq_0 = \qq_I$ and $\qq_T = \qq_F$. 
In the time-continuum limit $\ddt\rightarrow \dt$, the equations become a set of differential equations, which can 
be solved to yield a differentiable quantum path $\qq\ik$ from Eq.~\eqref{eq-original-diffeq}. This is the quantum 
trajectory given the most-likely record $\both u$. There can also exist multiple solutions, local minima and 
maxima of the PDF $\wp\c(\,\both{u}\,)$, where at least one of the solutions is the most-likely record $\both\bu_{\rm 
CDJ}^\star$. A thorough analysis for multiple solutions of the most-likely path can be found in Lewalle \etal~\cite{LewCha17}.

\subsection{Quantum state smoothing formalism}\label{sec-qss}
The final existing approach that uses past-future information for quantum systems is quantum state smoothing 
introduced by Guevara and Wiseman~\cite{Ivonne2015}. It was inspired by Ref.~\cite{Armen2009}, which applied a type of classical smoothing theory \cite{Sarkka13} (posterior decoding) to a quantum system, by treating the system semiclassically.  
Quantum state smoothing theory  was proposed as a way to generalize classical state smoothing  to the  fully  quantum realm, while guaranteeing a positive-semidefinite smoothed quantum state.  The smoothed quantum state is defined under a scenario of a partially observed quantum system, where the 
system of interest is subjected to two time-continuous monitoring channels. One is accessible by an observer 
(Alice), giving an \emph{observed} record $\bothp{\bo}$. The other is not accessible by her, but is assumed 
observed by an omniscient observer (Bob), giving an \emph{unknown} record $\bothp{\bu}^\true \equiv \bothp{\bu}$. 
Note that the `true' superscript indicates that the record is known to Bob, but unknown to Alice. The omniscient 
Bob also has access to Alice's record and therefore can construct a true quantum state of the system, $
\rho^\true_\tau = \rho_{\pasts{\bo}, \pasts{\bu}}$; as mentioned in Section~\ref{sec-RE}, a true quantum state can 
be computed using only the past portion of the complete records up until time $\tau$.  The central idea of quantum state smoothing~\cite{Ivonne2015} is that Alice uses her  future record $\futp{O}$ (as well as her past record $\past{O}$) to make inferences about Bob's past record $\past{U}$, 
and by appropriately averaging $\rho_{\pasts{\bo}, \pasts{\bu}}$ can obtain a smoothed quantum state, $\rho_{\bothps{O}}$, conditioned on the past-future information. This is explained in detail in the subsections below. 

The technique of quantum state smoothing has been applied to qubit  
systems~\cite{Ivonne2015,chantasri2019,LavCha2020b,GueWis20,LavGueWis21} 
as well as linear Gaussian quantum systems 
\cite{LavCha2019,LavCha2020a,LavCha2020b,Laverick21}. Numerous issues related to the quantum state smoothing theory have been investigated, including how the observers Alice and Bob should choose their unravellings to best demonstrate the efficacy of smoothing~\cite{chantasri2019,LavCha2020b}, 
the relation between the smoothed state, the true state, and classical smoothing~\cite{LavCha2020a,LavWarWis21}, and %looking at subensembles to illustrate 
 measures of closeness between the true state, filtered state, and smoothed state in individual stochastic trajectories \cite{GueWis20,LavGueWis21}.  
%how the future information can make the smoothed state very different from the filtered quantum state for particular instances of Bob's record~\cite{GueWis20}. 
Quantum  state  smoothing has also been illustrated in the simpler regime of a single discrete-time measurement by Budini~\cite{Budini2018b}. The same author also considered the case where, in place of an unknown record, there is a unknown true state of a classical stochastic system interacting with the quantum system of interest~\cite{Budini2017}, a sort of hybrid quantum--classical  state  smoothing. 
%systems, \ie, a quantum system attached to a classical register. In his theory, the classical register acts as an unobserved measurement record, analogous to the unknown measurement record (see the subsequent paragraph for a brief introduction) in quantum state smoothing.   

\subsubsection{Partially observed systems with observed and unknown records}

%The task for Alice, with only the observed record, is to estimate the true state of the system at any time $\tau$ at best she can. To simplify our notations for the rest of the paper, we will use a lower-case $\bothp u$ for a realization of the unknown record $\bothp \bu$.

The observer Alice can use the traditional quantum trajectory approach to compute the filtered state at time $
\tau$, which comes from tracing out the unknown bath that contains the unknown measurement records. 
Defining an additional measurement operation describing the backaction from the observed record, ${\cal M}
_{O_t} \bullet \equiv {\hat M}_{O_t} \bullet {\hat M}_{O_t}^\dagger$, and an unconditioned map for the unknown 
record, $e^{\ddt{\cal L}_{\rm u}} \bullet \equiv \int \dd u_t \, \wp\ost(u_t) {\cal M}_{u_t} \bullet$, the filtered state 
at time $\tau$ is given by
\begin{align}\label{eq-state}
 \rho\fil \equiv \rho_{\pasts{\bo}}  =  \frac{{\cal M}_{O_{\tau-\ddt}} e^{\ddt{\cal L}_{\rm u}}  \cdots {\cal M}_{O_{0}} e^{\ddt{\cal L}_{\rm u}}  \rho_0}{{\rm Tr} ({\cal M}_{O_{\tau-\ddt}} e^{\ddt{\cal L}_{\rm u}}  \cdots {\cal M}_{O_{0}} e^{\ddt{\cal L}_{\rm u}}  \rho_0)},
\end{align}
for the initial state $\rho_0$.
It can be shown~\cite{GamWis2005} that this is equivalent to
\begin{align}\label{eq-filstate}
\rho\fil = \int \!\! \dd\mu(\,\past{u}\,)\, \wp_{\pasts{\bo}}(\,\past{u}\,) \, \rho_{\pasts{\bo}, \pasts{u}}\,.
%\rho\fil \equiv \int \!\! {\cal D}\past u\, \wp_{\past \bo}(\past u) \rho_{\past\bo, \past u}.
\end{align}
That is, the filtered state can also be obtained by averaging the possible true states by integrating over all 
possible unknown records, with the weights given by the PDF conditioned on the past observed record 
$\wp_{\pasts{\bo}}(\,\past{u}\,) \equiv \wp\left(\past{u}\,| \,\past{\bo}\right)$. For the diffusive unknown record, the measure of the 
integration $\dd\mu(\,\past{u}\,) =\prod_{t=\tin}^{\tau-\ddt } \dd u_t$ is defined in a similar way as in Eq.~\eqref{eq-mlpprob}, but for the past record only. 

\subsubsection{From quantum state filtering to quantum state smoothing}

This form of filtered state Eq.~\eqref{eq-filstate}  naturally suggests that the observer could do better by using 
the past-future observed record to weight the true states. That is, it would be better to use the \emph{smoothed 
quantum state} at time $\tau$ defined as~\cite{Ivonne2015,chantasri2019},
\begin{align}\label{eq-qssstate}
\rho\sm \equiv \int \!\! \dd\mu(\,\past{u}\,)\, \wp_{\bothps{\bo}}(\,\past{u}\,)\,  \rho_{\pasts{\bo}, \pasts{u}}\,.
%\rho\sm \equiv \int \!\! {\cal D}\past u\, \wp_{\bothp \bo}(\past u) \rho_{\past \bo, \past u},
\end{align}
It has been shown numerically~\cite{Ivonne2015,chantasri2019,LavCha2019,GueWis20} that the smoothed quantum state 
gives an estimated state that is, on average, \emph{more pure} than the filtered quantum state. Moreover, it was shown analytically that this average purity for the smoothed (or filtered) state is equal to the average fidelity with the true state~\cite{Ivonne2015}. We note 
that the integrations in Eq.~\eqref{eq-filstate} and Eq.~\eqref{eq-qssstate} are defined in general and can be 
used for any type of unknown records. Also, the above can be easily generalized to the case where $
\bothp\bo$ and $\bothp{\bu}$ could represent multiple records obtained from different couplings to the system. 
% conditioned on the past-future observed record $\wp_{\bothp \bo}(\past u)$

\subsubsection{Calculational techniques}

The key problem of the quantum state smoothing is to compute the conditional PDF in Eq.~\eqref{eq-qssstate}. 
This can be calculated from an unnormalized true state ${\trho}_{\pasts{\bo},\pasts{u}}$ and the retrodictive 
effect~\cite{Ivonne2015,chantasri2019}.
The unnormalized state conditioned on measurement records (both observed and unknown) is calculated from 
a series of measurement operations applied to the system's initial state $\rho_0$, \ie, 
\begin{align}\label{eq-unnorm}
{\trho}_{\pasts{\bo}, \pasts{u}} = {\cal M}_{O_{\tau-\ddt}} {\cal M}_{u_{\tau-\ddt}} \cdots {\cal M}_{O_0} {\cal M}_{u_0} 
\rho_0.
\end{align}
This is similar to Eq.~\eqref{eq-qtrajmap}, but the map is defined with the measurement operations describing 
the backaction from both observed and unknown records, without the normalizing denominator. The retrodictive 
effect representing the statistics of the future record,  discussed in Section~\ref{sec-reltwostate}, %Eq.~\eqref{eq-retroprob}, 
is computed backward in time 
from the final identity matrix using adjoint operations, \ie,
\begin{align}\label{eq-povm}
{\hat E}_{\futps{\bo}} = e^{\ddt{\cal L}_{\rm u}^\dagger} {\cal M}_{O_\tau}^\dagger \cdots \, e^{\ddt{\cal L}_{\rm u}
^\dagger} {\cal M}_{O_{T-\ddt}}^\dagger  e^{\ddt{\cal L}_{\rm u}^\dagger} {\cal M}_{O_{T}}^\dagger \hat I,
\end{align}
for any time $\tau$. We have used the adjoint measurement operations for the observed record ${\cal M}
_{\bo\ik}^\dagger$ and an adjoint unconditioned map for the unknown record $e^{\ddt{\cal L}_{\rm u}^\dagger} 
\bullet$.

The point of using the unnormalized state is that its trace gives the probability of the conditioning measurement 
records, \ie, 
\begin{align}\label{eq-actualprob}
 \wp\left(\past{\bo}, \past{u}\right) \propto \wp\ost(\,\past{u}\,)\, {\rm Tr}\left( {\trho}_{\pasts{\bo},\pasts{u}} \right),
\end{align}
where $\wp\ost(\,\past{u}\,) = \prod_{t=\tin}^{\tau-\ddt} \wp\ost(u_t)$. Here $\wp\ost(u_t)$ is defined for diffusive 
measurements as in \erf{postend}, 
but there are analogous expressions for quantum jump trajectories. In \erf{eq-actualprob}, and subsequently, 
the proportionality sign ($\propto$) allows a factor that is independent of the unknown record, but may be a 
function of the observed record (which is fixed for any given estimation which Alice must perform). Using the 
Bayes' theorem, we obtain
\begin{align}
\label{eq-bothoprob1} \wp_{\pasts{\bo}}(\,\past{u}\,) =&\,\, \wp\left(\past{u} , \past{\bo}\right)/\wp(\past{\bo}\,)\nonumber \\
\propto &\,\,  \wp\ost(\,\past{u}\,) \, {\rm Tr}\left( {\trho}_{\pasts{\bo}, \pasts{u}} \right),
\end{align}
as the PDF of the past unknown record used for the filtered state in Eq.~\eqref{eq-filstate}, where the 
proportional factor is different from the one in Eq.~\eqref{eq-actualprob}. 

At this point, the reader may ask: Why bother with the second line of Eq.~\eqref{eq-bothoprob1} with its 
ostensible probabilities?
Why not use the first line, with the actual probabilities? That is, why not, for a given past observed record $\past{O}$, 
generate hypothetical past 
unobserved records $\past{u}$ simply by the usual quantum filtering theory (past-conditioned quantum 
trajectory theory)? 
The answer is that  this does not work~\cite{GamWis2005}.  If we were interested in only a single piece of record $O_t$ 
and $u_t$ in an infinitesimal 
time interval $[t,t+\ddt)$, then we could use the normalized quantum trajectory theory which simultaneously 
generates 
$O_t$ and $u_t$ with the correct statistics. However, while $\past{O}$ and $\past{u}$ are both  %notation $\past{O}$ and $\past{u}$ 
`past records' on an interval $[0, \tau)$,  %may hide the fact that 
parts of $\past{O}$ are in the future of parts of $\past{u}$. That is, to generate with the correct statistics 
$\past{u}$ given $\past{O}$, one would need to take into account how the later parts of $\past{O}$ affect the 
likelihood of the 
earlier parts of $\past{u}$, and standard quantum trajectories demonstrably does not do 
this~\cite{GamWis2005}. 
(If they did, they would automatically generate the smoothed quantum state, rather than the filtered quantum 
state as they do.) 

Thus it is necessary, whether using analytical or numerical methods, to consider   
% The answer is that when we want to generate a correct statistics of the past record given that the observed record is fixed. One has to be careful when considering statistics of a fictitious record (unknown record) conditioned on some known records (observed record)~\cite{GamWis2005}. This is because a fictitious result at any time $t$ is not only dependent on the known results in the past, but also strongly correlated with the known result at that same time. Therefore, the correct statistics for the unknown record can only be obtained via the 
unnormalized states, with the unobserved results generated according to an ostensible probability, with $u_t$ 
independent from $u_s$ for $t\neq s$. 
%  typically independently from each other. ] \
Once we have the filtered PDF $\wp_{\pasts{\bo}}(\pasts{u})$ from \erf{eq-bothoprob1}, the smoothed PDF is 
simply given by
\begin{align}
\label{eq-bothoprob2} \wp_{\bothps{\bo}}(\,\past{u}\,) = & \, \, \wp\left(\past{u} , \past{\bo}\right)\wp\left(\futp{\bo}\,|\, \past{u}, \past{\bo}\right)/\wp(\bothp{\bo}\,) \nonumber \\
 \propto & \,\,  \wp\ost(\,\past{u}\,)  \, {\rm Tr}\left( \op{E}_{\futps{\bo}}\, { \trho}_{\pasts{\bo}, \pasts{u}} \right), \nonumber \\
 \propto & \,\,  \wp_{\pasts{\bo}}(\,\past{u}\,) \, {\rm Tr}\left( \op{E}_{\futps{\bo}}\, { \rho}_{\pasts{\bo}, \pasts{u}} \right),
\end{align}
where %the last line is written in terms of the filtered PDF $\wp_{\past\bo}(\past u)$ and the normalized version of the true state 
${ \rho}_{\pasts{\bo}, \pasts{u}} = { \trho}_{\pasts{\bo}, \pasts{u}\,}/{\rm Tr}\left({ \trho}_{\pasts{\bo}, \pasts{u}}\right)$. 

There have been various techniques used in generating the conditional PDFs and in computing the integration 
in Eq.~\eqref{eq-qssstate} to obtain the smoothed quantum state. In the original work~\cite{Ivonne2015} (also 
\cite{chantasri2019}), the smoothed state for qubit examples was calculated numerically by simulating 
stochastic unnormalized states ${ \trho}_{\pasts{\bo}, \pasts{u}\,}$, with the appropriate ostensible probabilities (\eg, 
$\wp\ost(u\ik)$ for diffusive records). On the other hand, semi-analytical solutions for smoothed quantum states 
were possible for linear Gaussian systems as shown  by Laverick 
 \etal~\cite{LavCha2019},  where their closed-form solutions 
were derived and investigated. 
 
In this work, we introduce another method to compute the smoothed quantum state in Section~\ref{sec-example}. For a qubit system with particular dynamics and detection setups, the smoothed state can again be 
calculated via a semi-analytical approach, by directly computing a PDF of the unnormalized state conditioned 
on the past-future observed records. 

%The reason that this is true was shown in Ref.~\cite{LavGue2020}: the average infidelity with the true state is, for the estimates $\rho\fil$ and $\rho\sm$, equal to the cost function they minimize (this cost function will be revealed in the next section). Since $\bothp\bo$ contains strictly more information than $\past \bo$, and optimal estimate of the true state (for a given cost function) will, on average, be better if uses future information $\bothp\bo$.   

% It has also been shown that the smoothed state is an optimal estimator for the expected trace square deviation~\cite{LavGue2020}. 

%Therefore, we can use the definition of optimal state estimation in Section~\ref{sec-QSE} to show that the smoothed state is an estimator that minimizes an expected trace square deviation from true states, which will be discussed in the next section.

\section{Unifying theory of quantum state estimation}\label{sec-sevenQSE}

\begin{figure}[t]
\centering
\includegraphics[width=14cm]{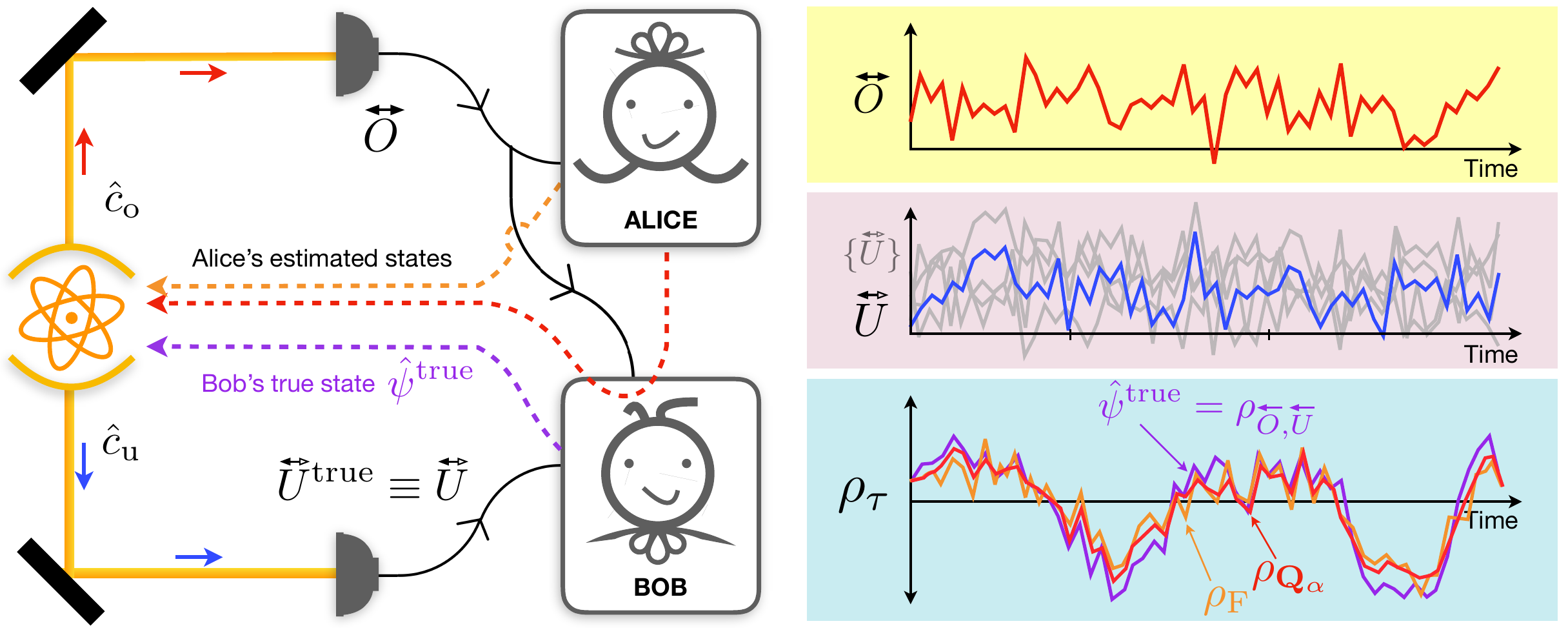}
\caption{ Schematic diagrams showing the Alice-Bob protocol for our unifying theory. (Left) A partially-observed quantum system (e.g., an atom) is coupled to two baths described by two Lindblad operators $\hat c_{\rm o}$ and $\hat c_{\rm u}$. Alice measures the first bath to obtain an observed record ${\protect\bothp{\bo}}$. Bob, who also has access to Alice's record, measures the second bath to obtain a record ${\protect \both{\bu}}$ (unobserved by Alice). Since Bob has complete information about the measurement records, he knows the true state of the quantum system (shown as the dashed purple arrow). Alice's task is to estimate Bob's true state as best she can, only using her observed record. The dashed orange arrow refers to Alice's estimation of the quantum state, without taking into account Bob's measurement settings (i.e.~the quantum filtered state in Eq.~\eqref{eq-filstate}). The dashed red arrow refers to Alice's estimation when she tries to estimate Bob's record and state using the past-future state estimation. (Right) Illustrative plots showing the observed record (the red signal in the top panel), Bob's true record (the blue signal in the middle panel), all possible records for Bob that Alice could have guessed (grey), Bob's true state (purple), the filtered state trajectory (orange), and some past-future state estimator (the red trajectory in the bottom panel).}
\label{fig-intro}
\end{figure}

Our unifying theory is based on the same scenario as in the quantum state smoothing formalism (see Figure~\ref{fig-intro}). That is, when a quantum system is partially observed by an observer, using the Alice-Bob protocol, we can build a framework that encompasses the existing formalisms, or at least generalizations of them. Given an observed record $\bothp{\bo}$, Alice's task is to try her best to guess the true state $\hat\psi^\true = |\psi_{\pasts{\bo},\pasts{\bu}}\ra \la \psi_{\pasts{\bo},\pasts{\bu}}| = \rho_{\pasts{\bo}, \pasts{\bu}}$, where the record $\both{\bu}$ is unknown to her. Note that we have excluded the measurement result at time $T$ in the unknown record for consistency with the most-likely path formalism. The best estimate of the true state is a state that optimizes some kind of cost function as in Section~\ref{sec-QSE} or \ref{sec-RE}. That is, the cost functions can be any functions (or functionals) defined for unknown variables, which include the unknown true state $\hat\psi^\true$, and the unknown record $\both{\bu}$. 

%\begin{figure}
%\centering
%\includegraphics[width=14cm]{Fig1-CostFDiagram2.pdf}
%\caption{Diagrams showing eight different expected costs, for $\textbf{Q}_1$--$\textbf{Q}_8$, which gives eight optimal estimators, connecting the existing formalisms (grey boxes): quantum state smoothing, the CDJ most-likely path formalism, and the Two-State vector formalism. \red The labels $(\textbf{Q}_1)$--$(\textbf{Q}_8)$ are colour-coded consistent with the colours used for estimators in the later plots and Table~\ref{tab-avecost}.. Expected costs in blue and pink boxes represent the costs defined in the quantum state space and the unknown record space, respectively. The connections are described in different lines and colored, following the discussion in the text. Since the observed record is fixed in the optimization, we omit the $\bo$-dependence in the definition of true states.}
%\label{fig-diagramQ}
%\end{figure}

\begin{figure}
\centering
\includegraphics[width=14cm]{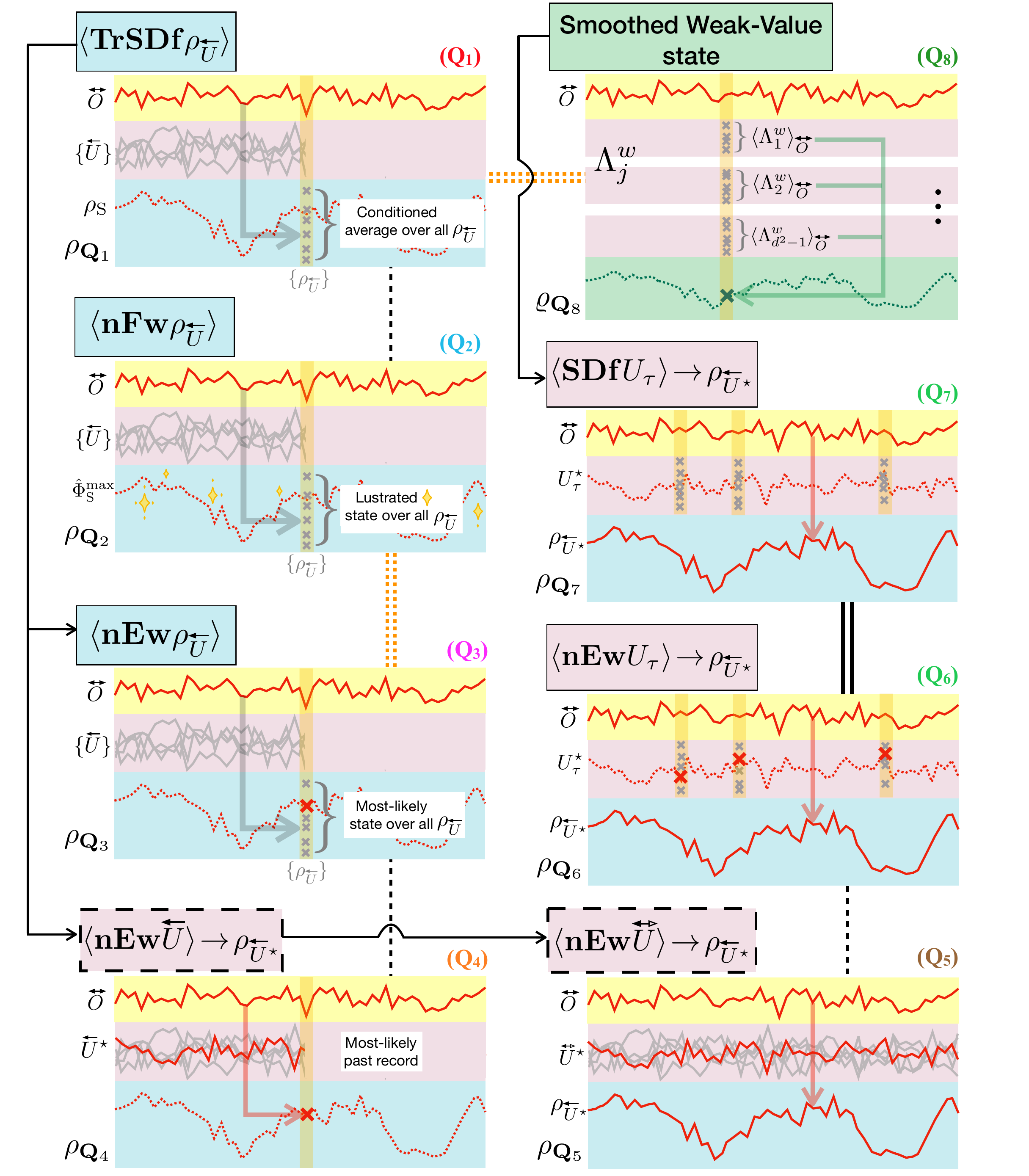}
\caption{ Diagrams showing eight different expected costs, $\textbf{Q}_1$--$\textbf{Q}_8$, and their schematic plots describing the calculations of the estimators. The labels $(\textbf{Q}_1)$--$(\textbf{Q}_8)$ are colour-coded consistent with the colours used for estimators in the later plots and Table~\ref{tab-avecost}. Yellow (top panels), pink (middle panels), and blue (bottom panels for $\textbf{Q}_1$--$\textbf{Q}_7$) backgrounds represent the observed record space, the unknown (unobserved) record space, and the quantum (state) estimator space, respectively. The green background (for $\textbf{Q}_8$) is used only for the SWV estimator, which is not necessarily a valid quantum state. The grey signals, $\{ {\protect \both{\bu} } \}$, indicate all possible past unobserved records. The grey crosses refer to all possible values (e.g, measurement results or states) at particular times. The red crosses refer to most-likely values. The dotted red state trajectories on blue backgrounds (bottom panels) for $\textbf{Q}_1$--$\textbf{Q}_4$ indicate that the estimated states are calculated locally at different times and are combined to get the whole trajectories. For $\textbf{Q}_5$--$\textbf{Q}_7$, the state trajectories are shown as solid red curves (in the bottom panels) indicating that they are calculated once from the record estimators (in the middle panels). The connections between different costs are described in different lines and colors, following Figure~\ref{fig-diagramQ} and the discussion in the text.  Although the trajectories and associated records look identical in each diagram, in reality they will be different in every case (except for $\textbf{Q}_6$ and $\textbf{Q}_7$, which yield the same).}
\label{fig-diagramQ2}
\end{figure}

We will focus on the most common distance measures as cost functions, and introduce in total eight quantum state estimators. We show, in Figure \ref{fig-diagramQ2}  (a more comprehensive version of Figure~\ref{fig-diagramQ}), a diagram summarizing our unifying theory. 
We explain the eight different state estimates operationally, in terms of the expected costs, labelled $\textbf{Q}_1$ to $\textbf{Q}_8$, and show how they are  connected or related (indicated by different types of connecting lines  or arrows).  We present the cost functions and their estimators in three subsections. The first three costs ($\textbf{Q}_1$--$\textbf{Q}_3$) are defined in the quantum state space (blue boxes  in Figure \ref{fig-diagramQ2}) and so  can be applied for any types of unknown records. The next four costs ($\textbf{Q}_4$--$\textbf{Q}_7$) are defined in the unknown record space (pink boxes  in Figure \ref{fig-diagramQ2})  and can be applied to only diffusive measurement records. The last subsection is the SWV state for $\textbf{Q}_8$, which is associated with a weak von Neumann measurement of an Hermitian observable. These costs are associated with distinct state estimators, making the complete connections among the three existing formalisms presented in the previous section.

%It might not be obvious how the formalisms can be unified under one theory. This is because they were all developed for quite different schemes. However, on a platform of a partially observed quantum system, the existing formalisms can be applied or used to arrive with state estimates, where each has its own cost function that it optimizes.

\subsection{Arbitrary-type unknown records ($\textbf{Q}_1$--$\textbf{Q}_3$)}
%\red [IF REF~\cite{LavGue2020} doesn't exist, we will need to say more here, but say about what?] 
Let us start with the smoothed quantum state on the top-left of the  diagrams
(\ie~both Figure~\ref{fig-diagramQ} and Figure \ref{fig-diagramQ2}).  
This was introduced in Ref.~\cite{Ivonne2015} as an analogue to the classical smoothed state.  
As prefigured in Section~\ref{sec-qss}, 
the smoothed quantum state, in Eq.~\eqref{eq-qssstate}, can be defined as the state that minimizes the expected Trace Square Deviation from the true state, \\ %~\cite{LavGue2020},\\
{\bf ($\textbf{Q}_1$) $\la$TrSDf$\rhoU\ra$:}
%\subsubsection{($\textbf{Q}_1$) $\la$TrSDf$\rhoU\ra$}
\begin{align}\label{eq-smt}
\est\rho_\tau = & \argmin_{\rho} \left\la {\rm Tr} \left [  \left(\rho - \rhoOU\right)^2  \right ] \right \ra_{\pasts{\bu} | \bothps{\bo}}\nonumber \\
= & \left\la \rhoOU \right\ra_{\pasts{\bu} | \bothps{\bo}} \equiv \rho\sm.
\end{align}
This follows from Eq.~\eqref{eq-q-sd}; see Ref.~\cite{LavGueWis21} for a more more detailed discussion, including another cost function that gives the same state. In Figure~\ref{fig-diagramQ2} (top left), we show illustrative plots explaining the calculation of the states $\rhoOU$ (grey crosses) from all possible $\pasts{\bu}$ (grey signals), where the smoothed state (an individual dot in the dotted red trajectory) is the conditioned average, $\left\la \rhoOU \right\ra_{\pasts{\bu} | \bothps{\bo}}$. 

Now, using the connection between two types of averages in Eq.~\eqref{eq-r-ave}, we can define the expected average of an arbitrary functional $\left\la {\cal A} \right\ra_{\pasts{\bu} | \bothps{\bo}}$ in two ways,
\begin{subequations}\label{eq-ave-alter}
\begin{align}
\label{eq-ave-alter1} \left\la {\cal A} \right\ra_{\pasts{\bu} | \bothps{\bo}} = & \int \!\! \dd\mu(\,\past{u}\,)\, \wp_{\bothps{\bo}}(\,\past{u}\,) \, {\cal A} \!\left[\rho_{\pasts{\bo}, \pasts{u}} \right]\\
\label{eq-ave-alter2}= & \int \!\! \dd \mu_{\rm H}(\hat\psi)\, \wp_{\bothps{\bo}}(\hat\psi) \, {\cal A}\!\left[\hat\psi \right].
\end{align}
\end{subequations}
The first line is a weighted average over all possible unknown records $\past{u}$, whereas the second line is a weighted average over all possible (pure) true states using a PDF of pure states conditioned on the past-future observed record, \ie,
\begin{align}\label{eq-bayes}
\wp_{\bothps{\bo}} (\hat\psi) = &\,\,   \wp\left(\hat\psi  | \past{\bo}\right)\wp\left(\futp{\bo}\, | \hat\psi\right)/\wp(\futp{\bo}\,)  \nonumber \\
\propto & \,\, \wp_{\pasts{\bo}}(\hat\psi)  \Tr{{\hat E}_{\futps{\bo}} \, \hat\psi},
\end{align}
as before in Eq.~\eqref{eq-bothoprob2}.
Therefore, the estimator of this trace-square-deviation cost function, the smoothed state, can also be written as an average over pure states,
\begin{align}\label{eq-qss-alter}
\rho\sm = \int \!\! \dd \mu_{\rm H}(\hat\psi)\, \wp_{\bothps{\bo}}(\hat\psi) \, {\hat\psi}.
\end{align}
We note that these two forms of conditioned averaged state, Eqs.~\eqref{eq-qssstate} and \eqref{eq-qss-alter}, will be useful when we define the estimators $\textbf{Q}_3$ and $\textbf{Q}_4$.

Following the dashed black line down from $\textbf{Q}_1$ in the diagrams of Figures~\ref{fig-diagramQ} and \ref{fig-diagramQ2}, since the trace square deviation is a distance measure between any two states, we can also consider another common distance measure, the quantum state fidelity, for a cost function. We use the Jozsa fidelity~\cite{Jozsa1994}, 
\begin{align}\label{eq-Jozsafid}
F[\rho\suba, \rho\subb] & \equiv \left[ \Tr{\sqrt{\rho\subb^{1/2} \rho\suba \rho\subb^{1/2}} } \right]^2,
\end{align}
since it has a direct connection to the fidelity for classical states Eq.~\eqref{eq-cjozsa} as will be discussed in Section~\ref{sec-classana}. Provided one of the state arguments in the fidelity function is a pure state, $\rho\suba = \hat\psi\suba$, the fidelity between $\rho\suba$ and $\rho\subb$ can be written as 
\begin{align}\label{eq-fidel}
F[\hat\psi\suba, \rho\subb] = {\rm Tr}\left(\hat\psi\suba \rho\subb \right).
\end{align}
Therefore, it is natural to define a new state estimator that minimizes a new cost function, the expected negative Fidelity with the true state,\\
{\bf ($\textbf{Q}_2$) $\la$nFw$\rhoU\ra$}:
\begin{align}\label{eq-th-q-nf}
\est\rho_\tau = & \argmin_{\rho} \left\la -F\!\left[\rho, \rhoOU \right] \right\ra_{\pasts{\bu} | \bothps{\bo}} \nonumber \\
= & \argmin_{\rho} \, [-{\rm Tr}\left( \rho \, \rho\sm\right)] \nonumber \\
= & \, |{\psi}\sm^{\rm max} \ra \la {\psi}\sm^{\rm max} | \equiv \hat {\psi}\sm^{\rm max}.
\end{align}
Here we have used Eq.~\eqref{eq-qss-alter} and \eqref{eq-fidel} to obtain the second line of Eq.~\eqref{eq-th-q-nf}, which leads to the optimal estimator that maximizes the trace ${\rm Tr}\left( \rho \, \rho\sm\right)$, namely the eigenprojector of the smoothed state $\rho\sm$ with the largest eigenvalue: $\rho\sm |{\psi}\sm^{\rm max} \ra = \lambda\sm^{\rm max}|{\psi}\sm^{\rm max} \ra$.  A more detailed treatment is given in  Ref.~\cite{LavGueWis21}, which calls $|{\psi}\sm^{\rm max} \ra$ the ``lustrated'' smoothed state. In Figure~\ref{fig-diagramQ2}, the illustrative plots for the calculation of the $\textbf{Q}_2$ state estimator  is very similar to that of $\textbf{Q}_1$; the only difference is that 
after it is calculated, the state estimator is made pure (lustrous) by the procedure just described.  Note, in the event that two or more eigenstates have equally large eigenvalues, any convex combination of the eigenstates will also be an optimal estimator for $\textbf{Q}_2$. %{ However, in this paper, we will only focus on cases with a unique pure-state solution.}
 Such cases form a set of measure zero;  in our numerical simulations, Sec.~\ref{sec-fpe},  they do not occur in any of the trajectories we show, and do not contribute to any of the averages we calculate. 

%This estimator was named the \emph{smoothed lustrated state} when it was introduced in Ref.~\cite{LavGue2020}. The estimator is a pure state and was coined a name \emph{lustrated smoothed state} for the first time in Ref.~\cite{Ivonnethesis}.

Moving down in the diagrams of Figures~\ref{fig-diagramQ} and \ref{fig-diagramQ2}, with the dotted orange line referring to an ``equal'' connection in classical state estimation (discussed later in Section~\ref{sec-classana}), we consider next the negative equality cost function in the state space. Using the definition of this cost in Eq.~\eqref{eq-q-ne}, we obtain an estimated state that minimizes an expected negative Equality with the true state,\\
{\bf ($\textbf{Q}_3$) $\la$nEw$\rhoU\ra$}:
\begin{align}\label{eq-th-q-ne}
\est\rho_\tau = & \argmin_{\rho} \left\la - \delta_{\rm H}\!\left[ \rho - \rhoOU \right] \right\ra_{\pasts{\bu} | \bothps{\bo}}\nonumber \\
= &  \argmin_{\hat\psi}\, [-\wp_{\bothps{\bo}}(\hat\psi)], 
\end{align}
which is the most-likely state using the PDF $\wp_{\bothps{\bo}} (\hat\psi)$ in Eq.~\eqref{eq-bayes}.
 In the event that the PDF $\wp_{\bothps{\bo}} (\hat\psi)$ has more than one most-likely state, any one of the states is an optimal solution to $\textbf{Q}_3$.  Unlike the case of degeneracy in $\textbf{Q}_2$ (above),  a convex combination of any of these states will be suboptimal.
We  stress that, even when they are non-degenerate, these cost functions do 
not give the same optimal state. That is, the most likely state (by the Haar measure) %in this case 
is not the same as the estimator in Eq.~\eqref{eq-th-q-nf} in general, even though they both are pure states.

In addition to the examples shown above, one could certainly come up with other types of distance measures in the true state space, and define new optimal estimators based on them. As stated before, in this work, we by no means aim to compile a comprehensive list of estimators. Thus it is time to move instead to estimators with optimality defined in the unknown measurement record space. 

\subsection{Diffusive-type unknown records ($\textbf{Q}_4$--$\textbf{Q}_7$)}\label{sec-sevenQSEdiff}
Given the two definitions of the smoothed state in Eq.~\eqref{eq-qssstate} and \eqref{eq-qss-alter}, if the weight in the state space $\wp_{\bothps{\bo}}(\hat\psi)$ can be used to find the most-likely state conditioned on the past-future observed record as in Eq.~\eqref{eq-th-q-ne}, we should also be able to use the weight in the record space $\wp_{\bothps{\bo}}(\,\past{u}\,)$ to find a similar ``most-likely'' estimator. Contrary to the averages, however, the peaks of a PDF are not preserved under a change of variables. Therefore, we do not expect the most-likely state to be the same as the state with the most-likely past record. Thus, the solid arrows from the smoothed state in $\textbf{Q}_1$, in the diagrams in Figures~\ref{fig-diagramQ} and \ref{fig-diagramQ2}, branch out to $\textbf{Q}_3$ (the former) and $\textbf{Q}_4$ (the latter). 

We formally define our new estimator as that which minimizes the expected negative Equality with the past Unknown record,\\
{\bf ($\textbf{Q}_4$) $\la$nEw$\past\bu\ra$}:
\begin{align}\label{eq-q4}
\est\rho_\tau =  \,\,  \rho_{\pasts{\bo},\est{\pasts{\bu}}} &, \text{  where}\nonumber \\
 \est{\past{\bu}}  \!\!\! =  \,\, \argmin_{\pasts{u}} & \left\la -\delta \left(\past{u} - \past{\bu} \right) \right\ra_{\pasts{\bu} | \bothps{\bo}} \nonumber \\
= \,\, \argmin_{\pasts{u}} & \, \,[-\wp_{\bothps{\bo}}(\,\past{u}\,)].
\end{align}
 Since the change of variables can affect the peaks of a PDF, one has to be clear to what measure of the PDF used in the $\argmax$ function in the last line of Eq.~\eqref{eq-q4}. For the most-likely state in Eq.~\eqref{eq-th-q-ne}, the Haar measure was chosen as a natural measure for pure quantum states. In the case of the unknown diffusive record, we choose the natural measure already defined in Eq.~\eqref{eq-mlpprob}. We also note that any constant scaling factors or offset added to the record does not affect the peaks of the PDF. 

% (\tfrac{1}{d} \hat I + \tfrac{1}{2} \hat{\bm \Lambda}\cdot \qq_{\tau})

%which is a quantum state with the most-likely past unknown record. 

To solve for the most-likely past record in Eq.~\eqref{eq-q4}, we generalize the CDJ technique of Section~\ref{sec-cdj}, where the boundary conditions are the initial-final states of the system, to boundary conditions that are functions of past-future observed records. Given that $\qq\ik$ represents the true state at time $t$, we realize that the conditional PDF for the unknown record is $\wp_{\bothps{\bo}}(\,\past{u}\,) \propto \wp(\,\past{u}, \past{\bo}\,) \wp(\futp{\bo} \,| \qq_\tau)$, where $\wp(\futp{\bo}\, | \qq_\tau) \propto {\rm Tr}[ {\hat E}_{\futps{\bo}\,} \hat S(\qq_\tau) ]$. Similarly to Eq.~\eqref{eq-mlpprob}, with discretized times, we write the probability function of the past unknown record, up to time $\tau$,
\begin{align}\label{eq-mlp1-prob}
\wp_{\bothps{\bo}} (\,\past{u}\,) &  \propto  \int \!\! \dd \mu(\,\past{\qq}\,) \,\, {\rm Tr}\left[ {\hat E}_{\futps{\bo}}\, \hat S(\qq_\tau) \right]  \prod_{t=\tin}^{\tau-\ddt} \wp( u\ik, \bo\ik | \qq\ik) \, \delta[\qq_{t+\ddt} - \bm{{\cal E}}(\qq\ik, u\ik, \bo\ik)],
\end{align}
allowing, as before, a proportional factor which is a function of the observed record. The measure of the integration is $\dd \mu(\,\past{\qq}\,)  = \prod_{t = \tin}^{\tau} \qq\ik$ and now, generalizing Ref.~\cite{Chantasri2013,chantasri2015stochastic} the boundary term can be identified as ${\cal B}\c =  \delta(\qq_0 -\qq_I) {\rm Tr}[ {\hat E}_{\futps{\bo}\,} \hat S(\qq_\tau) ]$. The condition on the past and future observed records is included in terms of $\bo\ik$ for $t\in \{ \tin, \ddt, ..., \tau-\ddt\}$ (past) and in the effect $\op{E}_{\futps{\bo}}$ for $\futp{\bo} = \{ O_t : t \in\{ \tau, \tau+\ddt, ..., T \}\}\tp$ (future), respectively. The quantum state $\qq\ik$ at any time needs to satisfy an update equation $\qq_{t+\ddt} =  \bm{{\cal E}}( \qq\ik , u\ik, \bo\ik )$, which describes the state evolution,
\begin{align}
{\rho}_{t+\ddt}=  \frac{{\cal M}_{O_t} {\cal M}_{u_t} \rho_t}{{\rm Tr}\big( {\cal M}_{O_t} {\cal M}_{u_t} \rho_t \big)},
\end{align}
modified from Eq.~\eqref{eq-stup} to include the measurement backaction from the observed record $\bo\ik$.

We follow the CDJ optimization process, maximizing the PDF under constraints. This gives the Lagrangian function as
\begin{align}
{\cal S} = & - \pp_{-\ddt}(\qq_0 - \qq_I) + \ln {\rm Tr}\left [  {\hat E}_{\futps{\bo}\,} \hat S(\qq_\tau) \right ]
+ \sum_{t=\tin}^{\tau-\ddt} \big\{ - \pp\ik  [ \qq_{t+\ddt} - \bm{{\cal E}}(\qq\ik, u\ik, \bo\ik) ] + \ln \wp(u\ik, \bo\ik | \qq\ik)  \big\},
\end{align}
with the Lagrange multipliers $\pp_t$. Extremizing the Lagrangian function, we arrive at a set of difference equations,
\begin{subequations}\label{eq-diffeq1}
\begin{align}
\qq_{t+\ddt} = & \,\, \bm{{\cal E}}( \qq\ik , u\ik, \bo\ik ), \\
\pp_{t-\ddt} =& \,\, \bm{\nabla}_{\qq\ik} \big[ \pp\ik \cdot \bm{{\cal E}}(\qq\ik, u\ik, \bo\ik ) + \ln \wp (u\ik, \bo\ik | \qq\ik)  \big], \\ 
\pp_{\tau-\ddt} = & \,\,\bm{\nabla}_{\qq_\tau} \ln {\rm Tr}\left[ {\hat E}_{\futps{\bo}\,} \hat S(\qq_\tau) \right],\\
0 = & \,\,   \frac{\partial}{\partial u\ik}  \big[ \pp\ik \cdot \bm{{\cal E}}(\qq\ik, u\ik, \bo\ik)+ \ln \wp( u\ik, \bo\ik | \qq\ik) \big] ,
\end{align}
\end{subequations}
which are slightly different from the original ones in Eq.~\eqref{eq-original-diffeq}.  The initial condition is fixed, $\qq_0 = \qq_I$, but the third line appears to be a new final boundary condition that depends on the observed record after time $\tau$ (instead of the final fixed state as in the original CDJ formalism~\cite{Chantasri2013,chantasri2015stochastic}). Solving the difference equations, Eqs.~\eqref{eq-diffeq1},  gives the most-likely past record $\{ u_0, u_{\ddt}, \cdots, u_{\tau-\ddt} \}$ and its associated state trajectory $\{ \qq_0, \qq_{\ddt}, \cdots, \qq_\tau \}$, for a particular value of $\tau$. 

It is important to note that the optimization in Eq.~\eqref{eq-q4} depends on the value of $\tau$, making the state estimator for $\textbf{Q}_4$ quite complicated to solve. The $\tau$-dependence is implicit in the definition of the past record $\past{\bu}$.  Its most-likely estimator varies in its length, from a single number for $\tau=\ddt$ to an entire record for $\tau = T$. Therefore, we have to solve Eqs.~\eqref{eq-diffeq1} and obtain the solutions of $\{ u_0, u_{\ddt}, \cdots, u_{\tau-\ddt} \}$ and $\{ \qq_0, \qq_{\ddt}, \cdots, \qq_\tau \}$ for every different value of $\tau$. The state estimator for $\textbf{Q}_4$ is then given by the collection of $\{ \qq_\tau : \tau \in\{ \ddt, 2\ddt, ..., T\} \}$, a final state of each solution of Eqs.~\eqref{eq-diffeq1}. In Figure~\ref{fig-diagramQ2}, the illustrative plots for $\textbf{Q}_4$ show the most-likely past record (red signal in the middle panel), which is used in calculating the state estimator (red cross) at time $\tau$.

Moving down in the diagram of Figure~\ref{fig-diagramQ} to the cost $\textbf{Q}_5$ takes us a step towards the original CDJ approach, where we optimize the entire unknown record, rather than the past portion of it as in $\textbf{Q}_4$. As a result, we have a state estimator that minimizes an expected negative Equality with entire Unknown records, \\
{\bf ($\textbf{Q}_5$) $\la$nEw$\both{\bu}\ra$}:
\begin{align}\label{eq-mlp}
\est\rho_\tau = & \,\, \rho_{\pasts{\bo}, \est{\pasts{\bu}}}, \text{  where}\nonumber \\ 
 \est{\past{\bu}} \!\!: \,\,\, \est{\both{\bu}}\!\!\! =  &\argmin_{\boths{u}} \, \left\la - \delta \left(\both{u} - \both{\bu} \right) \right\ra_{\boths{\bu} | \bothps{\bo}} \nonumber \\
 = &  \argmin_{\boths{u}} \, [-\wp_{\bothps{\bo}}(\,\both{u}\,)].
\end{align}
This is a direct generalization of the CDJ formalism, where the quantum dynamics now includes the measurement backaction of observed measurement records. 

 The conditional PDF for this case can be written as $\wp_{\bothps{\bo}}(\,\both{u}\,) \propto \wp\left(\both{u} , \both{O}\right)\wp(\bo_T | \qq_T)$. 
This gives
\begin{align}\label{eq-probmlp2}
\wp_{\bothps{\bo}}(\,\both{u}\,) & \propto \int \!\! \dd \mu (\,\both{\qq}\,)\,\, {\rm Tr}\left[ {\hat E}_T \hat S(\qq_T) \right]  \prod_{t=\tin}^{T-\ddt} \wp(u\ik, \bo\ik | \qq\ik) \delta[\qq_{t+\ddt} - \bm{{\cal E}}(\qq\ik, u\ik, \bo\ik)],
\end{align}
where we have used $\wp(\bo_T| \qq_T) = {\rm Tr}[ {\hat E}_T  \hat S(\qq_T) ]$ in terms of the effect $\hat E_T$ and state $\qq_T$ at the final time. The Lagrangian function for optimization is given by
\begin{align}\label{eq-action2}
{\cal S} = & - \pp_{-\ddt}(\qq_0 - \qq_I) + \ln {\rm Tr}\left[ {\hat E}_T \hat S(\qq_T) \right] + \sum_{t=\tin}^{T-\ddt} \big\{ - \pp\ik [ \qq_{t+\ddt} - \bm{{\cal E}}(\qq\ik, u\ik, \bo\ik) ]  + \ln \wp(u\ik, \bo\ik | \qq\ik)  \big\}.
\end{align}
 We then extremize the Lagrangian function to get
\begin{subequations}\label{eq-diffeq2}
\begin{align}
\qq_{t+\ddt} = &  \,\, \bm{{\cal E}}( \qq\ik , u\ik, \bo\ik ) ,\\
\pp_{t-\ddt} = & \,\,  \bm{\nabla}_{\qq\ik} \left[ \pp\ik \cdot \bm{{\cal E}}(\qq\ik, u\ik, \bo\ik )+ \ln \wp (u\ik, \bo\ik | \qq\ik)\right], \\  
\pp_{T-\ddt} = & \,\,\bm{\nabla}_{\qq_T} \ln {\rm Tr}\left[ {\hat E}_T \hat S(\qq_T) \right],\\
0 = & \,\, \frac{\partial}{\partial u\ik} \left[ \pp\ik \cdot \bm{{\cal E}}(\qq\ik, u\ik, \bo\ik) + \ln \wp( u\ik, \bo\ik | \qq\ik) \right],
\end{align}
\end{subequations}
where the third line serves as a final condition for this case. 

 We note that these difference equations, Eqs.~\eqref{eq-diffeq2}, are exactly the same as Eqs.~\eqref{eq-diffeq1} for $\tau = T$, when the past unknown record becomes the entire unknown record. Therefore, we only need to solve Eqs.~\eqref{eq-diffeq2} once and its solution $\{ \qq_0, \qq_{\ddt}, \cdots, \qq_T \}$ is the state estimator for $\textbf{Q}_5$ for all times $t \in \{ \tin, \ddt, ..., T\}$. This is in contrast to the laborious procedure required for ${\textbf{Q}_4}$. We show in Figure~\ref{fig-diagramQ2}  for ${\textbf{Q}_5}$ that the most-likely record (red signal in the middle panel) and its corresponding state trajectory (red trajectory in the bottom panel) can be calculated all at once (\ie,  we use solid curves) for all times.

Now that a clear connection between the smoothed quantum state in $\textbf{Q}_1$ and the generalized version of CDJ formalism in $\textbf{Q}_5$ has been established, we begin to extrapolate the cost function idea towards the two-state vector formalism. Moving up from $\textbf{Q}_5$ in the diagram of Figure~\ref{fig-diagramQ}, following the dashed black line up, we consider next the most-likely unknown record locally in time. In this case, the record estimator will be from optimizing a cost function independently at each time $\tau$, instead of the whole past record ($\textbf{Q}_4$) or the whole record ($\textbf{Q}_5$). We can then define a quantum state that is computed from a string of record estimators, where each minimizes an expected negative Equality with Unknown record at any time $\tau$, \\
{\bf ($\textbf{Q}_6$) $\la$nEw$\bu_\tau \ra$}:
\begin{align}\label{eq-q6}
\est\rho_\tau = & \,\, \rho_{\pasts{\bo}, \est{\pasts{\bu}}},  \text{   where}\nonumber \\
\est{\past{\bu}}\!\!\! :  \,\,\, \est{U}_\tau = & \argmin_{u_\tau}  \, \left\la - \delta \left(u_\tau - \bu_\tau \right) \right\ra_{\bu_\tau | \bothps \bo} \nonumber \\
= & \argmin_{u_\tau}\,[- \wp_{\bothps{\bo}}(u_\tau)].
\end{align}
Note the similarity in formulation to Eq.~\eqref{eq-mlp}. The PDF for an unknown result at time $\tau$ is given by
\begin{align}\label{eq-probut}
 \wp_{\bothps{\bo}}(u_\tau) = & \,  \wp\left(u_\tau |\,\past{\bo}\right)\wp\left(\futp{\bo}\,|u_\tau, \past{\bo}\right)/\wp(\,\futp{\bo}\,) \nonumber \\
 \propto & \,  \Tr{ {\op E}_{\futps{\bo}} {\hat M}_{u_\tau} \rho_{\pasts{\bo}} {\hat M}_{u_\tau}^\dagger},
\end{align}
where we have used the measurement operator ${\hat M}_{u_\tau}$ for a diffusive measurement result, defined in Eq.~\eqref{eq-moput}.%, and the proportionality factor is $1/\Tr{\op{E}_{\futp\bo} \rho_{\past\bo}}$. 

Having considered the negative equality cost for the local record, the next one, following up the double solid lines from $\textbf{Q}_6$ in the diagram, is the square deviation cost for the same parameters. We define a state path calculated from a string of record estimators that minimize an expected Square Deviation from Unknown records at any time $\tau$,\\
{\bf ($\textbf{Q}_7$) $\la$SDf$\,\bu_\tau\ra$}:
\begin{subequations}\label{eq-q7}
\begin{align}
\est\rho_\tau = & \,\, \rho_{\pasts{\bo}, \est{\pasts{\bu}}}, \text{   where}\nonumber \\
\est{\past{\bu}}\!\!\! :  \,\,\, \est{U}_\tau = & \, \argmin_{u_\tau}  \, \left\la \left(u_\tau - \bu_\tau \right)^2 \right\ra_{ \bu_\tau | \bothps{\bo}} \nonumber \\
= & \, \argmin_{u_\tau}  \, \left\la  - 2 u_\tau \bu_\tau + u_\tau^2 \right\ra_{ \bu_\tau | \bothps{\bo}} \label{eq-q-sd-ut-2} \\
= & \,\,  \la \bu_\tau \ra_{ \bu_\tau | \bothps{\bo}} \,,
\end{align}
\end{subequations}
where the average in the last line is defined with the same probability weight in Eq.~\eqref{eq-probut}. %The record estimator in this case is exactly what was used in the past quantum state formalism \cite{Gammelmark2013}, computing an estimate of a hidden measurement result. 

 The reason that we use the double solid lines connecting $\textbf{Q}_6$ and $\textbf{Q}_7$ is because they are equivalent for the continuous diffusive measurement. In the time-continuum limit, a record $u_\tau$ at any time $\tau$ is acquired during an infinitesimal time between $\tau$ and $\tau+\dt$ and is then regarded as a result of a weak measurement of an observable $\op U = \op c + \op c^\dagger$.  As mentioned in Section~\ref{sec-contmeas}, the record statistics is described by a measurement operation in Eq.~\eqref{eq-moput}, for a Lindblad operator $\hat c$ describing the coupling between the system and its bath. It can be shown that the mean and the mode (most-likely value) of the PDF $\wp_{\bothps{\bo}}(u_\tau)$ in Eq.~\eqref{eq-probut} for this weak measurement are the same to first order of $\dt$ (see~Appendix~\ref{sec-app-pdf} for its derivation). Therefore, the record  estimators of $\textbf{Q}_6$ and $\textbf{Q}_7$ coincide and are given by the real part of the weak value, 
\begin{align}\label{eq-wv-gen2}
\la \bu_\tau \ra_{\bu_\tau | \bothps{\bo}} \approx  \frac{\Tr{ \op{E}_{\futps{\bo}}\, \op{c} \, \rho_{\pasts{\bo}} + \rho_{\pasts{\bo}} \, \op{c}^\dagger \, \op{E}_{\futps{\bo}}}}{ \Tr{ \op{E}_{\futps{\bo}}\, \rho_{\pasts{\bo}}}},
\end{align}
 as per Eq.~\eqref{eq-wv-gen}, to leading order in $\dt$. In Figure~\ref{fig-diagramQ2}, we show in the illustrative plots for $\textbf{Q}_6$ and $\textbf{Q}_7$ that the record estimators (shown as the dotted red signals in the middle panels) are the most-likely and the mean measurement results, respectively. The state estimators for both cases are the state trajectories (solid red trajectories in the bottom panels) calculated from the record estimators.

\subsection{Unknown result of the weak von Neumann measurements ($\textbf{Q}_8$)}

For completeness of our proposed quantum state estimation theory, we construct a cost function that leads to the smoothed weak value (SWV) state defined in Eq.~\eqref{eq-wvs} for the weak value in Eq.~\eqref{eq-wv}. Considering the weak von Neumann measurement of an Hermitian observable $\op{X}$ at time $\tau$, the real part of the weak value, which can be written as an expectation of the SWV state,
\begin{align}
{\rm Re} \frac{\Tr{ \op{E}_{\futps{\bo}} \, \rho_{\pasts{\bo}} \op X}}{\Tr{ \op{E}_{\futps{\bo}} \,\rho_{\pasts{\bo}}}} = & \, \Tr{\varrho_{\rm SWV} \op X} \nonumber \\
= &  \, _{\op{E}_{\futpss{\bo}}}\,\!\!\la X^w \ra_{\rho_{\pastss{\bo}}} \equiv  \la X^w \ra_{X^w | \bothps{\bo}} , 
\end{align}
is also equivalent to the past-future conditioned average of the weak measurement results $X^w$. Therefore, we can consider the expectation value $\Tr{\varrho_{\rm SWV} \op X}$ as an estimator that minimizes an expected square deviation from true (but unknown) weak measurement results. However, knowing the expectation value for one observable $\op X$ cannot uniquely determine the full description of the SWV state. We need the weak values for all observables from the operator algebra for the Hilbert space. It suffices to consider the set of generalized Gell-Mann matrices $\{ \op\Lambda_j \}$ for $i = 1, 2, ..., d$ (where $d$ is the dimension of the quantum system), satisfying the orthonormal property $\Tr{{\op\Lambda}_i {\op\Lambda}_j} = 2 \delta_{i,j}$~\cite{NLevelBloch}. 

The SWV state is an estimator $\varrho \in { {\mathfrak G}'({\mathbb H})}$ that minimizes an expected square deviation, between its trace with an observable $\op{\Lambda}_j$ and that observable's weak measurement result $\Lambda_j^w$, summed over all observables in the set of generalized Gell-Mann matrices. That is,\\
{\bf ($\textbf{Q}_8$) SWV state}:
\begin{align}\label{eq-q8}
\est\varrho_\tau = & \argmin_{\varrho} \sum_{j=1}^{d^2-1}\left\la \left[ {\rm Tr}(\varrho \op{\Lambda}_j) - \Lambda^w_j \right]^2\right\ra_{\Lambda^w_j  | \bothps{\bo}}\,.
\end{align}
This cost function inside the angle brackets ensures that the complex-matrix estimator gives the correct real weak value $\Tr{\est\varrho_\tau \op{\Lambda}_j} = \la \Lambda^w_j \ra_{\Lambda^w_j  | \bothps{\bo}}$, leading to the SWV state, 
\begin{subequations}
\begin{align}\label{eq-wvsGell}
\est\varrho_\tau = &\,\, \tfrac{1}{d} {\hat 1} + { \tfrac{1}{2}}\! \sum_{j=1}^{d^2-1} \la \Lambda^w_j \ra_{\Lambda^w_j  | \bothps{\bo}}  \, {\op \Lambda}_j  \\
= & \label{eq-wvsO} \,\, \frac{\op{E}_{\futps{\bo}}\, \rho_{\pasts{\bo}} + \rho_{\pasts{\bo}}\,\op{E}_{\futps{\bo}}} {\Tr{\op{E}_{\futps{\bo}}\,\rho_{\pasts{\bo}} + \rho_{\pasts{\bo}}\, \op{E}_{\futps{\bo}}}}.
\end{align} 
\end{subequations}
One can also see that for the case without the postselection, or the future record, the solution of the optimization reduces back to the filtered state, giving the usual expectation value $\Tr{\rho_{\pasts{\bo}}\, \op{\Lambda}_j} = \la \Lambda^w_j \ra_{\Lambda^w_j  | \pasts{\bo}}$.  We show in Figure~\ref{fig-diagramQ2} for $\textbf{Q}_8$ that the SWV state (illustrated as a dotted green trajectory in the bottom panel) has its components being the conditioned expectation values of the orthonormal Gell-Mann observables. We chose the green color for this estimator because it is not necessarily a valid quantum state, to distinguish it from other estimators that are valid quantum states.

\subsection{Connections via classical state estimation}\label{sec-classana}

In the quantum state estimation diagram in Figure~\ref{fig-diagramQ}, we have included the double-dotted orange lines which indicate ``classically the same" connections. This indicates that when we replace everywhere the quantum state $\rho$ with the classical state $\wp(\bx)$ for a discrete configuration $\bx$, the two estimators in question yield the same optimal classical state. As we preluded in Section~\ref{sec-CSE}, we assume that the true classical state under continuous monitoring can be determined from complete records. That is the true state for a discrete configuration $\bx$ at time $\tau$ is given by $\wp_\tau^\true(\bx) = \wp_{\pasts{\bo},\pasts{\bu}}(\bx) =  \delta_{\bx,\bx_{\pastss{\bo},\pastss{\bu}}}$, where $\bx_{\pasts{\bo},\pasts{\bu}}$ is the configuration determined by a complete knowledge of both observed and unknown records. In this subsection, we present the mathematical justification for the two connections shown in the diagram: (1) between the quantum state smoothing (${\textbf{Q}_1}$) and the SWV state (${\textbf{Q}_8}$), and (2) between the negative Fidelity cost (${\textbf{Q}_2}$) and the negative Equality cost (${\textbf{Q}_3}$). While we present the results for the states conditioned on the past-future record, $\bothp{\bo}$, in fact these classical equalities result from the cost functions themselves, and so hold for conditioning on any data (D).

% The cost-optimization estimation for quantum states using past-future information presented in Section~\ref{sec-sevenQSE} can interestingly motivate new ideas for state estimators for special cases, such as when the condition on the observed record is not for both past and future portion, or when the quantum state can be reduced to the classical state (as mentioned in Section~\eqref{sec-CSE}).  Therefore, in this section, we briefly overview how the eight expected costs (and associated estimators) introduced for the quantum case can be reformulated in special cases and how they can be linked to existing theories. We consider three scenarios: state estimation unconditioned on observed records, state estimation using past-only information, and estimation of classical states.

%We also show that, as one would expect, the classical analogues of the smoothed quantum state $\rho_{\textbf{Q}_1}$ and that of the smoothed weak-value state $\rho_{\textbf{Q}_8}$ both reduce to the standard classical smoothed technique [REF].

We first consider a classical analog of ${\textbf{Q}_1}$. By replacing the trace with the sum over the basis states, the classical analogue of the trace square deviation cost Eq.~\eqref{eq-q-sd} is the sum square deviation cost as shown in Eq.~\eqref{eq-cstateBME1}. Therefore, conditioned on the past-future observed record, we obtain a classical state estimator that minimizes an expected Sum Square Deviation cost from the true state,\\
{\bf ($\textbf{C}_1$) $\la$\textSigma SDf\,$\wpU\ra$}:
\begin{align}\label{eq-c-smt}
\est\wp_\tau(\bx) &=  \argmin_{\wp} \left\la \sum_{\bx'} \left [ \wp(\bx') - \wpOU(\bx')\right ]^2 \right\ra_{\pasts{\bu} | \bothps{\bo}}\,\,\!\!\!\!(\bx) \nonumber \\
& =  \int \!\! \dd\mu(\,\past{u}\,)\, \wp_{\bothps{\bo}}(\,\past{u}\,) \, \delta_{\bx,\bx_{\pastss{\bo},\pastss{u}}} = \wp_{\bothps{\bo}}(\bx).
\end{align}
In the second line, the estimator is written in terms of the conditional average of possible true states, $\delta_{\bx,\bx_{\pastss{\bo},\pastss{u}}}$, with the weight $\wp_{\bothps{\bo}}(\,\past{u}\,)$. The integral with the $\delta$-function then effectively leads to the change of variables between the two PDFs, \ie, from $\wp_{\bothps{\bo}}(\,\past{u}\,)$ to $\wp_{\bothps{\bo}}(\bx)$. The end result is the \emph{classical smoothed state}, the Bayesian PDF of the system's configuration conditioned on $\bothp{\bo}$~\cite{Sarkka13}. 

Next, we consider a classical version of the expected cost for $\textbf{Q}_8$ and its optimal SWV state. For a system to be considered classical, its initial conditions, final conditions, and its dynamics can all be described probabilistically in a fixed basis. In that basis, the state matrix is diagonal. Let us consider Eq.~\eqref{eq-q8}, where we used the Gell-Mann matrices to represent observables in the Hilbert space. For the classical counterpart, we can keep the form of the estimator in terms of the Gell-Mann matrices as in Eq.~\eqref{eq-wvsGell}, by only keeping the diagonal ones. However, we can easily see how the estimator turns out to be exactly the classical smoothed state, by using projectors as observables in the fixed basis instead. The basis is the discrete configuration $\bx$, which can take any value in the countable set ${\mathbb X}$. 

Therefore, we replace the set of observables with the projectors $|\bx \ra \la \bx|$, where $\bx \in {\mathbb X}$, and replace the trace with the sum over the elements of ${\mathbb X}$. This means we can substitute $\op{\Lambda}_j$ in Eq.~\eqref{eq-q8} with $\Lambda_{\bx}(\bx') = \delta_{\bx,\bx'}$, which is a projector in the language of functions of $\bx'$. These observables can be measured weakly, and classically it means that the measurement is noisy, where a measurement result, $\Lambda^w_{\bx}$, includes additional uncorrelated noise from the detection procedure. Combining all these classical analogs, we then obtain a formula for the expected cost and the classical estimator,\\
{\bf ($\textbf{C}_8$) Classical SWV state $=$ ($\textbf{C}_1$) $\la$\textSigma SDf\,$\wpU\ra$}:
\begin{align}\label{eq-c8}
\est\wp_\tau(\bx) = & \argmin_{\wp} \sum_{\bx \in {\mathbb X}}\left\la \left[ \sum_{\bx' \in {\mathbb X}}\wp(\bx') \Lambda_{\bx}(\bx') - \Lambda^w_{\bx} \right]^2\right\ra_{\Lambda^w_{\bx}  | \bothps{\bo}} \nonumber \\
 = & \argmin_{\wp} \sum_{\bx \in {\mathbb X}}\left\la \left[ \wp(\bx)- \Lambda^w_{\bx} \right]^2\right\ra_{\Lambda^w_{\bx}  | \bothps{\bo}} \nonumber \\
 = & \,   \la \Lambda^w_{\bx} \ra_{\Lambda^w_{\bx}  | \bothps{\bo}}\,,
\end{align}
where we have substituted $\Lambda_{\bx}(\bx') = \delta_{\bx,\bx'}$ in the first line to get the second line. In the last line, we have obtained the estimator in a similar way as for the mean square deviation cost in Eq.~\eqref{eq-confestBME1}. We can then use the fact that an average result of a  noisy (non-disturbing) measurement is equal to an average result of a perfect measurement, given that the detection noise is unbiased and not correlated with the true observables. Then, we realize that the perfect measurement of an observable $\bx$ gives an outcome as either 1 or 0, depending on whether one finds the system to be in the state $x$ or not, respectively. Therefore, continuing from Eq.~\eqref{eq-c8}, we get 
\begin{align}
\est\wp_\tau(\bx)  = \la \Lambda^w_{\bx} \ra_{\Lambda^w_{\bx}  | \bothps{\bo}} = \la \delta_{\bx,\bx'} \ra_{\bx'  | \bothps{\bo}} = \wp_{\bothps{\bo}}(\bx),
\end{align}
as the classical version of the SWV state, which is exactly the same as the classical smoothed state in Eq.~\eqref{eq-c-smt}.

We can also construct the classical version of the SWV state directly from Eq.~\eqref{eq-wvsO} using state matrices corresponding to  classical states. A  state  matrix of a classical state can be represented by a diagonal matrix in a fixed basis. Therefore, in that basis, the filtered and retrofiltered states in Eq.~\eqref{eq-wvsO} can be written as $\rho\fil = \sum_{\bx} \wp_{\pasts{\bo}}(\bx) |\bx \ra \la \bx |$ and $\hat E_{\futps{\bo}} = \sum_{\bx} \wp(\futp{\bo}\,|\bx) |\bx\ra \la \bx |$~\cite{LavCha2020a}. The diagonal elements of $\rho\fil$ are the probabilities for the system to be found in configurations $\bx$, given the past observed record; while, the diagonal elements of the retrodictive effect are probabilities of the future record given configurations $\bx$. Given both matrices, we obtain,
\begin{align}
\varrho_{\rm SWV, \rm diag}  & = \,  \frac{ \sum_{\bx} \wp_{\pasts{\bo}}(\bx) \wp(\futp{\bo}\,|\bx) |\bx\ra \la \bx | }{\sum_{\bx} \wp_{\pasts{\bo}}(\bx) \wp(\,\futp{\bo}\,|\bx)}  \nonumber \\
  & = \sum_{\bx} \wp_{\bothps{\bo}}(\bx) |\bx \ra \la \bx |,
\end{align}
as a direct classical analogue of the SWV  state in Eq.~\eqref{eq-wvsO}.  This method of proving the equivalence was explicitly used in Refs.~\cite{Tsangsmt2009-2,LavCha2020a}.

%\begin{figure}
%\includegraphics[width=8.5cm]{CostFDiagram3.pdf}
%\caption{Diagrams showing eight different expected costs, for $\textbf{C}_1$--$\textbf{C}_8$ (the classical analogue of Figure~\ref{fig-diagramQ}), connecting the existing formalisms (grey boxes): classical state smoothing and the stochastic path integral formalism. Expected costs in blue and pink boxes represent the costs defined in the classical state space and the unknown record space, respectively.}
%\label{fig-diagramC}
%\end{figure}

We now turn to proving the  connection between the classical analogs of  the negative fidelity cost (${\textbf{Q}_2}$) and the negative equality cost (${\textbf{Q}_3}$). The expected negative equality cost for classical states was shown in Eqs.~\eqref{eq-cstate-ne} and \eqref{eq-cstate-necost}, where its optimal state is the pure state with the most-likely configuration, \ie, $\est\wp_{\rm nE}(\bx) = \delta_{\bx,\est\bx_{\rm MLE}}$.
For the negative fidelity, we then consider the classical fidelity 
\begin{align}\label{eq-cjozsa}
F\!\left[\wp_{\rm A}, \wp_{\rm B}\right] = \left(\sum_{\bx' \in \mathbb X}\sqrt{\wp_{\rm A}(\bx')\wp_{\rm B}(\bx')} \right)^2,
\end{align}
which quantifies the difference between two classical states, $\wp_{\rm A}(\bx)$ and $\wp_{\rm B}(\bx)$. This is the definition that motivated  Jozsa's fidelity~\cite{Jozsa1994} for quantum states in Eq.~\eqref{eq-Jozsafid}. Using this definition of fidelity in Eq.~\eqref{eq-cjozsa}, we can calculate an expected negative fidelity cost,
\begin{align}\label{eq-prove-nef}
\left\langle -F\!\left[\wp, \wp^\true\right] \right\rangle_{\wp^\true | \bothps\bo} = & -  \sum_{\bx' \in \mathbb X}  \wp_{\bothps\bo}(\bx') \, \left(\sum_{\bx'' \in \mathbb X} \sqrt{ \wp(\bx'')\,\delta_{\bx'',\bx'}} \right)^2 \nonumber \\
= & -\sum_{\bx'}  \wp_{\bothps\bo}(\bx') \, \left(\sum_{\bx''} \delta_{\bx'',\bx'}\sqrt{ \wp(\bx'')} \right)^2 \nonumber \\
= & -\sum_{\bx'}  \wp_{\bothps\bo}(\bx') \, \wp(\bx') \nonumber\\
\ge &  - \max_{\bx'}\, \wp_{\bothps\bo}(\bx'),
\end{align}
where properties of $\delta$-functions were used in getting the second and third lines. The lower bound on the last line of Eq.~\eqref{eq-prove-nef}  follows from  the positivity of $\wp_{\bothps\bo}(x)$ and the fact  that $\wp(x)$ is normalized. Obviously, we can saturate this lower bound, and hence obtain the optimal estimator, by choosing $\wp(\bx) = \delta_{\bx,\est\bx}$, where $\est\bx = \argmax_{\bx'} \wp_{\bothps\bo}(\bx')$.
%To obtain the last line of Eq.~\eqref{eq-prove-nef}, we use the convex sum property to show that the negative sum in the third line is minimized when $\wp(\bx')$ is a pure state, also with the most-likely configuration. That is, $\sum_{\bx'}  \wp\c(\bx') \, \wp(\bx') \le \argmax_{\bx'}\, \wp\c(\bx')$. 
Therefore, we have proved that \\
{\bf ($\textbf{C}_2$) $\la$nFw$\wpU\ra$ $=$ ($\textbf{C}_3$) $\la$nEw$\wpU\ra$}:\\
\begin{align}\label{eq-cstate-nFnE}
\est\wp_\tau(\bx) =&  \argmin_{\wp} \left\la -F\left[\wp, \wpOU \right] \right\ra_{\pasts{\bu} | \bothps{\bo}}\!(\bx) \nonumber \\
=& \argmin_{\wp} \left\la -\delta\!\left[\wp , \wpOU \right] \right\ra_{\pasts{\bu} | \bothps{\bo}}\!(\bx) \nonumber \\
= &\, \delta_{\bx, \est\bx_{\rm MLE}},
\end{align}
where
\begin{align}
\est\bx_{\rm MLE} = \argmax_{\bx}\,\wp_{\bothps{\bo}}(\bx),
\end{align}
is the most-likely configuration of the smoothed classical state. 

As was discussed in the quantum case, it is possible that the probability distribution $\wp_{\bothps{\bo}}(\bx)$ does not have a unique  maximum.  
While this is a set of measure zero, it is still worth discussing the  potential  differences between  the estimators from  $\textbf{C}_2$ and $\textbf{C}_3$ in the event of multiple  most-likely   configurations $\est\bx_{\rm MLE}$. For  $\textbf{C}_3$, any and only members of the set of $\delta$-functions $\{ \delta_{\bx,\est\bx_{\rm MLE}} \}$ will be an optimal classical state estimator. For $\textbf{C}_2$, all of these are optimal estimators, but, in addition, 
%if multiple optimal configurations exist, 
any state in the convex hull of this set will also be an optimal estimator.  
%$\{ \delta_{\bx,\bx^*_{\rm MLE}} \}$ will also be an optimal estimator.  As for , only the $\delta$-functions $\delta_{x,x^*_{\rm MLE}}$ are optimal. 
Thus, while it is possible to chose optimal estimators to break the equality between $\textbf{C}_2$ and $\textbf{C}_3$ in this degenerate case, it is also possible to choose the optimal estimators in a way that the equality is always satisfied.
%{\red However, we note that the above relationship, $(\textbf{C}_2) = (\textbf{C}_3)$, is true only when there is a unique most-likely configuration. In the case where $\wp_{\bothps{\bo}}(\bx)$ has multiple maxima, the optimal estimator for $\textbf{C}_3$ can be any one of the pure states, $\{ \delta_{\bx,\bx^*_{\rm MLE}} \}$, associated with the set of equally most-likely configurations. For $(\textbf{C}_2)$, the optimal estimator is any state of the convex hull of $\{ \delta_{\bx,\bx^*_{\rm MLE}} \}$.}

\subsection{Expected cost functions for the estimators}\label{sec-expcost}

Later, in Section~\ref{sec-eightestqubit}, we will investigate the eight estimators and their expected cost functions for a qubit example. By definition, each estimator should optimize its associated expected cost function. Nevertheless, we can still make use of the available estimator solutions and illustrate the optimality by evaluating expected costs for other (non-optimal) estimators to verify that the optimal solutions give the smallest values. Some of the cost functions are defined in the state space and some are defined in the unknown record space. Therefore, for the latter, we also need to solve for the past unknown record associated with each of the state estimators. We here consider seven out of eight expected costs. We exclude the expected cost of ${\textbf{Q}_4}$, for which one needs to search for past records that maximize the PDF, $\wp_{\bothps{\bo}}(\,\past{u}\,)$, for the non-optimal states at all times $\tau \in (\tin, T]$, which is extremely difficult to calculate. 

%and that of ${\textbf{Q}_8}$ which is defined on the complex-matrix space (with elements outside of the quantum state space), making it an unfair comparison to other seven state estimators

For the first three cost functions, defined in the state space, \ie, ${\textbf{Q}_1}$\,--${\textbf{Q}_3}$, their expected costs are straightforward to calculate. The expected costs,
\begin{align}
\label{eq-cq1}{\cal C}_1(\rho) = & \,\, \left\la {\rm Tr} \left [  \left(\rho - \rhoOU\right)^2  \right ] \right \ra_{\pasts{\bu} | \bothps{\bo}},\\
\label{eq-cq2}{\cal C}_2(\rho) = &  \,\, \left\la -F\!\left[\rho, \rhoOU \right] \right\ra_{\pasts{\bu} | \bothps{\bo}}, \\
\label{eq-cq3}{\cal C}_3(\hat\psi) = & \,\, -\wp_{\bothps{\bo}}(\hat\psi),
\end{align}
are simply the arguments of the $\argmin$ functions in Eqs.~\eqref{eq-smt}, \eqref{eq-th-q-nf}, and \eqref{eq-th-q-ne}, respectively. Note, however, that the third expected cost, ${\cal C}_3(\hat\psi)$, can only be applied to pure states.

For the next three cost functions defined in the record space, \ie, ${\textbf{Q}_5}$\,--${\textbf{Q}_8}$, their expected costs are dependent on the time resolution $\delta t$ (or $\dt$, which has to be specified as a small but finite value for the numerical evaluation). We therefore modify the expected costs in Eq.~\eqref{eq-mlp}, \eqref{eq-q6}, and \eqref{eq-q7} to obtain new definitions that are $\ddt$-independent, without affecting their optimization. For ${\textbf{Q}_5}$, the expected cost is the negative probability density, $-\, \wp_{\bothps{\bo}}(\,\both{u}\,)$, which has $\ddt$-dependent terms in the normalized factors of instantaneous-record PDFs, $\wp(u\ik, \bo\ik |\qq\ik)$, in Eq.~\eqref{eq-probmlp2}. Such terms can be eliminated by defining a new expected cost as a log ratio, \ie,
\begin{align}
\label{eq-cq5}{\cal C}_5(\both u)= \,\, \log[\wp_{\bothps{\bo}}(\,\both{u}_{{\textbf{Q}_5}})/\wp_{\bothps{\bo}}(\,\both{u}\,)],
\end{align}
which should have its minimum (zero) value when $\both{u} = \both{u}_{{\textbf{Q}_5}}$, the optimal record solution of Eqs.~\eqref{eq-diffeq2}. Similarly, the expected cost for ${\textbf{Q}_6}$ is $-\wp_{\bothps{\bo}}(u_\tau)$, which is a negative probability density of an unknown record at time $\tau$. The probability density has its variance depending on the time resolution $\ddt$. Therefore, we can define a new expected cost, 
\begin{align}
\label{eq-cq6}{\cal C}_6(u_\tau) = -2 v \log\left[ \! \sqrt{2\pi v} \, \wp_{\bothps{\bo}}(u_\tau)\right],
\end{align}
where the variance $v = \la \bu_\tau^2 \ra_{\bu_\tau | \bothps{\bo}} \sim 1/\dt$ is eliminated. 

Lastly, for ${\textbf{Q}_7}$ and ${\textbf{Q}_8}$, we can consider shifting the mean-square deviation to remove the terms which depend on the time resolution or weakness of the measurement, respectively. An example of such shifting is shown in Eq.~\eqref{eq-q-sd-ut-2}, which does not affect the optimization. Therefore, we  modify the expected costs for the last two categories to
\begin{align}
\label{eq-cq7}&{\cal C}_7(u_\tau) =  \,\, \left\la   u_\tau^2 - 2 u_\tau \bu_\tau \right\ra_{\bu_\tau | \bothps{\bo}}, \\
\label{eq-cq8}&{\cal C}_8(\varrho) = \,\, \sum_{j=1}^{d^2-1}\left\la {\rm Tr}(\varrho \op{\Lambda}_j)^2 - 2\, {\rm Tr}(\varrho \op{\Lambda}_j) \, \Lambda^w_j  \right\ra_{\Lambda^w_j  | \bothps{\bo}}.
\end{align}
The sum in Eq.~\eqref{eq-cq8} is over the set of generalized Gell-Mann matrices $\{ \op\Lambda_j \}$ as in Eq.~\eqref{eq-q8}. We can also show that ${\cal C}_6 = {\cal C}_7$ by writing
\begin{align}
{\cal C}_6(u_\tau) = u_\tau^2 - 2 u_\tau \la \bu_\tau \ra_{\bu_\tau | \bothps{\bo}} = {\cal C}_7(u_\tau),
\end{align}
given the form of $\wp_{\bothps{\bo}}(u_\tau)$ presented in Appendix~\ref{sec-app-pdf}.

%%%%%%%%%%%%%%%%%%%%%
% Example
%%%%%%%%%%%%%%%%%%%%%
\section{Example: single qubit with bosonic baths}\label{sec-example} 
In this section, we implement our unified theory of quantum state estimation for the example of a resonantly driven two-level system (qubit) spontaneously emitting photons to two independent vacuum bosonic baths~\cite{Ivonne2015,chantasri2019}. The qubit system has two eigenstates given by $|e\ra$ (excited state) and $|g\ra$ (ground state), with an excitation energy $\omega$ (taking $\hbar = 1$). In a frame rotating at the excitation frequency, the qubit is driven to rotate around the $x$-axis in the Bloch sphere with a Rabi frequency $\Omega$. The qubit-bath coupling operator responsible for the qubit's spontaneous decay is the lowering operator $\op{\sigma}_- = |g\ra\la e|$. The baths can be measured, for example, via photon detection or homodyne detection \cite{Wiseman1993-2}.

%Its quantum state at any time $t$ can be represented by a state matrix $\rho_t = \tfrac{1}{2}(\hat I + \qq_t \cdot \hat {\bm \sigma})$, as in Eq.~\eqref{eq-gellmann}, where the Gell-Mann matrices become the Pauli matrices $\hat {\bm \sigma} = \{ \hat \sigma_x , \hat \sigma_y , \hat \sigma_z\}$ and $\qq_t$ is a three-dimensional Bloch vector. 

Let us consider the partially-observed (Alice-Bob) scenario where the qubit's coupling to the respective observers' baths is described by Lindblad operators $\op{c}_{\rm o}$ and $\op{c}_{\rm u}$, which are both proportional to the lowering operator $\hat\sigma_{-}$. The lower-case letter subscripts `o' and `u' are to label the channels observed and unknown, respectively, by the observer Alice. Under the strongest Markov assumption~\cite{LiLi2018}, the unconditional dynamics $\mrho(t)$ of the qubit (no conditioning on measurement records) is described by the Lindblad master equation, 
\begin{align}\label{eq-master}
\dd \mrho(t) &=  - i  \, \dt [ \tfrac{\Omega}{2} \,\op{\sigma}_x , \mrho(t)] +  \dt \,{\cal D}[\op{c}_{\rm u}]\mrho(t) + \dt \,{\cal D}[\op{c}_{\rm o}]\mrho(t),  \nonumber \\
& =    \, \dt\, {\cal L}_{\rm u} \mrho(t) + \dt \,{\cal D}[\op{c}_{\rm o}]\mrho(t), 
\end{align}
where the system's Hamiltonian is ${\hat H} = (\Omega/2){\hat \sigma}_x$ and the superoperator is defined as ${\cal D}[\hat c] \bullet = \op{c}\bullet \op{c}^\dagger - \tfrac{1}{2}(\op{c}^\dagger \op{c} \bullet + \bullet \op{c}^\dagger \op{c})$. In the second line of Eq.~\eqref{eq-master}, we write the master equation in terms of the superoperator ${\cal L}_{\rm u}$ from Eq.~\eqref{eq-povm} for consistency.

%Assuming there is another omniscient observer named Bob, who can measure the bath unknown by Alice and also has access to Alice's record, he can have a complete information about the system and assign a true state to the system. Alice's task is then to try her best to estimate Bob's (true) state. 

The Lindblad evolution, Eq.~\eqref{eq-master}, can be \emph{unravelled} to different stochastic pure state trajectories, depending on the type of measurement applied on the baths. In this work, we consider the specific scenario where Alice measures her bath using photon detection and Bob measures his bath using a $y$-homodyne measurement (with a local oscillator phase $\Phi = \pi/2$). Therefore, we can denote the Lindblad operators for Alice's and Bob's channels by $\op{c}_{\rm o} = \sqrt{\gamma_{\rm o}}\,\op{\sigma}_-$ and $\op{c}_{\rm u} = \sqrt{\gamma_{\rm u}}\,\op{\sigma}_- e^{-i \pi/2}$, respectively. The coupling rates $\gamma_{\rm o}$ and $\gamma_{\rm u}$ sum to give the total system-bath coupling rate, $\gamma_{\rm o}+\gamma_{\rm u} = \gamma = 1/T_\gamma$, where $T_\gamma$ is the system's decay time. In this scenario, the unravelled qubit's state is always confined to the $y$--$z$ great circle of the Bloch sphere, which significantly simplifies the analysis.

 Following the presentation in sections~\ref{sec-cdj} and \ref{sec-qss}, 
we describe the conditional dynamics of the monitored quantum system by measurement operations.  For the photon detection on Alice's side, at most one photon can be detected during a sufficiently short time $\delta t$.  Thus, there are two operations associated with two possible outcomes (1 or 0 detected photon). If a photon is detected, the operation describing the state-update is ${\cal N}_{1} \rho  = \op{c}_{\rm o} \, \rho \,  \op{c}_{\rm o}\dg \delta t$. If no photon is detected, then the state evolves by a measurement operation, ${\cal N}_0 \rho = {\hat N}_0 \rho {\hat N}_0\dg$, where ${\hat N}_0 = {\hat 1}- \tfrac{1}{2} \op{c}_{\rm o}\dg \op{c}_{\rm o} \, \delta t  $.  Because ${\cal N}_{1}$ causes a finite change in the state---from $\rho$ to $\op{c}_{\rm o} \, \rho \,  \op{c}_{\rm o}\dg / \Tr{\op{c}_{\rm o}\dg \op{c}_{\rm o}\rho}$ when normalized---in a time interval of arbitrarily short duration $\delta t$, it can be called a {\em quantum jump}~\cite{BookCarmichael,Plenio1998,BookWiseman}. 

The above jump and no-jump evolution is in contrast to the dynamical measurements we have considered hitherto, of the homodyne type. This can be called  {\em quantum diffusion} because in the limit $\delta t \to \dt$ the state evolution is continuous but not differentiable in time~\cite{BookCarmichael,Plenio1998,BookWiseman}. This is the type of measurement we consider here for Bob's side; specifically, a  %For the 
$y$-homodyne detection %on Bob's side, 
for which 
the measurement result $u_t$ can take any real value. The measurement backaction for this diffusive measurement is described by the operation ${\cal M}_{u_t} \rho = {\hat M}_{u_t} \rho {\hat M}_{u_t}\dg$ where ${\hat M}_{u_t}$ is given by Eq.~\eqref{eq-moput}, which includes the unitary evolution for the short time step $\delta t$.  It is worth remarking that the term ``quantum jump'' is sometimes also used for a rapid, but actually continuous, transition that can occur in a particular limit of a diffusive measurement, as studied experimentally in Ref.~\cite{MinMun2019}; this could be contrasted with the experimental study of a quantum jump of the type considered here, induced by a  single-photon detection, in Ref.~\cite{smiRei2002}.  

 It can be shown that the above operations, ${\cal N}_{\bo\ik}$ and ${\cal M}_{u\ik}$, unravel the Lindblad master equation, Eq.~\eqref{eq-master}, by writing the unconditional state at time $t+\ddt$,
\begin{align}\label{eq-masterunnorm}
 {\mrho}(t+\ddt) = & \,\, \sum_{O_t} \!\int\!\! {\rm d} u_t \, \wp\ost(u_t)  \trho_{\bo\ik,u\ik} \nonumber \\
=& \, \,  \mrho(t)  + \ddt {\cal L}_{\rm u} \mrho(t) + \ddt {\cal D}[{\hat c}_{\rm o}]\mrho(t), 
 \end{align}
where $\trho_{\bo\ik,u\ik} = {\cal N}_{O_t}{\cal M}_{u_t} \mrho(t)$ is the unnormalized state at time $t+\ddt$ conditioned on the results $\bo\ik$ and $u_t$. This agrees with a more conventional approach,
\begin{align}\label{eq-masternorm} 
\mrho(t+\ddt) = & \,\, \sum_{O_t} \!\int\!\! {\rm d} u_t \, \wp(u_t, \bo_t | \rho(t)) \rho_{\bo\ik,u\ik} \nonumber  \\
= & \sum_{O_t=0,1} \!\!\int\!\! {\rm d} u_t \, \wp\ost(u_t)  {\rm Tr}( \trho_{\bo\ik,u\ik} ) \rho_{\bo\ik,u\ik},  
\end{align}
where the unconditional state is obtained from summing (integrating) over all possible normalized states $\rho_{\bo\ik,u\ik} = \trho_{\bo\ik,u\ik}/{\rm Tr}( \trho_{\bo\ik,u\ik} )$, with the actual probability $\wp(u\ik, \bo\ik | \rho(t) ) = \wp\ost(u_t)  {\rm Tr}( \trho_{\bo\ik,u\ik} )$.

 These two ways of writing the unconditional state, Eq.~\eqref{eq-masterunnorm} and Eq.~\eqref{eq-masternorm}, give us an idea that, to obtain correct statistics for the Lindblad master equation, we can either generate normalized states with the actual probabilities, or generate unnormalized states with ostensible probabilities, \eg, treating $u\ik$ as a random variable with the Gaussian distribution $\wp\ost(u_t)$. This latter method is essential for estimation with past-future information as discussed in Section~\ref{sec-qss}. We introduce a stochastic master equation for the unnormalized state in the next subsection.

\subsection{Estimating dynamics between two jumps}\label{sec-btw}

\begin{figure}[t]
\centering
\includegraphics[width=14cm]{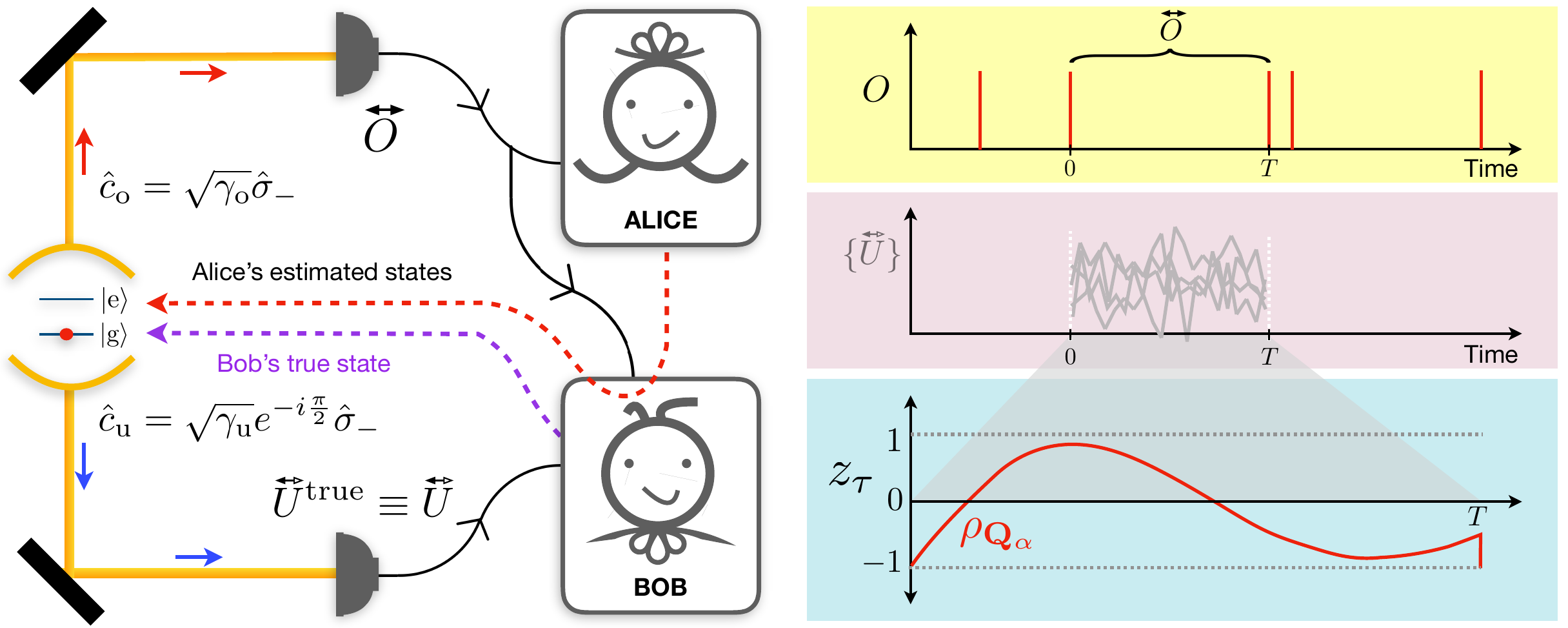}
\caption{ Schematic diagrams showing the Alice-Bob protocol for the two-level system (qubit) coupled to two bosonic baths via the Lindblad operators $\hat c_{\rm o}$ and $\hat c_{\rm u}$ and measured by Alice and Bob, respectively (similar to Figure~\ref{fig-intro}). Here, Alice's measurement is photon detection, where its record is a set of `clicks' at random times. Bob's measurement is $y$-homodyne detection, where its record is a noisy photocurrent.  Considering Alice estimating the state dynamics of the qubit between any two clicks (jumps), where the time $t$ is redefined to be between $t=0$ and $t=T$, Bob's possible unobserved records are shown as grey records in the middle right panel. Alice's estimated state dynamics (here represented by one Bloch-vector component, $z_\tau$) for some cost function is shown as the red curve in the enlarged time interval in the bottom right panel.}
\label{fig-introqubit}
\end{figure}

Consider the Alice-Bob scenario  as shown in Figure~\ref{fig-introqubit}, where Alice has access to only the observed record and her task is to estimate Bob's true state. Alice's record is from the photon detection, so her record (during a time period of interest) is a series of $J$ detections (jumps) at times denoted by $T_j \in \{ T_1, T_2, ..., T_J \}$. After each detection, the qubit state deterministically collapses to the ground state, regardless of its preceding state $ \rho(T_j)$ right before the jumps. That is,
\begin{align}
\rho(T_j+\ddt) = \frac{ {\cal N}_{1} \rho(T_j)}{{\rm Tr}[ {\cal N}_1 \rho(T_j) ]} = |g\ra\la g|,
\end{align}
using the measurement operation for jumps defined earlier.
Since the jumps reset the qubit state to its ground state, we can consider separately its dynamics between any two consecutive jumps (or the initial time and $T_1$, or $T_J$ and the final time). The first and last intervals are negligible for the long-time averages. We therefore consider a \emph{block} of qubit's evolution between any two jumps, where, for the rest of this Section, we will redefine $t$ as the time from  the ground state after a jump at time $T_j$, running from $t=0$ to $t = T = T_{j+1} - T_j$. The observed record in this block is thus given by $\bothp{\bo} = \{ \bo_0 , \bo_{\ddt}, ..., \bo_{T-\ddt} ,\bo_T \}\tp = \{ 0,0,..., 0,1\}\tp$; that is, no jump during the interval and one jump at the end.  The jump $\bo_T$ at the end of the interval is included in the observed record, even though the unobserved record is still defined as $\both{u} = \{ u_0 , u_{\ddt}, ..., u_{T-\ddt} \}\tp$. This is because the jump resets the state to the ground state and the unobserved record at the final time is not relevant.

For Alice to estimate the state dynamics between jumps, she needs to know the possible dynamics of the true state. The true state is conditioned on both $\bothp{\bo}$ (observed) and $\both{u}$ (unknown) records. Following the unnormalized state conditioned on both records in Eq.~\eqref{eq-unnorm}, the state update for the unnormalized true state $\trho\god(t) \equiv \trho_{\pasts{\bo}, \pasts{u}}$ is
\begin{align}\label{eq-updatetrue}
\trho\god(t+\ddt) = &\, {\cal N}_{0} {\cal M}_{u_t} \trho\god(t), \nonumber \\
= &\, \trho\god(t) - i \, \ddt\, [\hat H, \trho\god(t)]  - \tfrac{1}{2}(\ddt - u_t^2\ddt^2) \bar{\cal H}[\hat c^2_{\rm u}]\trho\god(t) \\
 &  +\, u_t^2 \ddt^2 \hat c_{\rm u} \trho\god(t) \hat c_{\rm u}^\dagger - \ddt \bar{ \cal H}[\tfrac{1}{2} \hat c_{\rm u}^\dagger \hat c_{\rm u}] \trho\god(t)   + \, u_t \ddt \, {\bar {\cal H}}[{\hat c}_{\rm u}] \trho\god(t)  -  \ddt\, {\bar{ \cal H}}\left[\tfrac{1}{2} {\hat c}_{\rm o}^{\dagger}{\hat c}_{\rm o} \right]\trho\god(t), \nonumber
\end{align}
where ${\bar {\cal H}}[\hat c] \bullet = \hat c \bullet + \bullet \hat c^{\dagger}$. We note the importance of keeping terms of the order $u_t^2\ddt^2$, in order to see how various differential equations in the time-continuum limit arise from this one map. Depending on the assumed statistical properties of the record $u\ik$, Eq.~\eqref{eq-updatetrue} can lead to two different differential equations for the qubit state, which will be used in the Fokker-Planck technique in Section~\ref{sec-fpe} and the most-likely path technique in Section~\ref{sec-mlp}.

From the map Eq.~\eqref{eq-updatetrue}, we can derive a stochastic master equation (SME) for the unnormalized state, which averages to the unconditioned master equation. By referring to Eq.~\eqref{eq-masterunnorm} and the discussion surrounded, we should treat the measurement result $u_t$ as having a zero-mean Gaussian statistics in order to generate a correct statistics for the unnormalized conditioned states. Therefore, we can derive the SME in the It\^o interpretation~\cite{BookGardiner2,BookJacobsSto} by taking $\lim_{\ddt \rightarrow \dt} u_t^2 \ddt^2 = \dt$, which gives
\begin{align}\label{eq-untruequbit}
\dd \trho\god = & - i\, \dt\, [\hat H, \trho\god] + \dt\, {\cal D}[{\hat c}_{\rm u}]\trho\god + u_t \dt \, {\bar{\cal H}}[{\hat c}_{\rm u}]\trho\god -  \dt\, {\bar{ \cal H}}\left[\tfrac{1}{2} {\hat c}_{\rm o}^{\dagger}{\hat c}_{\rm o} \right]\trho\god.
\end{align}
Here we are omitting the $t$-dependence argument for the state whenever possible. Moreover, since the initial state $\rho\god(t=0)$ is a pure state, the true state at any time is always pure, and Eq.~\eqref{eq-untruequbit} could be replaced with a stochastic Schr\"odinger equation (SSE) \cite{BookGardiner1,BookWiseman} or, indeed, something even simpler as we will see below in an equation for the qubit's angle $\theta$.

%We also use the fact that, for the fluorescence channel, $\hat c_{\rm u} \propto \hat\sigma_{-}$, we have $\hat c_{\rm u}^2 = ( \hat c_{\rm u}^\dagger)^2 = 0$, making the term $\bar{\cal H}[\hat c^2_{\rm u}]$ in Eq.~\eqref{eq-updatetrue} vanish.

For completeness, we also give the normalized version of the SME in Eq.~\eqref{eq-untruequbit}:
\begin{align}\label{eq-truequbit}
\dd \rho\god =& - i\, \dt\, [\hat H, \rho\god] + \dt\, {\cal D}[{\hat c}_{\rm u}]\rho\god  -  \dt\, {\cal H}\left[\tfrac{1}{2} {\hat c}_{\rm o}^{\dagger}{\hat c}_{\rm o} \right]\rho\god + \dt\{ u_t - {\rm Tr}[ (\hat c_{\rm u} + \hat c_{\rm u}^\dagger) \rho\god ] \} \, {\cal H}[{\hat c}_{\rm u}]\rho\god.
\end{align}
This is the usual It\^o SME for a normalized state conditioned on both Alice's no-jump and Bob's diffusive records, where ${ {\cal H}}[\hat c] \bullet = \hat c \bullet + \bullet \hat c^{\dagger} - {\rm Tr}(  \hat c \bullet + \bullet \hat c^{\dagger} )\bullet$. Note that the statistics of the unknown record $u\ik$ in this case is governed by its actual conditional PDF $\wp_{\pasts{\bo}}(u\ik)$. The PDF can be approximated as a Gaussian function with a mean given by $\la \bu\ik \ra_{\bu\ik |\bothps{\bo}} = {\rm Tr}[ (\hat c_{\rm u} + \hat c_{\rm u}^\dagger)\rho\god]$; see Eq.~\eqref{eq-app-probupast} in Appendix~\ref{sec-app-pdf}. Therefore, the SME for the normalized state in 
Eq.~\eqref{eq-truequbit} is usually seen written in terms of an innovation, $\dd w\god(t) \equiv u_t \dt - {\rm Tr}[ (\hat c_{\rm u} + \hat c_{\rm u}^\dagger) \rho\god ]\dt$, which has the same statistics as the Wiener increment, \ie, $\la \dd w\god \ra = 0$ and $\la \dd w\god^2 \ra = \dt$. %The second and third terms represent the diffusive ($y$-homodyne) dynamics, and the last term describes the backaction from no-jump observed record. We defined the Lindblad operators ${\hat c}\subp = \sqrt{\Gamma} \,{\hat \sigma}_- e^{-i \Phi} $ and ${\hat c}\subn = \sqrt{\gamma} \,{\hat \sigma}_-  $ for the diffusive and jump processes, with the coupling rate $\Gamma$ and $\gamma$, respectively. The diffusive measurement record and the Wiener increment $\dd W$ is related by $\dd J_{\Phi,t}= {\rm Tr}\{({\hat c}\subp + {\hat c}\subp^\dagger)\rho\god(t)\} \dt+ \dd W(t)$. We note that the possible true state $\rho\god$ at any time is always pure, if the initial state $\rho\god(t=0) = \rho\god(t_j)$ is a pure state.

Alice, with only the record of observed jumps, can calculate her filtered state trajectory $\rho\fil$ conditioned on her past observed record following Eq.~\eqref{eq-filstate}. This can also be obtained by averaging over all possible unknown records from Eq.~\eqref{eq-truequbit}, which is equivalent to averaging the equation over the innovation $\dd w\god$, yielding
\begin{align}\label{eq-filterqubit}
\dd \rho\fil = \!- i\, \dt\, [{\hat H}, \rho\fil] + \dt\, {\cal D}[{\hat c}_{\rm u}]\rho\fil -  \dt\, {\cal H}\left[\tfrac{1}{2} {\hat c}_{\rm o}^{\dagger}{\hat c}_{\rm o} \right]\rho\fil,
\end{align}
 which is Eq.~\eqref{eq-truequbit} with its last term removed.
Contrary to the true state dynamics, this filtered state does not have unit purity, even though the initial state is pure. However, for a better estimate of the true state, the observer can use the entire observed record (both past-future information) in the estimation, following the estimators based on the cost functions presented in Section~\ref{sec-sevenQSE}. 

 Since, the true qubit state in Eq.~\eqref{eq-truequbit} is limited to the $y$--$z$ great circle of the Bloch sphere, other states, such as the unconditioned state Eq.~\eqref{eq-master}  or the filtered state Eq.~\eqref{eq-filterqubit}, should always be on the $y$--$z$ plane. Therefore, we can reparametrize the qubit states with two parameters, $\theta$ and $R$, where $z = R \cos \theta$ and $y= R \sin \theta$ are the two coordinates of the Bloch vector (for the pure true state, $R = 1$). The excited state is $\theta = 0$ and the ground state is $\theta = \pi$. The normalized and unnormalized true states at any time $t$ are given by $\rho\god(t) = (1/2)( \hat I + \sin\theta_t \hat \sigma_y + \cos\theta_t \hat \sigma_z)$ and $\trho\god(t)= (\qp_t/2)( \hat I + \sin\theta_t \hat \sigma_y + \cos\theta_t \hat \sigma_z)$, respectively, where $\qp_t \equiv {\rm Tr}[\trho\god(t)]$ takes care of the state's norm. We then obtain the state update Eq.~\eqref{eq-updatetrue} for these new variables,
\begin{subequations}\label{eq-updatetheta}
\begin{align}
\theta_{t+\ddt} = &\,\, \theta_t - \ddt\, \Omega + \ddt \tfrac{\gamma}{2} \sin\theta_t + (\ddt-u_t^2\ddt^2)\tfrac{\Gamma}{2}\sin\theta_t - u_t^2 \ddt^2\tfrac{\Gamma}{2}  \cos\theta_t \sin\theta_t - u_t \ddt \sqrt{\Gamma}(\cos\theta_t+1), \\
\lambda_{t+\ddt} =& \,\, \lambda_t - \ddt \tfrac{\gamma}{2} ( \cos\theta_t + 1 ) \lambda - u_t \ddt \sqrt{\Gamma} \sin\theta_t \lambda_t  - (\ddt- u_t^2\ddt^2)\tfrac{\Gamma}{2} \lambda_t - (\ddt- u_t^2\ddt^2)\tfrac{\Gamma}{2} \cos\theta_t \lambda_t ,
\end{align}
\end{subequations}
 with terms up to the order $u\ik^2 \ddt^2$. These equations will be used in deriving two types of differential equations in the next subsections, as mentioned above.

In the following, we present the calculation methods for the eight estimators, and the numerical simulation results, in three subsections. Each subsection contains the estimators that are calculated with similar mathematical techniques, which are the Fokker-Planck equation, the CDJ most-likely path technique, and the local optimal records. We finish the section with the calculation of the expected cost functions and their averages for all possible jump times, where possible.

%The first subsection involves the calculation of the first three estimators, for which their cost functions are all defined in the state's \red state  matrix space. We proposed a technique using a \emph{quasi} Fokker-Planck equation to solve for a probability density function for true states parametrized in a qubit's angle variable. For the second subsection,..  Because there is a mapping between a Langevin equation to the Fokker-Planck equation, we will use this technique to simplify our problem, and not having to simulate trajectories as in [REF]. We note that the calculations are only for the qubit's evolution between any two jumps, since the jump records reset the qubit to the ground state making the evolution between jumps independent from each other.

\subsection{State estimators with Fokker-Planck equation}\label{sec-fpe}
The first three cost functions, $\textbf{Q}_1$--$\textbf{Q}_3$, are defined in the space of unknown true states at any time, from Eq.~\eqref{eq-qss-alter}, Eq.~\eqref{eq-th-q-nf} and Eq.~\eqref{eq-th-q-ne}. The three estimators are, respectively, the mean, the max-eigenstate of the mean, and the mode, of the PDF $\wp_{\bothps{\bo}}(\hat\psi)$ of the true state conditioned on the past-future observed record. One could use the stochastic simulation to find the PDF of the true state. However, in this subsection, the true state can be parametrized by the one-dimensional variable, $\theta$. This allows us to solve the PDF $\wp_{\bothps{\bo}}(\theta)$ via a semi-analytical method similar to solving a Fokker-Planck equation.

The dynamical equations for the true state are shown in Eq.~\eqref{eq-untruequbit} (unnormalized) and Eq.~\eqref{eq-truequbit} (normalized version) and one might be tempted to think that the equation for the normalized true state could be enough. However, as mentioned in the quantum state smoothing formalism, Section~\ref{sec-qss}, the correct statistics of true states or unknown records, conditioned on the fixed observed record, cannot be obtained from the normalized state and should be derived from the unnormalized state equations and their norms. Therefore, let us refer back to the PDF of the unobserved record written in terms of the unnormalized state in Eq.~\eqref{eq-bothoprob1}. From this, we can derive a general form of the PDF $\wp_{\pasts{\bo}}(\hat\psi)$ of the true state:
\begin{align}\label{eq-findpdf}
\wp_{\pasts{\bo}}(\hat\psi) \propto & \int \!\! \dd\mu(\,\past{u}\,)  \delta[\hat\psi - \rho_{\pasts{\bo}, \pasts{u}}] \wp\ost(\,\past{u}\,) {\rm Tr}\left( {\trho}_{\pasts{\bo}, \pasts{u}} \right), \nonumber \\
= & \int \!\! \dd\mu(\,\past{u}\,) \!\! \int\!\! \dd \lambda\, \delta[\lambda - {\rm Tr}(\trho_{\pasts{\bo}, \pasts{u}})] \,\delta[\hat\psi - \rho_{\pasts{\bo}, \pasts{u}}] \, \wp\ost(\,\past{u}\,) \lambda.
\end{align}
In the second line, we added the $\delta$-function mapping the state norm to the variable $\lambda$. This is so that we can separately define an ostensible PDF, 
\begin{align}
Q_{\pasts{\bo}}(\lambda, \hat\psi) =  \int \!\! \dd\mu&(\,\past{u}\,)\,  \delta[\lambda - {\rm Tr}(\trho_{\pasts{\bo}, \pasts{u}})]\,  \delta[\hat\psi - \rho_{\pasts{\bo}, \pasts{u}}]\, \wp\ost(\,\past{u}\,),
\end{align}
which has an advantage that it can be solved for using the dynamical equation for the unnormalized state and its norm, for example, as in Eq.~\eqref{eq-updatetheta}. 

It is important to note that the ostensible PDF $Q_{\pasts{\bo}}(\lambda, \hat\psi)$ is not the actual joint PDF of $\lambda$ and $\hat\psi$. However, the actual PDF of the true state can be obtained via
\begin{align}\label{eq-quasi-int}
\wp_{\pasts{\bo}}(\hat\psi) \propto \tilde\wp_{\pasts{\bo}}(\hat\psi) =  \int \!\! \dd \lambda \, Q_{\pasts{\bo}}(\lambda, \hat\psi) \lambda,
\end{align}
where the tilde indicates an unnormalized PDF, \ie, $\wp_{\pasts{\bo}}(\hat\psi) = \tilde\wp_{\pasts{\bo}}(\hat\psi) / \int \!\!\dd \mu_{\rm H}(\hat\psi) \wp_{\pasts{\bo}}(\hat\psi)$. Note the multiplication by $\lambda$ in the integrand in Eq.~\eqref{eq-quasi-int}. The past-future conditional PDF $\wp_{\bothps{\bo}}(\hat\psi)$ is then given by the Bayesian rule in Eq.~\eqref{eq-bayes}. For the particular example of the qubit pure state parametrized by the Bloch angle $\theta$, we can use $\dd \mu_{\rm H}(\hat\psi) = \dd \theta$ as the Haar measure and replace 
\begin{align}
\hat\psi = \hat S(\theta) = (1/2)( \hat I + \sin\theta \hat \sigma_y + \cos\theta \hat \sigma_z),
\end{align}
where the PDF of the pure state becomes $\wp_{\bothps{\bo}}(\theta)$.

In order to solve for the ostensible PDF $Q_{\pasts{\bo}}(\lambda, \theta)$, we start with a Langevin-type equation for the unnormalized state from Eqs.~\eqref{eq-updatetheta}. As we mentioned in the previous subsection, the statistics of $u\ik$ are governed by the ostensible PDF $\wp\ost(u\ik)$, appropriate to the unnormalized state. Therefore, we can take $\dd v = \lim_{\ddt \rightarrow \dt} u_t\ddt $ to have the same statistics as the Wiener increment. This then leads to the It\^o rule, $ \lim_{\ddt \rightarrow \dt} u_t^2\ddt^2 = \dt$, which makes the terms with $(\ddt - u_t^2\ddt^2)$ in Eqs.~\eqref{eq-updatetheta} disappear. We obtain a two-dimensional Langevin equation for $\theta$ and $\qp$,
\begin{subequations}
\begin{align}\label{eq-langevin}
\dd \theta = & \, A_{\theta}(\theta) \dt  + B_{\theta}(\theta) \dd v,\\
\dd \qp  = & A_\qp(\qp, \theta) \dt + B_\qp(\qp,\theta) \dd v,
\end{align}
\end{subequations}
where we have defined
\begin{subequations}
\begin{gather}
A_\theta(\theta) = - \Omega  + \tfrac{\gamma}{2}  \sin\theta - \tfrac{\Gamma}{2}  \cos\theta \sin\theta, \quad B_\theta(\theta) = - \sqrt{\Gamma} (\cos\theta + 1),\\
A_\qp(\qp,\theta) = - \tfrac{\gamma}{2}   (\cos\theta+1)\, \qp,  \quad B_\qp(\qp, \theta) = -  \sqrt{\Gamma} \sin\theta \, \qp.
\end{gather}
\end{subequations}
From the Langevin equation, we can write a corresponding Fokker-Planck equation for $Q_{\pasts{\bo}}(\qp,\theta)$, describing the diffusion in the two-dimensional space $(\theta,\qp)$,
\begin{align}\label{eq-FPE}
\partial_t Q_{\pasts{\bo}}&(\qp, \theta) = - \sum_{j} \partial_j [ A_j(\qp,\theta) \, Q_{\pasts{\bo}}(\qp,\theta )] +  \sum_{i,j} \tfrac{1}{2}\partial_i\partial_j \left[ B_i (\qp,\theta) B_j (\qp,\theta) \, Q_{\pasts{\bo}}(\qp,\theta) \right],
\end{align}
where the summations are over the coordinate labels, $i, j \in \{ \qp, \theta \}$. %which can be easily normalized by dividing it with a norm ${\cal N}(t) = \int \! \dd \theta\, \twp_{\past\bo}(\theta,t)$.

The unnormalized true PDF in the $\theta$ variable is defined in Eq.~\eqref{eq-quasi-int}.
Substituting this into Eq.~\eqref{eq-FPE}, we obtain a partial differential equation for the unnormalized true PDF,
\begin{align}\label{eq-PDEsol}
\partial_t \twp_{\pasts{\bo}}(\theta) = &- \tfrac{\gamma}{2}(1+\cos\theta)\twp_{\pasts{\bo}}(\theta)  + \partial_\theta \left\{  \left[ \sqrt{\Gamma}\sin\theta \, B_\theta(\theta)- A_\theta(\theta) \right] \twp_{\pasts{\bo}}(\theta)\right\} +   \tfrac{1}{2} \partial^2_\theta\left[ B_\theta^2(\theta) \twp_{\pasts{\bo}}(\theta) \right].
\end{align}
This can be solved numerically with an appropriate initial condition and a periodic boundary condition for the variable $\theta$. Thus we can finally obtain the normalized past-only and past-future conditional PDFs of the true state,
\begin{align}
\wp_{\pasts{\bo}}(\theta)  = &  \frac{\twp_{\pasts{\bo}}(\theta)}{\int\!  \dd \theta  \, \twp_{\pasts{\bo}}(\theta)},\\
\wp_{\bothps{\bo}}(\theta) =  & \frac{\twp_{\pasts{\bo}}(\theta) {\rm Tr} \big [ \hat E_{\futps{\bo}}\, \hat S(\theta) \big ] }{\int\!  \dd \theta\, \twp_{\pasts{\bo}}(\theta) {\rm Tr}\big[\hat E_{\futps{\bo}} \,\hat S(\theta) \big ]}.
\end{align}
We show in Appendix~\ref{sec-app-retroeff} how the retrofiltered matrix ${\hat E}_{\futps{\bo}}$ can be simply calculated from the no-observed-jump evolution. These PDFs above are then used to compute the smoothed state (the estimator for ${\textbf{Q}_1}$): 
\begin{align}
 \rho_{\textbf{Q}_1} = &\, \rho\sm =   \int\!\! \dd \theta \, \hat S(\theta) \,\wp_{\bothps{\bo}}(\theta).
\end{align}
Moreover, following Eqs.~\eqref{eq-th-q-nf} and \eqref{eq-th-q-ne}, the other two estimators in this subsection, $\rho_{\textbf{Q}_2}$ and $\rho_{\textbf{Q}_3}$, are found as
\begin{align}
 \rho_{\textbf{Q}_2} & =  \hat{\psi}\sm^{\rm max} = \hat S(\theta_{\textbf{Q}_2}),\\
 \rho_{\textbf{Q}_3} & = \hat S(\theta_{\textbf{Q}_3}).
\end{align}
The first line corresponds to the max-eigenvalue eigenvector of the smoothed state, \ie, $ \rho\sm  |{\psi}\sm^{\rm max} \ra = \lambda\sm^{\rm max}|{\psi}\sm^{\rm max} \ra$, which leads to $\theta_{\textbf{Q}_2} = \arg {\rm Tr}[(\hat \sigma_z + i \hat \sigma_y)\rho\sm]$. The second line is the most-likely state obtained from the most-likely angle, $\theta_{\textbf{Q}_3} = \argmax_{\theta} \wp_{\bothps{\bo}}(\theta)$. 

\begin{figure}[t]
\centering
\includegraphics[width=16.6cm]{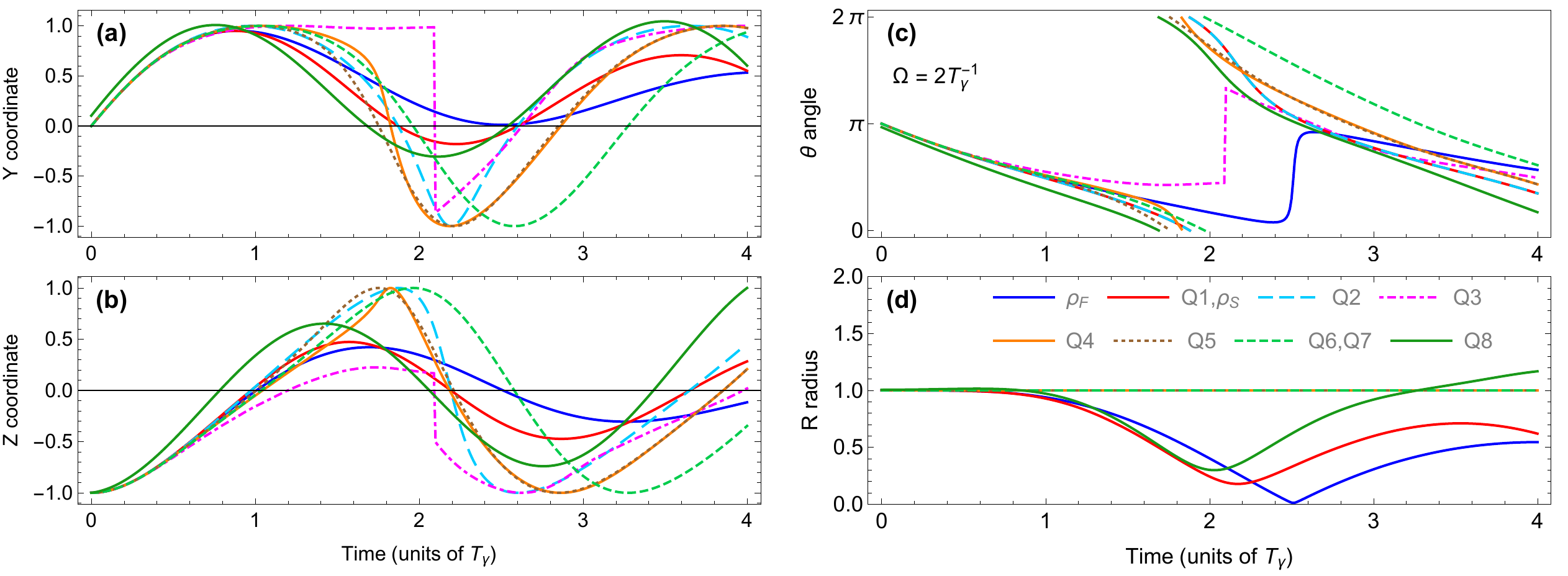}
\caption{Dynamics of the eight estimators (seven estimators of the true states, $\rho_{\textbf{Q}_1}$\,--\,$\rho_{\textbf{Q}_7}$, and the SWV state $\varrho_{\textbf{Q}_8}$) for the example of a resonantly driven and damped qubit. The estimators at any time $t$ are conditioned on the past-future observed measurement record, which is a string of photon detection events. The time interval shown, $[0, T)$, is the block of time between two such detections (jumps), 
 not including the state after the second jump (which is the same as its initial state, except for the case $\varrho_{\textbf{Q}_8}$).   The eight estimators are shown in: (a) $y$-coordinate, (b) $z$-coordinate, (c) qubit's angle $\theta$ around the $x =0$ plane, and (d) qubit's radius $R$. The usual filtered (past-only) quantum trajectory $\rho\fil$ is also shown in blue curves for comparison. The parameters used in the numerical simulation are: $T = 4 T_\gamma$ and $\Omega = 2 T_\gamma^{-1}$, in the units of the system's total decay time  $T_\gamma$.  The parameters are chosen such that we can see an interesting feature: the discontinuity in the most-likely state $\rho_{\textbf{Q}_3}$ right after the time $t = 2.09 T_\gamma$.}
\label{fig-trajs}
\end{figure}

\begin{figure}[t]
\centering
\includegraphics[width=11cm]{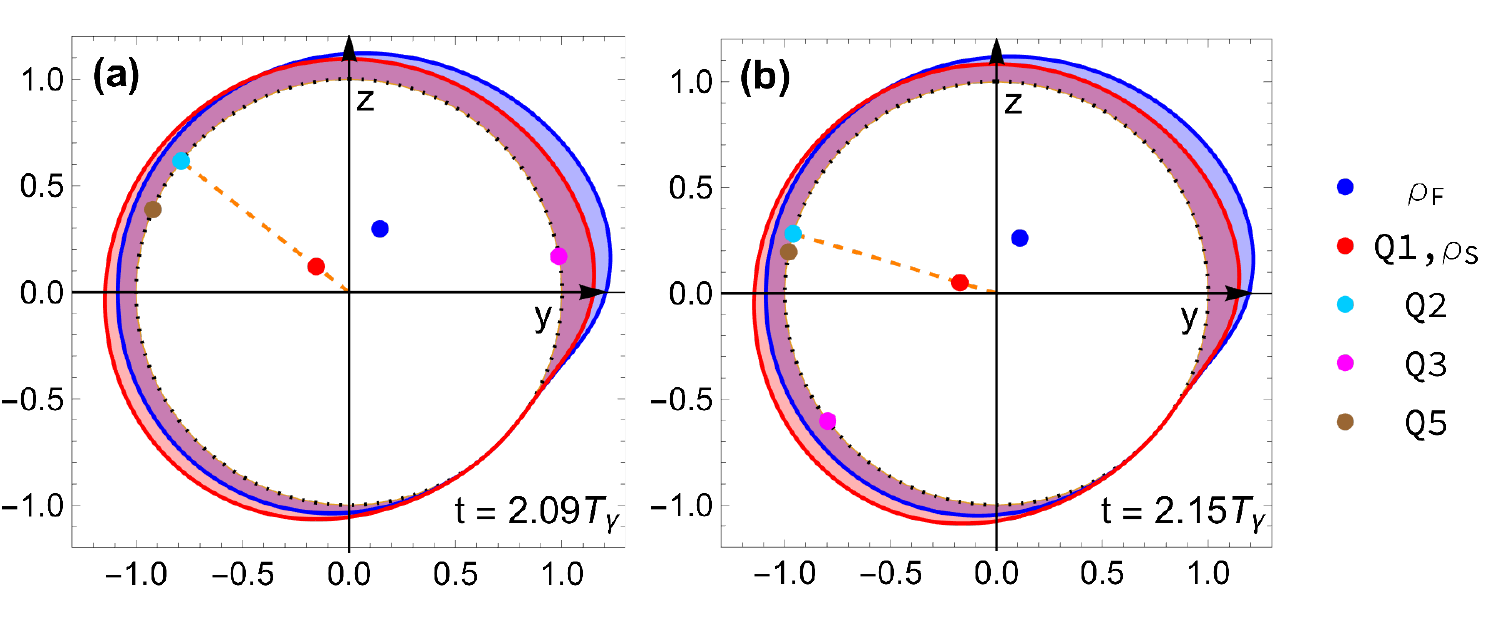}
\caption{The Bloch great circles (dotted black) and the PDFs of (pure) true states, $\wp_{{\protect \pasts{\bo}}}(\theta)$ (solid blue curves) and $\wp_{{\protect \bothps{\bo}}}(\theta)$ (solid red curves), at any qubit's angle $\theta \in (0,2\pi]$, for two different times: (a) $t = 2.09 T_\gamma$ and (b) $t = 2.15 T_\gamma$. The `height' of the probability density is measured in the radial direction. Colored dots show some qubit's state estimators ($\rho\fil$, $\rho\sm$, $\rho_{\textbf{Q}_2}$, $\rho_{\textbf{Q}_3}$ and $\rho_{\textbf{Q}_5}$) for both times, using the same color scheme as in Figure~\ref{fig-trajs}. The time of the final jump, and other parameters, are the same as in Figure~\ref{fig-trajs}. Note the interesting large jump that happens for the most-likely state $\rho_{\textbf{Q}_3}$ (magenta) between the two times. The orange dashed lines are to show that the estimator $\rho_{\textbf{Q}_2}$ (cyan) always has exactly the same Bloch angle as the smoothed state (red).} 
\label{fig-blochprob}
\end{figure}

We show in Figure~\ref{fig-trajs} the numerical results of qubit state estimators between two jumps, compared with the regular filtered quantum trajectory $\rho\fil$ obtained from Eq.~\eqref{eq-filterqubit} (blue curves). We solve the partial differential equation in Eq.~\eqref{eq-PDEsol} (using Mathematica's `NDSolve' package) with a delta-function initial state replaced with an approximated narrow-width Gaussian function $\wp_{t=0} (\theta) = (2\pi \sigma^2)^{-1/2}\exp[ - (\theta-\pi)^2/2\sigma^2]$, for the initial ground state $\theta_0 = \pi$. We chose $\sigma = 0.01$, which gives the value at its peak to be $\wp_{t=0}(\pi) \approx 39.89$ instead of infinity.

The first three estimators, $\rho_{\textbf{Q}_1} = \rho\sm$, $\rho_{\textbf{Q}_2}$, and $\rho_{\textbf{Q}_3}$, are shown in red, dashed light blue, and dot-dashed magenta curves, respectively. The smoothed state $\rho\sm$ is not necessarily pure, while the other two estimators are pure states with unit radius; see Figure~\ref{fig-trajs}(d). At the final time, $T=4 T_\gamma$, one might expect that $\rho\fil(T) = \rho\sm(T)$ since the conditioning on the past record should be the same as the conditioning on the whole record at the final time. However, this is not true for our example because the last jump is not included in the past record $\past{\bo}$. One could include the last jump by considering the state estimators with one more time step, but all estimators would simply jump to the ground state at the new final time. Also note that, in Figure~\ref{fig-trajs}(c), the estimator $\rho_{\textbf{Q}_2}$ always has exactly the same Bloch angle as the smoothed state  (also shown as orange dashed lines in Figure~\ref{fig-blochprob}). This is because $\rho_{\textbf{Q}_2}$, in Eq.~\eqref{eq-th-q-nf}, is simply the purest state possible with the same Bloch angle as the smoothed state.

We have chosen the parameter regime (detail in Figure~\ref{fig-trajs}'s caption) such that we can see an interesting feature: the discontinuity in the most-likely state $\rho_{\textbf{Q}_3}$ at $t = 2.09 T_\gamma$. This occurs because the most-likely state follows the highest point of the PDF of true states Eq.~\eqref{eq-bayes} at any local time. The PDF can have multiple competing peaks, which results in most-likely states jumping between local maxima as the PDF changes in time. We show in Figure~\ref{fig-blochprob} how the PDFs and the jumping states (magenta dots) look, right before the jump time, $t = 2.09 T_\gamma$, and not long after the jump, $t = 2.15 T_\gamma$. Also, in Figure~\ref{fig-blochprob}(a)-(b), we can see that the smoothed state and $\rho_{\textbf{Q}_2}$ have the same angle on the Bloch sphere at both times.

%\begin{align}
%\label{eq-maptheta} \bm{{\cal E}}(\theta\ik, u\ik, \bo\ik = 0) = &\,\, \theta\ik -\dt \, \Omega + \dt \tfrac{\Gamma + \gamma}{2}  \sin\theta\ik \nonumber \\
%& - u\ik \dt  \sqrt{\Gamma}(\cos\theta\ik+1) ,
%\end{align}

\subsection{State estimators with CDJ optimization}\label{sec-mlp}
For the estimators of $\textbf{Q}_4$ and $\textbf{Q}_5$, we need to optimize the PDFs of the unknown measurement record conditioned on the observed record. We follow the optimization approach presented in Section~\ref{sec-sevenQSE}. As in the previous subsection, we are interested in the block of qubit's dynamics between any two jumps, where the observed record is given by $\bothp{\bo} = \{ \bo_0 , \bo_{\ddt}, ..., \bo_{T-\ddt} ,\bo_T \}\tp = \{ 0,0,..., 0,1\}\tp$ and the unknown record $u\ik$ is from the $y$-homodyne measurement. 

Following the CDJ optimization, we need to solve the difference equations in Eqs.~\eqref{eq-diffeq1} and Eqs.~\eqref{eq-diffeq2}, for the estimators $\rho_{\textbf{Q}_4}$ and $\rho_{\textbf{Q}_5}$, respectively. These difference equations have to be derived from the state update for normalized states, given by the $\theta$ equation in Eqs.~\eqref{eq-updatetheta}, and the PDFs of time-local unknown measurement results. For the state update, since a solution of the difference equations gives smooth optimal unknown records and their corresponding smooth state paths, we need to treat the variable $u_t$ as zeroth order in $\ddt$, instead of ${\cal O}(\ddt^{-1/2})$ as in the Fokker-Planck approach. Therefore, the state mapping $\theta_{t+\ddt} = \bm{{\cal E}}(\theta\ik, u\ik, \bo\ik =0)$ is given by
\begin{align}
\theta_{t+\ddt} = &\, \theta_t - \ddt\, \Omega + \ddt \tfrac{\gamma+\Gamma}{2} \sin\theta_t - u_t \ddt \sqrt{\Gamma}(\cos\theta_t+1),  \\
\label{eq-lambdt}\lambda_{t+\ddt} =& \, \left[1 - \ddt \tfrac{\gamma+\Gamma}{2} ( \cos\theta_t + 1 ) - u_t \ddt \sqrt{\Gamma} \sin\theta_t \right ] \lambda_t.
\end{align}
This is the update equation for the true state Eqs.~\eqref{eq-updatetheta} keeping up to first order in $\ddt$. For the PDF of the unknown result, we can use the equation of the state norm,
\begin{align}
\wp(u\ik, \bo\ik = 0 | \theta\ik) =&\, \, \wp\ost(u_t) {\rm Tr}[{\cal N}_0 {\cal M}_{u\ik} \hat S(\theta\ik) ] \nonumber \\
= & \,\, \wp\ost(u_t)\lambda_{t+\ddt}(\theta\ik, u\ik,\lambda_t)/\lambda_t,
\end{align}
with the form of $\lambda_{t+\ddt}$ given in Eq.~\eqref{eq-lambdt}. This leads to the logarithm of the probability to first order in $\ddt$,
%\begin{align}
%\wp(u\ik, \bo\ik = 0 | \theta\ik) \approx & \exp\left\{ - \frac{\dt}{2} \left(u\ik + \sqrt{\Gamma}\sin\theta\ik \right)^2 \right\} \nonumber \\ 
%& \times \exp\left\{- \frac{\dt}{2}\gamma\,(1+\cos\theta\ik) \right\},
%\end{align}
%where the measurement operation is defined with a measurement operator ${\hat M}_{u\ik,O\ik=0} = (\dt/2\pi)^{1/4}\exp(u_t^2 \dt/4)\{\hat{1} - \tfrac{1}{2}({\hat c}\subn^\dagger {\hat c}\subn + {\hat c}\subn^2 + {\hat c}\subp^\dagger {\hat c}\subp + {\hat c}\subp^2) \dt +{\hat c}\subp u_t \dt + {\cal O}(\dt^2)\}$~\cite{Wong2020} (see Appendix~\ref{sec-app-ppdf} for derivations).
\begin{align}
\ln \wp(u\ik, & \bo\ik = 0 | \theta\ik) = \tfrac{1}{2} \ln (\ddt/2\pi) -  \ddt \left[ \tfrac{1}{2} u_t^2 +  \tfrac{\gamma+\Gamma}{2} ( \cos\theta_t + 1 ) + u_t  \sqrt{\Gamma} \sin\theta_t \right].
\end{align}
Putting all the components together, we can construct the difference equations as in Eqs.~\eqref{eq-diffeq1} or \eqref{eq-diffeq2}. 
Taking the continuum limit $\ddt \rightarrow \dt$, we obtain a set of ordinary differential equations (ODEs) for $\theta_t$, its conjugate $p_t$, and an optimal record $u_t$,
\begin{subequations}\label{eq-diffeqcont}
\begin{align}
\partial_t \theta_t  = & - \Omega  + \tfrac{\Gamma+\gamma}{2} \sin\theta_t -  \sqrt{\Gamma}(\cos\theta_t+1) u_t, \label{eq-thetaeq}\\
\partial_t p_t = &  -  \tfrac{\Gamma+\gamma}{2}\,p_t\, \cos\theta_t - \sqrt{\Gamma} \, p_t\, \sin\theta_t\, u_t 
+ \sqrt{\Gamma}\cos\theta_t\, u_t - \tfrac{\Gamma + \gamma}{2}  \sin\theta_t ,\\
u_t = & -  \sqrt{\Gamma}\left( p_t + p_t \cos\theta_t + \sin\theta_t \right).
\end{align}
\end{subequations}
The initial condition for these ODEs is the qubit's initial state $\theta_0 = \pi$ (ground state). However, the final condition for the ODEs will depend on the problem of interest; that is, whether Eqs.~\eqref{eq-diffeqcont} will be used for $\textbf{Q}_4$ or $\textbf{Q}_5$.

Let us first consider the simpler case, $\textbf{Q}_5$, where the optimal unknown record and its corresponding state estimator $\rho_{\textbf{Q}_3}$ are calculated from the ODEs just once with the final boundary condition at time $T$. Following Eqs.~\eqref{eq-diffeq2}, the final condition is in the third line. Using the final observed jump record, we get ${\hat E}_T = {\hat c}_{\rm o}^{\dagger} {\hat c}_{\rm o} \dt = \gamma \dt |e\ra \la e |$ and $\wp(\bo_T = 1 | \theta_T) = {\rm Tr}[ {\hat c}_{\rm o}^{\dagger} {\hat c}_{\rm o} \hat S(\theta_T) ] \dt  = \tfrac{\gamma}{2}(1+\cos\theta_T)\dt$. This leads to the final condition for the conjugate variable,
\begin{align}\label{eq-finalbc2}
 p_T =  \,  - \frac{\sin\theta_T}{1+\cos\theta_T},
\end{align}
where we have approximated $p_T = \lim_{\ddt \rightarrow \dt} p_{T-\delta t}$ for the time-continuum limit. Solving Eq.~\eqref{eq-diffeqcont} with the final condition Eq.~\eqref{eq-finalbc2}, we obtain the optimal $\both{u} = \{ u_t : t \in [0, T)\}\tp$ and its state path $\{ \theta_t : t \in [0, T]\}\tp$, where the latter is exactly the qubit state estimator $\rho_{\textbf{Q}_5}$.

For the estimator of $\textbf{Q}_4$, the calculation is more involved, since we need to solve the ODEs for every intermediate time $t=\tau$ during the time period of interest $\tau \in (0, T]$. Following Eqs.~\eqref{eq-diffeq1}, we numerically solve the ODEs, Eqs.~\eqref{eq-diffeqcont}, for discrete values of $\tau \in \{ \ddt, 2\ddt, ..., T\}$, with the final conditions,
\begin{align}
 p_\tau = & \,\,\frac{\partial}{\partial \theta_\tau} \ln {\rm Tr} [ {\hat E}_{\futps{\bo}}\, \hat S(\theta_\tau) ]  \nonumber \\
 = &\, \frac{  - \zeta \sin\theta_\tau +  \beta \cos\theta_\tau}{\alpha + \zeta \cos\theta_\tau +  \beta \sin\theta_\tau},
\end{align}
for any time $\tau$. Here we have defined the retrofiltered matrix's elements as
\begin{align}
{\hat E}_{\futps{\bo}}=  \alpha\, \hat 1 + \beta \, \hat\sigma_y + \zeta \, \hat\sigma_z,
\end{align}
where all elements are functions of $\tau$ (see Appendix~\ref{sec-app-retroeff}). In contrast to the previous $\rho_{\textbf{Q}_5}$, the state estimator $\rho_{\textbf{Q}_4}$ is a series of $\theta_\tau$, each of which is a `final' qubit's state via the solution of the ODEs, Eqs.~\eqref{eq-diffeqcont}, for each value of $\tau \in \{ \ddt, 2\ddt, ..., T\}$. As noted before, the qubit state estimator $\rho_{\textbf{Q}_4}$ at the final time $\tau = T$ is exactly the same as $\rho_{\textbf{Q}_5}$.

We solve for $\rho_{\textbf{Q}_4}$ and $\rho_{\textbf{Q}_5}$ using the numerical method and Python codes developed by Lewalle \emph{et~al.}~\cite{LewCha17,Lewalle2018chaos}. Since the ODEs are non-linear, there can be multiple solutions, which serve as multiple optimal paths (in a similar way that a PDF could have multiple peaks). We therefore solve for all candidate most probable solutions and pick the one which does maximize the joint probability density of the path, \eg calculated from Eqs.~\eqref{eq-mlp1-prob} or \eqref{eq-probmlp2}. The solutions of $\rho_{\textbf{Q}_4}$ and $\rho_{\textbf{Q}_5}$ are shown as solid orange and dotted brown curves, respectively, in Figure~\ref{fig-trajs}. They are similar in shape, have unit purities, and the same final states $\rho_{\textbf{Q}_4}(T) = \rho_{\textbf{Q}_5}(T)$ as expected, but they are not exactly the same. In Figure~\ref{fig-blochprob}, only $\rho_{\textbf{Q}_5}$ is shown as the brown dots. The corresponding optimal record for $\rho_{\textbf{Q}_5}$ will be discussed more in Section~\ref{sec-eightestqubit}.

\subsection{State estimators using local optimal records}
For the last three estimators of $\textbf{Q}_6$--$\textbf{Q}_8$, their cost functions are defined with the time-local unknown records. We have assumed that the unknown record was from a continuous diffusive measurement ($y$-homodyne measurement for the qubit example). Therefore, the measurement can be thought of as a series of infinitely weak measurements, each with an infinitesimal duration $\dt$. As mentioned in Section~\ref{sec-sevenQSEdiff}, the maximum value of the PDF in Eq.~\eqref{eq-probut} is the same as its mean value given by the weak value in Eq.~\eqref{eq-wv-gen2}. Thus we can calculate the state estimators $\rho_{\textbf{Q}_6} = \rho_{\textbf{Q}_7}$ from the state update Eqs.~\eqref{eq-updatetheta} using the weak value record in place of $u\ik$. For the last estimator $\varrho_{\textbf{Q}_8}$, we use Eq.~\eqref{eq-wvsO} to compute the SWV state. It is important to note that since, in our case, the Lindblad operator for the unknown measurement record ${\hat c}_{\rm u}$ is not Hermitian, the SWV state is not associated with the weak value defined in Eq.~\eqref{eq-wv-gen}, \ie, $\la \bu_\tau \ra_{\bothps{\bo}} \ne {\rm Tr} [ ({\hat c}_{\rm u} + {\hat c}^\dagger_{\rm u}) \varrho_{\textbf{Q}_8}  ] $, given the definition of the mean in Eq.~\eqref{eq-wv-gen2}. This would be true if ${\hat c}_{\rm u}$ were Hermitian, when Eqs.~\eqref{eq-wv-gen} and \eqref{eq-wv} coincide. 

We show the numerical results for the last three estimators in Figure~\ref{fig-trajs} as dashed green (for $\rho_{\textbf{Q}_6} = \rho_{\textbf{Q}_7}$) and solid dark green (for the SWV state $\varrho_{\textbf{Q}_8}$) curves. As expected, we see in Figure~\ref{fig-trajs}(d) that the former estimators are pure quantum states with unit Bloch radius, while the the SWV `state' is not even constrained within the valid quantum state space as it can have a Bloch radius larger than one.

\begin{figure}
\centering
\includegraphics[width=16.6cm]{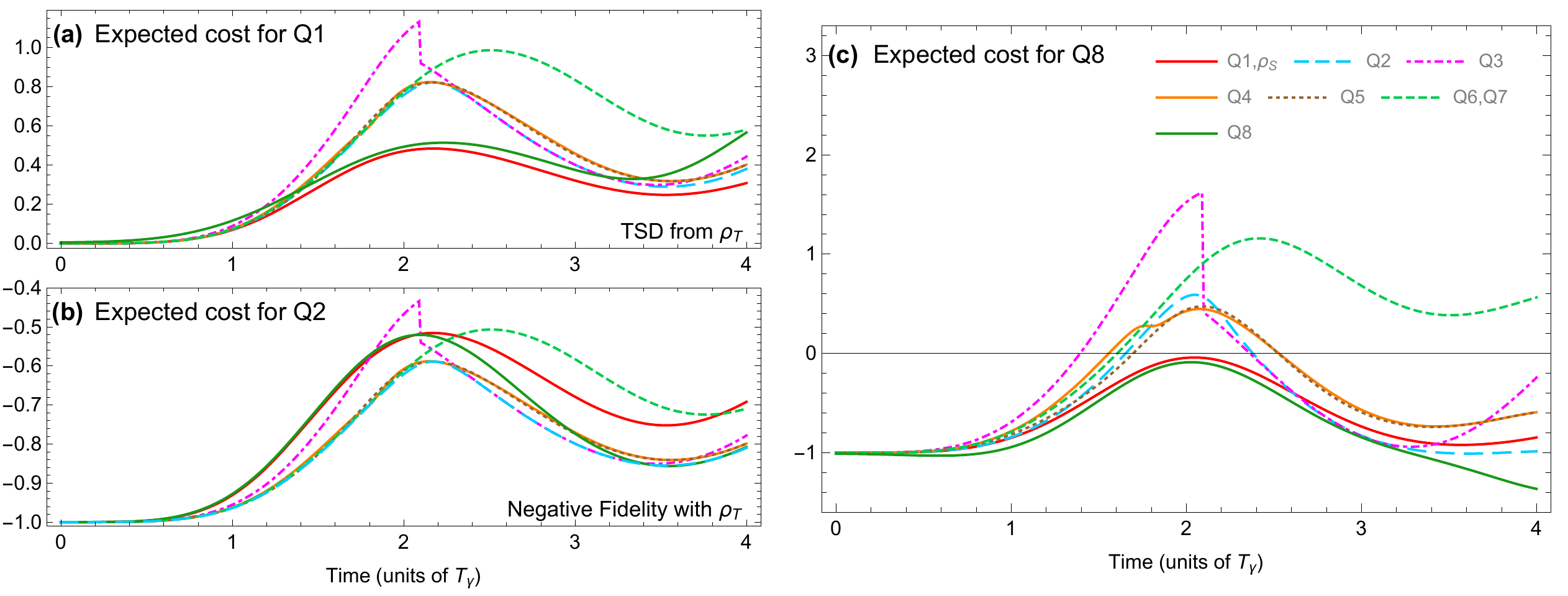}
\caption{Expected costs for ${\textbf{Q}_1}$, ${\textbf{Q}_2}$, and ${\textbf{Q}_8}$ (in panels (a)-(c), respectively) are numerically evaluated for all eight estimators. The first two expected costs are defined in the true state space: the trace square deviation (${\cal C}_1$) and the negative fidelity (${\cal C}_2$) from the true state. The optimal estimators for the two costs are the smoothed quantum state, $\rho_{\textbf{Q}_1}$, and the max-eigenvalue eigenvector of the smoothed state, $\rho_{\textbf{Q}_2}$. The last expected cost, ${\cal C}_8$, in panel (c), is defined in Eq.~\eqref{eq-cq8}, for which the SWV state, $\varrho_{\textbf{Q}_8}$, is the optimal estimator.}
\label{fig-cost1}
\end{figure}

%For the last, the expected cost is the negative probability density, $-\wp_{{\protect \bothp\bo}}(\theta,t)$, plotted on a log scale for the absolute value. This is numerically solved using a narrow-width Gaussian function as its initial state, so that $-\wp_{{\protect \bothp\bo}}(\theta = \pi,t=0) \approx -39.89$ \hmw{The plot seems to show it starting at $-50$. Also here's an idea to consider: just truncate the plot at $-5$ because they are all practically identical there, so that we can see the separation better. Q1 should not be in the legend.} instead of minus infinity at the initial time. The results show that the associated estimator for each of the costs gives the smallest values, as expected.

\subsection{Expected costs for qubit estimators}\label{sec-eightestqubit}

From the eight estimators presented in Figure~\ref{fig-trajs}, we can also show that they do optimize their corresponding expected cost functions compared to all the other estimators. We do this by evaluating, where possible, the expected costs (or equivalent versions theoreof) discussed in Section~\ref{sec-expcost} for the eight qubit estimators. We first numerically calculate the local-state expected costs that can be applied to all eight estimators. The expected costs are ${\cal C}_1$, ${\cal C}_2$, and ${\cal C}_8$ and the results are shown in Figure~\ref{fig-cost1}(a), (b), and (c), respectively. We see that the estimators $\rho_{\textbf{Q}_1}$, $\rho_{\textbf{Q}_2}$, and $\varrho_{\textbf{Q}_8}$ give the minimum values for the expected costs ${\cal C}_1$, ${\cal C}_2$, and ${\cal C}_8$, respectively.

\begin{figure}
\centering
\includegraphics[width=16.6cm]{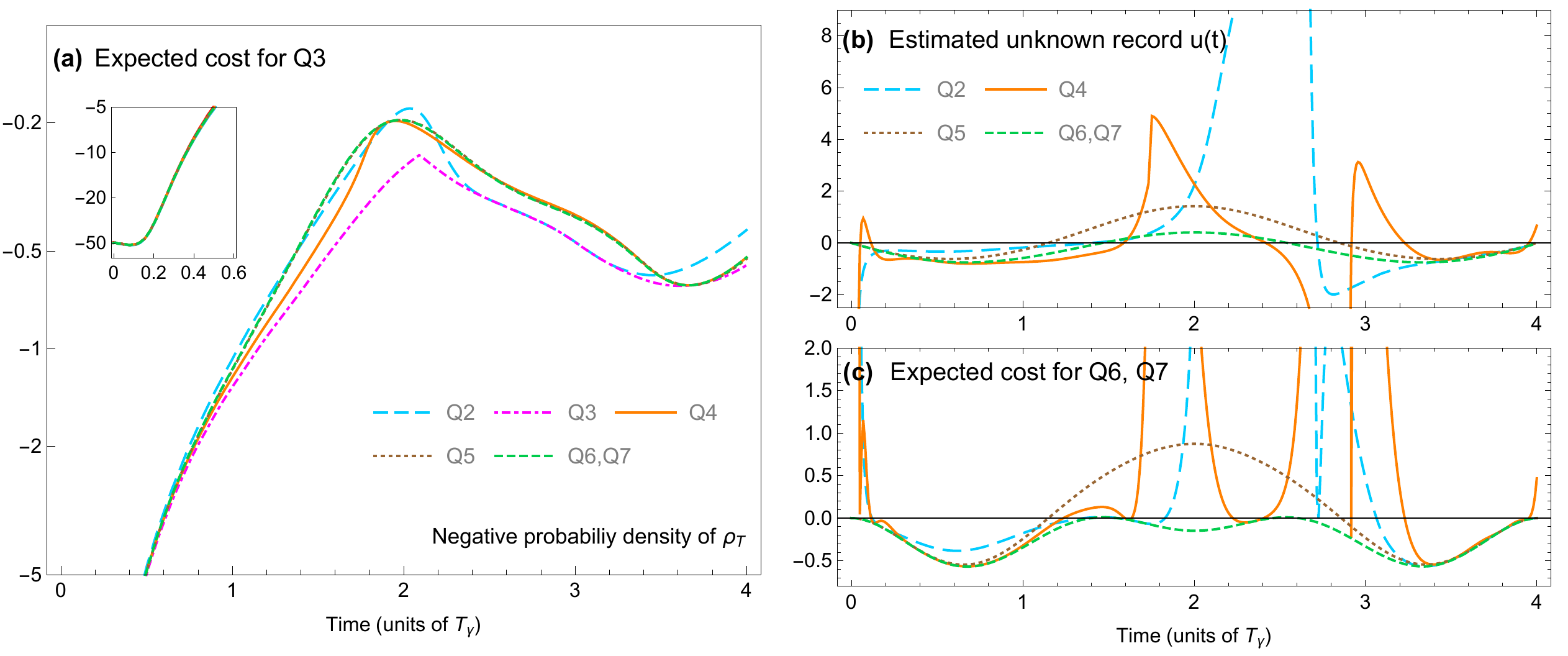}
\caption{Expected costs for ${\textbf{Q}_3}$, ${\textbf{Q}_6}$ and ${\textbf{Q}_7}$ are numerically evaluated for relevant optimal and sub-optimal state estimators (pure-state estimators only). We show in (a) the expected costs ${\cal C}_3$ which can be applied to all six pure-state estimators: $\rho_{\textbf{Q}_2}$\,--$\rho_{\textbf{Q}_7}$, showing that the most-likely state, ${\rho}_{\textbf{Q}_3}$ is optimal. In panel (b), we show the associated (estimated) unknown records solved for $\rho_{\textbf{Q}_2}$, $\rho_{\textbf{Q}_4}$\,--$\rho_{\textbf{Q}_7}$, excluding $\rho_{\textbf{Q}_3}$ because it has a discontinuity in the state evolution (see texts). The modified expected costs ${\cal C}_6 = {\cal C}_7$ in Eqs.~\eqref{eq-cq6} and \eqref{eq-cq7} are shown in panel (c).}
\label{fig-cost2}
\end{figure}

For ${\cal C}_3$, the expected cost (negative equality with the true state) can only be applied to pure-state estimators. This is because we assume the true state is pure and thus the likelihood for a mixed state to be the true state is zero. The results are shown in Figure~\ref{fig-cost2}(a), where the expected cost ${\cal C}_3$ is evaluated for the pure-state estimators: $\rho_{\textbf{Q}_2}$\,--$\rho_{\textbf{Q}_7}$. 
 The most-likely state $\rho_{\textbf{Q}_3}$ gives the lowest value for the negative probability density at all time. As we have discussed in Section~\ref{sec-expcost}, the expected cost ${\cal C}_4$ is excluded from our analysis because it is extremely difficult to calculate.
%\hmw{We should be clear: that means that the cost for the others is higher. They are not possible true states, so the probability is zero, which would be infinitely high on that log-scale plot. We should say that.}
%\hmw{Even though it is out of order, I think Q8 should go here. It is, after all, a local-state expected cost. If you want to avoid jumping backwards and forwards you could even start with Q8. After all, the plot looks closest to Q1, which is not surprising since classically (or for filtering only) they are the same estimator. So Fig. 4 would have 3 sublots, and Fig. 5 would have 3 also, including Q3, or 4, including a narrow rectangle for Q5. That way it is (almost) the same set of estimators being plotted in each subfigure for 4 and for 5. And so rewrite. But actually, how come 4(c) has a legend that says all 8 of the states are plotted there, when it should only be 5 of them. In fact, the red curve should definitely not be there.}

For the expected costs defined in the unknown record space, ${\cal C}_5$--${\cal C}_7$, 
the relevant estimators have to also be pure states, because we need to solve for corresponding unknown records that reproduce them as possible evolutions of the pure true state.  The calculation for the unknown records can be done in different ways, but it is not guarantee that any pure-state evolution will have a physical corresponding unknown record. For $\rho_{\textbf{Q}_2}$, $\rho_{\textbf{Q}_3}$, and $\rho_{\textbf{Q}_4}$, their records can be calculated by reversing the dynamics in Eq.~\eqref{eq-thetaeq}. For $\rho_{\textbf{Q}_5}$ and $\rho_{\textbf{Q}_6}$ (or $\rho_{\textbf{Q}_7}$), their associated unknown records can be determined in the process of calculating the state estimators in Eq.~\eqref{eq-diffeq2} and \eqref{eq-wv-gen2}. However, for our driven-qubit example, the state estimator $\rho_{\textbf{Q}_3}$ exhibits the discontinuity (as shown in Figure~\ref{fig-trajs}) at $T = 2.09 T_{\gamma}$, which leads to an unphysical infinite measurement result at that particular time. Therefore, we only show the estimated unknown records for the five pure-state estimators in Figure~\ref{fig-cost2}(b), excluding the estimator $\rho_{\textbf{Q}_3}$.

Using the estimated unknown records, we can calculate the expected costs ${\cal C}_5$--${\cal C}_7$. The expected cost, ${\cal C}_5$, is the log ratio of the probability densities of the entire unknown record, as shown in Eq.~\eqref{eq-cq5}. This quantity is non-local in time and therefore is a single number associated with each of the state estimators: $1.8\times 10^{16}$ for $\rho_{\textbf{Q}_2}$, $8.7\times 10^{10}$ for $\rho_{\textbf{Q}_4}$, $0$ for $\rho_{\textbf{Q}_5}$, and $3.3\times 10^{1}$ for $\rho_{\textbf{Q}_6}$ and $\rho_{\textbf{Q}_7}$. The expected costs ${\cal C}_6$ and ${\cal C}_7$ in Eqs.~\eqref{eq-cq6} and \eqref{eq-cq7} are exactly the same and are presented in Figure~\ref{fig-cost2}(b), showing that the estimators $\rho_{\textbf{Q}_6}=\rho_{\textbf{Q}_7}$ give the minimum value.

%Lastly, as 
The expected cost functions we have presented so far are dependent on the duration of time $T$ between two jumps. 
 However, in some cases, 
we can also calculate an average of the expected cost over all possible jump times, $T$. This can be done by defining a jump-time-average expected cost as
\begin{align}\label{eq-avetimecost}
\bar{\cal C}_\kappa (\bullet)  \equiv & \int^\infty_0 \!\!\! \dd T \, \wp_{\rm wait}(T) \,  {T}^{-1} \!\! \int_0^T\!\!\! \dt \, \, {\cal C}_\kappa (\bullet_t),
\end{align}
where the first time-integral is weighted by the waiting-time PDF, %probability  for the wait time (for jumps), $P_{\rm wait}(T)$, 
normalized as $\int^\infty_0 \! \dd T \, \wp_{\rm wait}(T) = 1$. The second time-integral, $T^{-1}\int_0^T \!\dt\, \bullet$, is to average over the local time, $t$, which is applied to the expected cost that is $t$-dependent.  The waiting-time PDF can be calculated from,  
\begin{align}\label{eq-jumptimeavg}
\wp_{\rm wait}(T) & = {\rm Tr}\left[ \hat{c}_{\rm o}^\dagger \hat{c}_{\rm o}\, \rho_{\pasts{\bo}}(T) \right] {\rm Tr}\left[ {\trho}_{\pasts{\bo}}(T)\right]  \nonumber \\
& = {\rm Tr}\left[ \hat{c}_{\rm o}^\dagger \hat{c}_{\rm o}\, \trho_{\pasts{\bo}}(T) \right],
\end{align}
where the unnormalized filtered state is given by
\begin{align}
\trho_{\pasts{\bo}}(T) =  {\cal M}_{O_{T-\ddt}} e^{\ddt{\cal L}_{\rm u}}\cdots \,  {\cal M}_{O_{0}} e^{\ddt{\cal L}_{\rm u}}\rho_0.
\end{align}
The first line in Eq.~\eqref{eq-jumptimeavg} is a product of the probability of an observed jump occurred at time $T$ (given that the system's state is $\rho\fil(T)$) multiplied by the probability of having no jumps before that (for the length of time $T$). The latter is given by the trace of the unnormalized filtered state, which tells us the probability of the observed no-jump record from $t=0$ to $t = T$.

We approximate the integrals in Eq.~\eqref{eq-jumptimeavg} by using numerical summations (details are in Appendix~\ref{sec-app-numer}). The results are shown in Table~\ref{tab-avecost}, where the jump-time-average expected costs are calculated for ${\textbf{Q}_1}$, ${\textbf{Q}_2}$, ${\textbf{Q}_3}$, and ${\textbf{Q}_8}$, applied to relevant estimators. The minimal values of the average are shown in bold  font.  These results show that the optimal estimators minimize their expected costs,  
not just for the value $T=4 T_\gamma$. We note that we did  not include ${\textbf{Q}_5}$ in our calculation for the jump-time-average expected cost, because solving the whole-record optimization becomes extremely resource-extensive for large values of $T$. The same reason holds for why we did not apply the calculation to $\rho_{\textbf{Q}_4}$ and $\rho_{\textbf{Q}_5}$. For ${\textbf{Q}_6}$ and ${\textbf{Q}_7}$, there were problems in reversing the dynamics to get reasonable estimated unknown records, and records could be infinite in some cases.  (Evidence of this divergence can be seen for the estimated record for $\rho_{\textbf{Q}_2}$ and $\rho_{\textbf{Q}_4}$ shown in Figure~\ref{fig-cost2}(b).)

{\renewcommand{\arraystretch}{1.5}
\renewcommand{\tabcolsep}{0.2cm}
\begin{table}[t]
\centering
\begin{tabular}{|c|c|c|c|c|c|}
\hline
\backslashbox{Cost function}{Estimator} & $\textcolor{colq1}{\rho_{\textbf{Q}_1}}$  & $\textcolor{colq2}{\rho_{\textbf{Q}_2}}$ & $\textcolor{colq3}{\rho_{\textbf{Q}_3}}$ & $\textcolor{colq67}{\rho_{6,7}}$ & $\textcolor{colq8}{\varrho_{\textbf{Q}_8}}$ \\
\hline
{\bf $\la$TrSDf$\rhoU\ra$}: $\textcolor{colq1}{{{\cal C}_1}}$, Eq.~(\ref{eq-cq1}) &  {\bf 0.31} & 0.46 & 0.48 & 0.62 & 0.34 \\
{\bf $\la$nFw$\rhoU\ra$}: $\textcolor{colq2}{{{\cal C}_2}}$, Eq.~(\ref{eq-cq2}) &  -0.69 & {\bf -0.77} & -0.76 & -0.69 & -0.70 \\
{\bf $\la$nEw$\rhoU\ra$}: $\textcolor{colq3}{{{\cal C}_3}}$, Eq.~(\ref{eq-cq3}) &  NA & -2.76 & {\bf -2.85} & -2.69 & NA \\
{\bf SWV}: $\textcolor{colq8}{{{\cal C}_8}}$, Eq.~(\ref{eq-cq8}) &  -0.44 & -0.18 & -0.07 & 0.22 & {\bf -0.51} \\
\hline
\end{tabular}

\caption{Numerical results for the jump-time-average expected costs ${\bar {\cal C}}_\kappa(\rho)$ for $\kappa = 1,2,3,8$ applied to relevant estimators: $\rho_{\textbf{Q}_1}$, $\rho_{\textbf{Q}_2}$, $\rho_{\textbf{Q}_3}$, $\rho_{\textbf{Q}_6}=\rho_{\textbf{Q}_7}$, and $\varrho_{\textbf{Q}_8}$.  `NA' indicates `Not applicable'. The bold numbers are to emphasize the minimum values in each row. Details of the numerical calculation are presented in Appendix~\ref{sec-app-numer}. }
\label{tab-avecost}
\end{table}}

%%%%%%%%%%%%%%%%%%%
% Discussion
%%%%%%%%%%%%%%%%%%%%%
\section{Discussion}\label{sec-discussion}
%Discuss about the generalization to true mixed states and which cost functions would not apply.
% [Discuss: what we did, how can this be useful and impact on the field, and what are the leftover questions to be asked ] 

We have presented the cost-optimized quantum state estimation theory, unifying various existing formalisms that use the past-future information for quantum systems. The three main existing theories are the smoothed quantum state (minimizing the trace square deviation cost from true states), the most-likely path (minimizing the negative equality with entire true unknown records), and the smoothed weak-value state (minimizing the square deviation of an expectation value from true weak measurement results). As shown in Figure~\ref{fig-diagramQ}, the theory outlines explicit connections of possible estimators using various choices of cost functions, where the latter can be defined either in the unknown state space or the unknown measurement record space. 
%We also used the classical-quantum state correspondence and showed that the classical analogue of such state estimation theory can be defined. Interesting connections were shown in Figure~\ref{fig-diagramC}, where the quantum state smoothing and the smoothed weak-value state both reduce to the classical state smoothing as expected.

In addition to clarifying the connection among existing formalisms, the theory suggested six new quantum state estimators ($\rho_{\textbf{Q}_2}$\,--$\rho_{\textbf{Q}_8}$) that could be used in estimating unknown quantum states in different contexts. It is noted that, for problems with unknown quantities, there is no single best guess for an unknown, but there is an optimal guess with respect to a chosen cost function. The theory was then applied to the coherently driven qubit coupled to bosonic baths, showing that all but two of the estimators are distinct, even though they all represent an optimal estimate of the unknown qubit's state. We do not claim to have compiled all possible cost functions in this work; however, we have built the framework that allows other estimators to be introduced.

One obvious generalization of our proposed formalism would be to problems where the true states are \emph{mixed}. Indeed, this is necessary for experimental tests or  practical applications of the formalism, in which Bob is a real party with a real measurement record, not a hypothetical omniscient observer. That is because an actual experiment performed by Bob will not be able to garner all of the information unavailable to Alice. Some information will be irrevocably lost to the environment, causing the state jointly conditioned on Alices's and Bob's measurement records to be mixed.   %only a portion of it.   
%since the unknown true states in real experiments are generally mixed. 
For most cases, except $\textbf{Q}_2$, the form of the estimators presented here still hold, after replacing the true states, $\rhoOU = \psiOU$, with pure mixed states. For $\textbf{Q}_2$, the fidelity simplification in Eq.~\eqref{eq-fidel} will need to be modified for mixed states, which will likely result in an optimal estimator different from Eq.~\eqref{eq-th-q-nf}.

The central concern of this paper was conditioning on the past-future information. 
However, the idea of different optimal estimates resulting from different cost function can also be applied to  conditioning only on past observation, or even no observation at all. Some estimators can be easily surmised. For example, for $\textbf{Q}_1$, the quantum smoothed state will be replaced by the quantum filtered state (for the past-only conditioning), or by a solution of the Lindblad master equation (for the no observation conditioning). However, for some other estimators, the generalization is quite involved, and this is a topic for future investigation.

\section{Acknowledgments}
We would like to thank Philippe Lewalle and John Steinmetz for providing Python software used in numerically solving the most-likely paths. We acknowledge the traditional owners of the land on which this work was undertaken 
at Griffith University, the Yuggera people. This research is supported by the Australian Research Council 
Centre of Excellence Program CE170100012 and Mahidol University (Basic Research Fund: fiscal year 2021) Grant Number BRF1-A29/2564.  A.C.~acknowledges the support of the Griffith University 
Postdoctoral Fellowship scheme and the Australia Awards Endeavour Scholarships and Fellowships.



\appendix

\setcounter{section}{0}% Reset numbering for sections
\renewcommand{\thesection}{\Alph{section}}

\section{ Probability density function for time-local unknown records}\label{sec-app-pdf}
Here we discuss some properties of the PDF of the time-local unknown record, $\wp_{\bothps{\bo}}(u_\tau)$, and show that the mean and the mode of both distributions are the same for a weak measurement with a Lindblad operator $\op{c} = \op{c}_{\rm u}$. We note that the derivation below is applied to the past-only conditional PDF, $\wp_{\pasts{\bo}}(u_\tau)$, by taking $\op E_{\futps{\bo}} = \hat I$. 

From the definition of the PDF in Eq.~\eqref{eq-probut}, we obtain
\begin{align}
 \wp_{\bothps{\bo}}(u_\tau) = & \,  \frac{\Tr{ {\op E}_{\futps{\bo}}\,{\hat M}_{u_\tau} \rho_{\pasts{\bo}}\, {\hat M}_{u_\tau}^\dagger}}{\int \!\!\dd u_t \, \Tr{ {\op E}_{\futps{\bo}}\,{\hat M}_{u_t} \rho_{\pasts{\bo}}\, {\hat M}_{u_t}^\dagger}},
\end{align}
where ${\hat M}_{u\ik} = (\ddt/2\pi)^{1/4}\exp(u_t^2\ddt/4) [  \hat{1} - \tfrac{1}{2}( {\hat c}^\dagger {\hat c} + {\hat c}^2) \ddt +{\hat c}\, u_t \ddt  + {\cal O}(\ddt^2) ]$ (similar to Eq.~\eqref{eq-moput} in the main text, but without the Hamiltonian and only keeping to first order in $\ddt$) is the measurement operator for the unknown record $u\ik$ at any time $t$. We can show that the PDF $\wp_{\bothps{\bo}}(u_\tau)$ can be approximated as a Gaussian function, correct up to first order in $\ddt$,
\begin{align}\label{eq-app-probuboth}
\wp_{\bothps{\bo}}(u_\tau) \approx & \left(\frac{\ddt}{2\pi}\right)^{1/2} \!\!\!\! \exp\left(-u_\tau^2 \ddt/2\right) \left[   \op{1} + u_\tau \frac{\Tr{\op E_{\futps{\bo}}\,  \op c  \, \rho_{\pasts{\bo}} + \rho_{\pasts{\bo}}\, \op c^\dagger\, \op E_{\futps{\bo}}}}{\Tr{\op E_{\futps{\bo}} \,\rho_{\pasts{\bo}}}}  \ddt +{\cal O}(\ddt^2) \right] .
%\wp_{\past\bo}(u_\tau) \approx \left(\frac{\dt}{2\pi}\right)^{\tfrac{1}{2}} \exp\left\{ - \left(u_\tau + \sqrt{\Gamma}\sin\theta \right)^2\right\},
\end{align}
The PDF definitely has the same mean and mode, \ie, 
\begin{align}\label{eq-app-probupast}
\la u_\tau \ra_{\bothps{\bo}} = \argmax_{u_\tau} \wp_{\bothps{\bo}} (u_\tau) =\frac{\Tr{\op E_{\futps{\bo}}\,  \op c  \, \rho_{\pasts{\bo}} + \rho_{\pasts{\bo}}\, \op c^\dagger\, \op E_{\futps{\bo}}}}{\Tr{\op E_{\futps{\bo}} \,\rho_{\pasts\bo}}},
\end{align}
which is exactly the real part of the weak value in Eq.~\eqref{eq-wv-gen2}.
The second moment of this distribution is dominated by $1/\ddt$, that is $\la u_\tau^2 \ra_{\bothps{\bo}} \approx 1/\ddt$, therefore the distribution can be approximated as a Gaussian function with variance $1/\ddt$. Similarly, the mean and mode for the filtering PDF, $\wp_{\pasts{\bo}}(u_\tau)$, are then simply $\la u_\tau \ra_{\pasts{\bo}} = \argmax_{u_\tau} \wp_{\pasts{\bo}} (u_\tau) = {\rm Tr} [ (  \op c  + \op c^\dagger) \rho_{\pasts{\bo}} ]$.

\section{ Retrodictive effect matrix for the driven qubit example}\label{sec-app-retroeff}
The POVM $\hat E_{\futps{\bo}}$ can be computed semi-analytically for the coherently-driven qubit example used in the main text. Here, we rewrite the POVM for the partially observed system, Eq.~\eqref{eq-povm} in the main text,
\begin{align}\label{eq-povm-app}
{\hat E}_{\futps{\bo}} = e^{\ddt{\cal L}_{\rm u}^\dagger} {\cal M}_{O_\tau}^\dagger \cdots e^{\ddt{\cal L}_{\rm u}^\dagger} {\cal M}_{O_{T-\ddt}}^\dagger \hat E_T,
\end{align}
which can be used to compute a PDF of the future record given a state $\rho_\tau$ at time $\tau$,
\begin{align}
\wp(\,\futp{\bo}\,|\rho_\tau) = \Tr{ {\hat E}_{\futps{\bo}}\, \rho_\tau}.
\end{align}
In the main text, we are interested in the observed record $\bothp{\bo} = \{ \bo_0 , \bo_{\ddt}, ..., \bo_{T-\ddt} ,\bo_T \}\tp = \{ 0,0,..., 0,1\}\tp$ including the final jump at time $T$. Therefore, the final POVM is given by
\begin{align}
{\hat E}_T = {\hat c}_{\rm o}^{\dagger} {\hat c}_{\rm o} \ddt = \gamma \ddt |e\ra \la e |,
\end{align}
and the POVM at other times can be calculated from a dynamical equation
\begin{align}
{\hat E}_{t-\ddt}  = e^{\ddt{\cal L}_{\rm u}^\dagger} {\cal N}^\dagger_0 {\hat E}_t,
\end{align}
where we have replaced all ${\cal M}^\dagger_{O_t}$ for $t \in [\tau, T)$ in Eq.~\eqref{eq-povm-app} by the adjoint of the no-jump map ${\cal N}^\dagger_0$. We can take the continuum limit $\ddt \rightarrow \dt$ and write the dynamical equation in the form, ${\hat E}_{t-\dt}  = {\hat E}_t -  {\rm d}{\hat E}$, where ${\rm d}{\hat E}$ is given by
\begin{align}
- {\rm d}{\hat E} =& + i \dt [\tfrac{\Omega}{2} \hat \sigma_x, {\hat E}_t] + \dt {\cal D}^{\dagger}[{\hat c}_{\rm u}]{\hat E}_t - \dt\, \frac{1}{2} \left( {\hat c}_{\rm o}^{\dagger} {\hat c}_{\rm o} {\hat E}_t + {\hat E}_t{\hat c}_{\rm o}^{\dagger} {\hat c}_{\rm o}\right).
\end{align}
The first two terms and the last term are from expanding the operations $e^{\ddt{\cal L}_{\rm u}^\dagger}$ and ${\cal N}^\dagger_0$ to first order in $\ddt$ (or $\dt$), respectively. 

Given the form of the Hermitian POVM matrix
\begin{align}
{\hat E}_{\futps{\bo}}=  \left( \begin{matrix} \alpha + \zeta & -i \beta \\ i \beta & \alpha - \zeta\end{matrix} \right), 
\end{align}
where  $\alpha$, $\beta$, and $\zeta$  are real numbers as functions of time (one can check that the real part of its diagonal elements has to be zero).  From this we obtain a set of differential equations for the matrix elements,
\begin{align}  %-   (I removed a minus from both sides -- it is the same eqn as before.
-\frac{{\rm d}}{\dt}\left( \begin{matrix} \alpha \\  \beta \\  \zeta \end{matrix} \right) = -\frac{1}{2}\left( \begin{matrix} {\gamma} &  0 & \gamma+2\Gamma\\
0& \gamma+\Gamma &-2\Omega \\ 
\gamma&  2\Omega & \gamma+2\Gamma\end{matrix} \right) 
\left( \begin{matrix} \alpha \\ \beta \\ \zeta \end{matrix} \right),  
\end{align} 
which can be solved backwards in time (as indicated by the sign of the left hand side)  to get $\hat E_{\futps{\bo}}$ at time $\tau$ from the final condition $\hat E_T =  \gamma \ddt |e\ra \la e |$,   \ie, $\alpha=\zeta = \gamma \ddt /2$, $\beta=0$.

\section{ Numerical calculation for time-average expected costs}\label{sec-app-numer}

The jump-time-average expected cost and the waiting-time PDF are given in Eqs.~\eqref{eq-avetimecost} and \eqref{eq-jumptimeavg}, respectively. However, the time integral with the infinite wait time limit is impractical, given that one needs to calculate the state estimators for any wait time $T$ numerically. We therefore approximate the wait time integral with a finite summation such that an average of an arbitrary function $f(T)$ is given by
\begin{align}
\int_0^\infty \!\!\! \dd T \, \wp_{\rm wait}(T) f(T) & = \int_0^1\!\!\! \dd x \, f( G^{-1}(x)),\\
& = \Delta x \sum_{j = 1}^J  f( G^{-1}(j \Delta x)),
\end{align}
where $G(\tau) \equiv \int_0^\tau \! \dd T \, \wp_{\rm wait}(T)$ and $\Delta x = 1/J$. For simplicity, let us denote a set of jump times by $\{ T_j \}$ where each element $T_j \equiv  G^{-1}(j \Delta x)$, for $j = 1, 2,...,J$, we can write the time-average expected cost Eq.~\eqref{eq-avetimecost} as
\begin{align}
\bar{\cal C}_\kappa(\bullet) \approx  \frac{\sum_{j=1}^J  \Delta t \sum_{k=0}^{K_j}  \,  {\cal C}_{\textbf{Q}_\kappa}(\bullet_k)   }{\sum_{j=1}^J  \sum_{k=0}^{K_j} \Delta t},
\end{align}
where we have also discretized the second time integral in Eq.~\eqref{eq-avetimecost} with a step size $\Delta t$ and $K_j$ is the time steps given the jump time $T_j = K_j \Delta t$. The numerical results shown in Table~\ref{tab-avecost} are calculated with $\Delta x = 0.1$ and $\Delta t = 0.05$.

\begin{figure}[h]
\centering
\includegraphics[width=16.6cm]{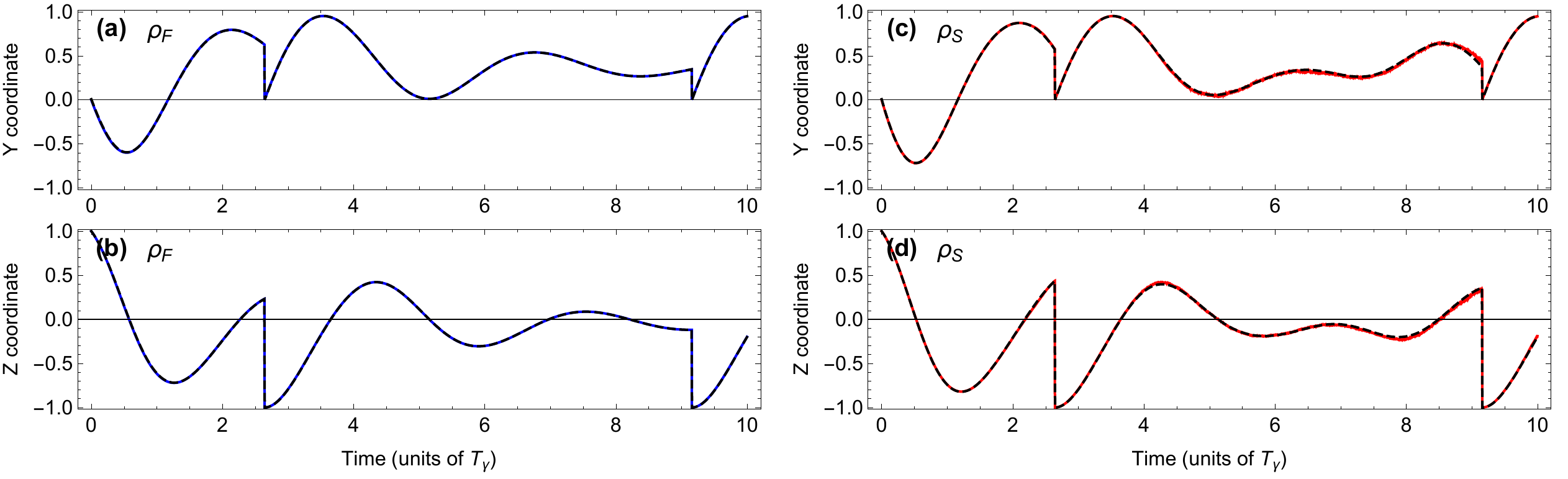}
\caption{Comparison of the filtered and smoothed states calculated using two different techniques. The dashed curves are from the Fokker-Planck techniques presented in Section~\ref{sec-fpe} and the solid colored are from trajectory simulation in Ref.~\cite{Ivonne2015,chantasri2019}.}
\label{fig-compare}
\end{figure}

Moreover, in order to show that our technique solving the filtered and smoothed states using the Fokker-Planck equation Section~\ref{sec-fpe} are correct, we show in Figure~\ref{fig-compare} the comparison of the states with solutions from simulating all possible true states and averaging them as in Eq.~\eqref{eq-filstate} and \eqref{eq-qssstate}. We use the numerical techniques and C++ programming codes used  in Refs.~\cite{Ivonne2015,chantasri2019,GueWis20}.

\section{ Glossary of acronyms and abbreviations}
\label{Glossary}

\begin{table}[h]
\centering
\begin{tabular}{|c|c|}
\hline
Abbreviation & Meaning\\
\hline
SD & Square Deviation\\[1mm]
nE & negative Equality\\[1mm]
BME & Bayesian Mean Estimator\\[1mm]
MLE & Maximum Likelihood Estimator\\[1mm]
PDF & Probability Density Function\\[1mm]
$\Sigma$SD & sum Square Deviation\\[1mm]
TrSD & Trace Square Deviation\\[1mm]
TSVF & Two-State Vector Formalism\\[1mm]
POVM & Positive Operator-Valued Measure\\[1mm]
SWV & Smoothed Weak-Value\\[1mm]
QED & Quantum Electrodynamics\\[1mm]
CDJ & Chantasri, Dressel and Jordan\\[1mm]
ODE & Ordinary Differential Equation\\[1mm]
\hline
\end{tabular}
\caption{List of abbreviations (in order of appearance)}
\end{table}

\begin{table}[h]
\centering
\setlength{\extrarowheight}{8pt}
\begin{tabular}{|p{6mm}|p{0.32\textwidth}|c|c|}
\hline
& (Expected) Cost Function & Acronym & Definition\\\hline
${\bf Q}_1$ & {\bf Tr}ace-{\bf S}quare {\bf D}eviation \newline {\bf f}rom the true state & {\bf $\la$TrSDf$\rhoU\ra$} & $\left\la {\rm Tr} \left [  \left(\rho - \rhoOU\right)^2  \right ] \right \ra_{\pasts{\bu} | \bothps{\bo}}$\\\hline
${\bf Q}_2$ & {\bf n}egative {\bf F}idelity \newline {\bf w}ith the true state & {\bf $\la$nFw$\rhoU\ra$} & $\left\la -F\!\left[\rho, \rhoOU \right] \right\ra_{\pasts{\bu} | \bothps{\bo}}$\\\hline
${\bf Q}_3$ &  {\bf n}egative {\bf E}quality \newline {\bf w}ith the true state & {\bf $\la$nEw$\rhoU\ra$} & $\left\la - \delta_{\rm H}\!\left[ \rho - \rhoOU \right] \right\ra_{\pasts{\bu} | \bothps{\bo}}$\\\hline
${\bf Q}_4$ & {\bf n}egative {\bf E}quality \newline {\bf w}ith the past unknown record & {\bf $\la$nEw$\past\bu\ra$} & $\left\la -\delta \left(\past{u} - \past{\bu} \right) \right\ra_{\pasts{\bu} | \bothps{\bo}}$\\\hline
${\bf Q}_5$ & {\bf n}egative {\bf E}quality \newline {\bf w}ith the entire unknown record & {\bf $\la$nEw$\both{\bu}\ra$} & $\left\la - \delta \left(\both{u} - \both{\bu} \right) \right\ra_{\boths{\bu} | \bothps{\bo}}$\\\hline
${\bf Q}_6$ & {\bf n}egative {\bf E}quality \newline {\bf w}ith the unknown record at $\tau$ & {\bf $\la$nEw$\bu_\tau \ra$} & $\left\la - \delta \left(u_\tau - \bu_\tau \right) \right\ra_{\bu_\tau | \bothps \bo}$\\\hline
${\bf Q}_7$ & {\bf S}quare {\bf D}eviation \newline {\bf f}rom the unknown record at $\tau$ & {\bf $\la$SDf$\,\bu_\tau\ra$} & $\left\la \left(u_\tau - \bu_\tau \right)^2 \right\ra_{ \bu_\tau | \bothps{\bo}}$\\\hline
${\bf Q}_8$ & Sum Square Deviation \newline from the weak measurement result ({\bf S}moothed {\bf W}eak-{\bf V}alue state) & {\bf SWV}& $\displaystyle{\sum_{j=1}^{d^2-1}\left\la \left[ {\rm Tr}(\varrho \op{\Lambda}_j) - \Lambda^w_j \right]^2\right\ra_{\Lambda^w_j  | \bothps{\bo}}}$\\
\hline
\end{tabular}
\caption{List of cost functions and definitions of expected cost functions}
\end{table}

\begin{table}
\centering
\begin{tabular}{|c|c|l|}
\hline
Notation & Definition & Description\\
\hline
${\bf x}$ & & The configuration, a vector of $d$ classical variables.
\\[2mm]
$\mathbb{X}$ & & The set of possible configurations ${\bf x}$ for the classical system.
\\[2mm] 
$\wp({\bf x})$ & & \makecell{A probability density function of the configuration ${\bf x}$. \\ Also referred to as the classical state.}
\\[4mm]
$\wp_{\rm D}({\bf x})$ & $\wp({\bf x}|{\rm D})$ & \makecell{The probability density function of the configuration ${\bf x}$ \\ conditioned  on $\rm D$. Also referred to as the conditioned \\ (classical) state.}
\\[5mm]
$\langle\bullet\rangle_{A|{\rm D}}$ & $\displaystyle{\int}\dd\mu(A) \wp_{\rm D}(A) \bullet$ & \makecell{The expectation value, averaging over $A$ and conditioned on $\rm D$, \\ where the integration measure $\dd\mu(A)$ will vary depending on $A$.}
\\[4mm]
$\est A_{\rm C}$ & & \makecell{An optimal estimator, given $D$, for cost function \\ with abbreviation ${\rm C}$.}
\\[4mm]
$\mathbb{P}$ & $\{\wp({\bf x}): {\bf x} \in \mathbb{X} \}$ & \makecell{The set of all possible normalized probability density functions \\ of the configuration.}
\\[4mm]
$\rho$ &  & A density operator/quantum state.
\\[2mm]
$\tilde{\rho}$ & & An unnormalized quantum state.
\\[2mm]
$\rho_{\rm D}$ & & A quantum state conditioned on $\rm D$.
\\[2mm]
$\hat{\psi}$ & $\ket{\psi}\bra{\psi}$ & A pure quantum state.
\\[2mm]
$\varrho$ & &  An indefinite Hermitian matrix with unit trace.
\\[2mm]
$\hat{E}$ & & A POVM element/quantum effect.
\\[2mm]
$\hat{E}_{\rm D}$ & & A quantum effect conditioned on D.
\\[2mm]
$\mathbb{H}$ & & The Hilbert space of the quantum system.
\\[2mm]
$\mathfrak{G}({\mathbb{H}})$ & & The set of all valid density operators in the Hilbert space.
\\[2mm]
$\mathfrak{G}'({\mathbb{H}})$ & & The set of unit-trace, Hermitian operators in the Hilbert space.
\\[2mm]
$\dd\mu_{\rm H}(\hat{\psi})$ & & The Haar measure over pure states.
\\[2mm]
$\past{R}$ & $\{R_t: t\in[0, \tau)\}\tp$ & The measurement results $R_t$ {\em prior} to the estimation time 
$\tau$.
\\[2mm]
$\fut{R}$ & $\{R_t: t\in[\tau, T)\}\tp$ & \makecell{The measurement results $R_t$ including at and {\em posterior} \\ to the estimation time $\tau$, {\em excluding} the result at the final time $T$. }
\\[4.5mm]
$\futp{R}$ & $\{R_t: t\in[\tau, T]\}\tp$ & \makecell{The measurement results $R_t$ including at and {\em posterior} \\ to the estimation time $\tau$, {\em including} the result at the final time $T$. }\\[4.5mm]
$\both{R}$ & $\{R_t: t\in[0, T)\}\tp$ & \makecell{The set of all measurement results $R_t$ both {\em prior} and {\em posterior} \\ to the estimation time $\tau$, {\em excluding} the result at the final time $T$. }\\[4.5mm]
$\bothp{R}$ & $\{R_t: t\in[0, T]\}\tp$ & \makecell{The set of all measurement results $R_t$ both {\em prior} and {\em posterior} \\ to the estimation time $\tau$, {\em including} the result at the final time $T$. } \\[4.5mm]
$ _{\hat f}\langle X^w\rangle_{\hat i}$ & $\displaystyle{\lim_{\epsilon \to 0}\int_{-\infty} ^{\infty}\dd x \,x\,\wp(x|\hat{i},\hat{f}) }$ & \makecell{The weak-value associated with pre- ($\hat{i}$) and post-selecting ($\hat{f}$) \\  on the  outcome of a weak measurement of $\hat{X}$; \\ $\epsilon$ is the measurement strength.} \\[5mm]
\hline
\end{tabular}
\caption{List of notations and descriptions}
\end{table}

\clearpage 

\bibliographystyle{elsarticle-num} 
%\bibliography{UnifyingBib}

\begin{thebibliography}{100}
\expandafter\ifx\csname url\endcsname\relax
  \def\url#1{\texttt{#1}}\fi
\expandafter\ifx\csname urlprefix\endcsname\relax\def\urlprefix{URL }\fi
\expandafter\ifx\csname href\endcsname\relax
  \def\href#1#2{#2} \def\path#1{#1}\fi

\bibitem{BookJazwinski}
A.~H. Jazwinski, {Stochastic processes and filtering theory}, Chapman and Hall,
  New York, 1970.

\bibitem{KalBuc1961}
R.~E. Kalman, R.~S. Bucy, {New results in linear filtering and prediction
  theory}, Journal of Basic Engineering, Transactions of the ASME 83~(1) (1961)
  95--108.
\newblock \href {http://dx.doi.org/10.1115/1.3658902}
  {\path{doi:10.1115/1.3658902}}.

\bibitem{Kushner1964}
H.~J. Kushner, {On the differential equations satisfied by conditional
  probability densities of Markov processes, with applications}, Journal of the
  Society for Industrial and Applied Mathematics Series A Control 2~(1) (1964)
  106--119.
\newblock \href {http://dx.doi.org/10.1137/0302009}
  {\path{doi:10.1137/0302009}}.

\bibitem{FraPot1969}
D.~C. Fraser, J.~E. Potter, {The optimum linear smoother as a combination of
  two optimum linear filters}, IEEE Transactions on Automatic Control 14~(4)
  (1969) 387--390.
\newblock \href {http://dx.doi.org/10.1109/TAC.1969.1099196}
  {\path{doi:10.1109/TAC.1969.1099196}}.

\bibitem{BookWienerSmt}
N.~Wiener, {Extrapolation, interpolation and smoothing of stationary time
  series: with engineering applications}, Martino Fine Book, 2013.

\bibitem{Wheatley2010}
T.~A. Wheatley, D.~W. Berry, H.~Yonezawa, D.~Nakane, H.~Arao, D.~T. Pope, T.~C.
  Ralph, H.~M. Wiseman, A.~Furusawa, E.~H. Huntington, {Adaptive optical phase
  estimation using time-symmetric quantum smoothing}, Physical Review Letters
  104~(9) (2010) 093601.
\newblock \href {http://dx.doi.org/10.1103/PhysRevLett.104.093601}
  {\path{doi:10.1103/PhysRevLett.104.093601}}.

\bibitem{Ivonne2015}
I.~Guevara, H.~Wiseman, {Quantum state smoothing}, Physical Review Letters
  115~(18) (2015) 180407.
\newblock \href {http://dx.doi.org/10.1103/PhysRevLett.115.180407}
  {\path{doi:10.1103/PhysRevLett.115.180407}}.

\bibitem{Huang2018}
Z.~Huang, M.~Sarovar, {Smoothing of Gaussian quantum dynamics for force
  detection}, Physical Review A 97~(4) (2018) 042106.
\newblock \href {http://dx.doi.org/10.1103/PhysRevA.97.042106}
  {\path{doi:10.1103/PhysRevA.97.042106}}.

\bibitem{Laverick2018}
K.~T. Laverick, H.~M. Wiseman, H.~T. Dinani, D.~W. Berry, {Adaptive estimation
  of a time-varying phase with coherent states: Smoothing can give an unbounded
  improvement over filtering}, Physical Review A 97~(4) (2018) 042334.
\newblock \href {http://dx.doi.org/10.1103/PhysRevA.97.042334}
  {\path{doi:10.1103/PhysRevA.97.042334}}.

\bibitem{LavCha2019}
K.~T. Laverick, A.~Chantasri, H.~M. Wiseman, {Quantum state smoothing for
  linear Gaussian systems}, Physical Review Letters 122~(19) (2019) 190402.
\newblock \href {http://dx.doi.org/10.1103/PhysRevLett.122.190402}
  {\path{doi:10.1103/PhysRevLett.122.190402}}.

\bibitem{BookHelstrom}
C.~W. Helstrom, {Quantum detection and estimation theory (Vol 123 of
  Mathematics in Science and Engineering)}, Academic Press, New York, 1976.

\bibitem{Tsangsmt2009}
M.~Tsang, {Time-symmetric quantum theory of smoothing}, Physical Review Letters
  102~(25) (2009) 250403.
\newblock \href {http://dx.doi.org/10.1103/PhysRevLett.102.250403}
  {\path{doi:10.1103/PhysRevLett.102.250403}}.

\bibitem{Wise2002}
H.~M. Wiseman, {Weak values, quantum trajectories, and the cavity-QED
  experiment on wave-particle correlation}, Physical Review A 65~(3) (2002)
  032111.
\newblock \href {http://dx.doi.org/10.1103/PhysRevA.65.032111}
  {\path{doi:10.1103/PhysRevA.65.032111}}.

\bibitem{Budini2017}
A.~A. Budini, {Smoothed quantum-classical states in time-irreversible hybrid
  dynamics}, Physical Review A 96~(3) (2017) 032118.
\newblock \href {http://dx.doi.org/10.1103/PhysRevA.96.032118}
  {\path{doi:10.1103/PhysRevA.96.032118}}.

\bibitem{LuisPsmooth2017}
L.~P. Garc{\'{i}}a-Pintos, J.~Dressel, {Past observable dynamics of a
  continuously monitored qubit}, Physical Review A 96~(6) (2017) 062110.
\newblock \href {http://dx.doi.org/10.1103/PhysRevA.96.062110}
  {\path{doi:10.1103/PhysRevA.96.062110}}.

\bibitem{Budini2018b}
A.~A. Budini, {Entropic relations for retrodicted quantum measurements},
  Physical Review A 97~(1) (2018) 012132.
\newblock \href {http://dx.doi.org/10.1103/PhysRevA.97.012132}
  {\path{doi:10.1103/PhysRevA.97.012132}}.

\bibitem{BookWiseman}
H.~M. Wiseman, G.~J. Milburn, {Quantum measurement and control}, Cambridge
  University Press, UK, 2010.

\bibitem{Watanabe1955}
S.~Watanabe, {Symmetry of physical laws. Part III. Prediction and
  retrodiction}, Reviews of Modern Physics 27~(2) (1955) 179.
\newblock \href {http://dx.doi.org/10.1103/RevModPhys.27.179}
  {\path{doi:10.1103/RevModPhys.27.179}}.

\bibitem{Vaidman2017}
L.~Vaidman, {Weak value controversy}, Philosophical Transactions of the Royal
  Society A: Mathematical, Physical and Engineering Sciences 375~(2106) (2017)
  20160395.
\newblock \href {http://dx.doi.org/10.1098/rsta.2016.0395}
  {\path{doi:10.1098/rsta.2016.0395}}.

\bibitem{Tsang2019}
M.~Tsang, {Quantum analogs of the conditional expectation for retrodiction and
  smoothing: a unified view}, arXiv:1912.02711 [quant-ph].

\bibitem{ABL1964}
Y.~Aharonov, P.~G. Bergmann, J.~L. Lebowitz, {Time symmetry in the quantum
  process of measurement}, Physical Review 134~(6B) (1964) B1410.
\newblock \href {http://dx.doi.org/10.1103/PhysRev.134.B1410}
  {\path{doi:10.1103/PhysRev.134.B1410}}.

\bibitem{AV2002}
Y.~Aharonov, L.~Vaidman, {The two-state vector formalism of quantum mechanics},
  in: Muga J.G., Mayato R.S., Egusquiza I.L. (Eds.), Time in quantum mechanics,
  vol 72., Springer, Berlin, 2002, pp. 369--412.
\newblock \href {http://dx.doi.org/10.1007/3-540-45846-8_13}
  {\path{doi:10.1007/3-540-45846-8_13}}.

\bibitem{Gammelmark2013}
S.~Gammelmark, B.~Julsgaard, K.~M{\o}lmer, {Past quantum states of a monitored
  system}, Physical Review Letters 111~(16) (2013) 160401.
\newblock \href {http://dx.doi.org/10.1103/PhysRevLett.111.160401}
  {\path{doi:10.1103/PhysRevLett.111.160401}}.

\bibitem{AAV1988}
Y.~Aharonov, D.~Z. Albert, L.~Vaidman, {How the result of a measurement of a
  component of the spin of a spin-1/2 particle can turn out to be 100},
  Physical Review Letters 60~(14) (1988) 1351.
\newblock \href {http://dx.doi.org/10.1103/PhysRevLett.60.1351}
  {\path{doi:10.1103/PhysRevLett.60.1351}}.

\bibitem{Tsangsmt2009-2}
M.~Tsang, {Optimal waveform estimation for classical and quantum systems via
  time-symmetric smoothing}, Physical Review A 80~(3) (2009) 033840.
\newblock \href {http://dx.doi.org/10.1103/PhysRevA.80.033840}
  {\path{doi:10.1103/PhysRevA.80.033840}}.

\bibitem{Ohki15}
K.~Ohki, {A smoothing theory for open quantum systems: The least mean square
  approach}, Proceedings of the IEEE Conference on Decision and Control 54th
  IEEE~(CDC) (2015) 4350--4355.
\newblock \href {http://dx.doi.org/10.1109/CDC.2015.7402898}
  {\path{doi:10.1109/CDC.2015.7402898}}.

\bibitem{Ohki19}
K.~Ohki, {Quantum smoother for open quantum systems driven by quantum
  jump-diffusion processes}, Proceedings of the ISCIE International Symposium
  on Stochastic Systems Theory and its Applications (2019) 25--28\href
  {http://dx.doi.org/10.5687/sss.2019.25} {\path{doi:10.5687/sss.2019.25}}.

\bibitem{ZeiDem1987}
O.~Zeitouni, A.~Dembo, {A maximum a posteriori estimator for trajectories of
  diffusion processes}, Stochastics 20~(3) (1987) 221--246.
\newblock \href {http://dx.doi.org/10.1080/17442508708833444}
  {\path{doi:10.1080/17442508708833444}}.

\bibitem{DurBac1978}
D.~D{\"{u}}rr, A.~Bach, {The Onsager-Machlup function as Lagrangian for the
  most probable path of a diffusion process}, Communications in Mathematical
  Physics 60~(2) (1978) 153--170.
\newblock \href {http://dx.doi.org/10.1007/BF01609446}
  {\path{doi:10.1007/BF01609446}}.

\bibitem{DAD2014}
D.~A. Dutra, B.~O.~S. Teixeira, L.~A. Aguirre, {Maximum a posteriori state path
  estimation: Discretization limits and their interpretation}, Automatica
  50~(5) (2014) 1360--1368.
\newblock \href {http://dx.doi.org/10.1016/j.automatica.2014.03.003}
  {\path{doi:10.1016/j.automatica.2014.03.003}}.

\bibitem{Dykman1994}
M.~I. Dykman, E.~Mori, J.~Ross, P.~M. Hunt, {Large fluctuations and optimal
  paths in chemical kinetics}, The Journal of Chemical Physics 100~(8) (1994)
  5735--5750.
\newblock \href {http://dx.doi.org/10.1063/1.467139}
  {\path{doi:10.1063/1.467139}}.

\bibitem{Chantasri2013}
A.~Chantasri, J.~Dressel, A.~N. Jordan, {Action principle for continuous
  quantum measurement}, Physical Review A 88~(4) (2013) 042110.
\newblock \href {http://dx.doi.org/10.1103/PhysRevA.88.042110}
  {\path{doi:10.1103/PhysRevA.88.042110}}.

\bibitem{chantasri2015stochastic}
A.~Chantasri, A.~N. Jordan, {Stochastic path-integral formalism for continuous
  quantum measurement}, Physical Review A 92~(3) (2015) 032125.
\newblock \href {http://dx.doi.org/10.1103/PhysRevA.92.032125}
  {\path{doi:10.1103/PhysRevA.92.032125}}.

\bibitem{chantasri2019}
A.~Chantasri, I.~Guevara, H.~M. Wiseman, {Quantum state smoothing: why the
  types of observed and unobserved measurements matter}, New Journal of Physics
  21~(8) (2019) 083039.
\newblock \href {http://dx.doi.org/10.1088/1367-2630/ab396e}
  {\path{doi:10.1088/1367-2630/ab396e}}.

\bibitem{GueWis20}
I.~Guevara, H.~M. Wiseman, {Completely positive quantum trajectories with
  applications to quantum state smoothing}, Physical Review A 102~(5) (2020)
  052217.
\newblock \href {http://dx.doi.org/10.1103/PhysRevA.102.052217}
  {\path{doi:10.1103/PhysRevA.102.052217}}.

\bibitem{LavCha2020a}
K.~T. Laverick, A.~Chantasri, H.~M. Wiseman, {General criteria for quantum
  state smoothing with necessary and sufficient criteria for linear Gaussian
  quantum systems}, Quantum Studies: Mathematics and Foundations 8~(1) (2021)
  37--50.
\newblock \href {http://dx.doi.org/10.1007/s40509-020-00225-7}
  {\path{doi:10.1007/s40509-020-00225-7}}.

\bibitem{holevo2001statistical}
A.~S. Holevo, {Statistical structure of quantum theory Vol. 67}, Springer
  Science {\&} Business Media, 2001.

\bibitem{LavCha2020b}
K.~T. Laverick, A.~Chantasri, H.~M. Wiseman, {Linear Gaussian quantum state
  smoothing: How to optimally `unobserve' a quantum system}, arXiv:2008.13348
  [quant-ph].

\bibitem{Eddington1928}
A.~S. Eddington, {The Nature of the Physical World}, Cambridge University
  Press, London, 1928.

\bibitem{Born26}
M.~Born, Quantenmechanik der sto{\ss}vorg{\"a}nge, Zeitschrift f{\"u}r Physik
  38~(11-12) (1926) 803--827.
\newblock \href {http://dx.doi.org/10.1007/BF01397184}
  {\path{doi:10.1007/BF01397184}}.

\bibitem{Watanabe1956}
S.~Watanabe, {Symmetry in time and Tanikawa's Method of superquantization in
  regard to negative energy fields}, Progress of Theoretical Physics 15~(6)
  (1956) 523--535.
\newblock \href {http://dx.doi.org/10.1143/ptp.15.523}
  {\path{doi:10.1143/ptp.15.523}}.

\bibitem{VAA1987}
L.~Vaidman, Y.~Aharonov, D.~Z. Albert, {How to ascertain the values of
  $\sigma_x$, $\sigma_y$, and $\sigma_z$ of a spin-1/2 particle}, Physical
  Review Letters 58~(14) (1987) 1385.
\newblock \href {http://dx.doi.org/10.1103/PhysRevLett.58.1385}
  {\path{doi:10.1103/PhysRevLett.58.1385}}.

\bibitem{AhaRoh91}
Y.~Aharonov, D.~Rohrlich, {TOWARDS A TWO VECTOR FORMULATION OF QUANTUM
  MECHANICS}, in: Quantum Coherence, World Scientific, 1991, pp. 221--231.
\newblock \href {http://dx.doi.org/10.1142/9789814439251_0018}
  {\path{doi:10.1142/9789814439251_0018}}.

\bibitem{AV91}
Y.~Aharonov, L.~Vaidman, {Complete description of a quantum system at a given
  time}, Journal of Physics A: General Physics 24~(10) (1991) 2315.
\newblock \href {http://dx.doi.org/10.1088/0305-4470/24/10/018}
  {\path{doi:10.1088/0305-4470/24/10/018}}.

\bibitem{Aharonov1998}
Y.~Aharonov, L.~Vaidman, {On the two-state vector reformulation of quantum
  mechanics}, Physica Scripta Vol.~T76 (1998) 85--92.

\bibitem{Qi2010}
S.~Qi, H.~Zheng, W.~Qiaoyan, L.~Wenmin, {Quantum blind signature based on
  Two-State Vector Formalism}, Optics Communications 283~(21) (2010)
  4408--4410.
\newblock \href {http://dx.doi.org/10.1016/j.optcom.2010.06.061}
  {\path{doi:10.1016/j.optcom.2010.06.061}}.

\bibitem{Yang2013}
C.~W. Yang, T.~Hwang, Y.~P. Luo, {Enhancement on ``Quantum blind signature
  based on two-state vector formalism''}, Quantum Information Processing 12~(1)
  (2013) 109--117.
\newblock \href {http://dx.doi.org/10.1007/s11128-012-0362-2}
  {\path{doi:10.1007/s11128-012-0362-2}}.

\bibitem{DanVaid13}
A.~Danan, D.~Farfurnik, S.~Bar-Ad, L.~Vaidman, {Asking photons where they have
  been}, Physical Review Letters 111~(24) (2013) 240402.
\newblock \href {http://dx.doi.org/10.1103/PhysRevLett.111.240402}
  {\path{doi:10.1103/PhysRevLett.111.240402}}.

\bibitem{Aharonov2014}
Y.~Aharonov, E.~Cohen, E.~Gruss, T.~Landsberger, {Measurement and collapse
  within the two-state vector formalism}, Quantum Studies: Mathematics and
  Foundations 1~(1-2) (2014) 133--146.
\newblock \href {http://dx.doi.org/10.1007/s40509-014-0011-9}
  {\path{doi:10.1007/s40509-014-0011-9}}.

\bibitem{CampagneI2014}
P.~Campagne-Ibarcq, L.~Bretheau, E.~Flurin, A.~Auff{\`{e}}ves, F.~Mallet,
  B.~Huard, {Observing interferences between past and future quantum states in
  resonance fluorescence}, Physical Review Letters 112~(18) (2014) 180402.
\newblock \href {http://dx.doi.org/10.1103/PhysRevLett.112.180402}
  {\path{doi:10.1103/PhysRevLett.112.180402}}.

\bibitem{Hashmi2016}
F.~A. Hashmi, F.~Li, S.~Y. Zhu, M.~S. Zubairy, {Two-state vector formalism and
  quantum interference}, Journal of Physics A: Mathematical and Theoretical
  49~(34) (2016) 345302.
\newblock \href {http://dx.doi.org/10.1088/1751-8113/49/34/345302}
  {\path{doi:10.1088/1751-8113/49/34/345302}}.

\bibitem{Nowakowski2018}
M.~Nowakowski, E.~Cohen, P.~Horodecki, {Entangled histories versus the
  two-state-vector formalism: Towards a better understanding of quantum
  temporal correlations}, Physical Review A 98~(3) (2018) 032312.
\newblock \href {http://dx.doi.org/10.1103/PhysRevA.98.032312}
  {\path{doi:10.1103/PhysRevA.98.032312}}.

\bibitem{BookVon1932}
J.~von Neumann, {Mathematical Foundations of Quantum Mechanics, English
  translation (Princeton University Press, Princeton, 1955)}, Springer, Berlin,
  1932.

\bibitem{Kofman2012}
A.~G. Kofman, S.~Ashhab, F.~Nori, {Nonperturbative theory of weak pre-and
  post-selected measurements}, Physics Reports 520~(2) (2012) 43--133.
\newblock \href {http://dx.doi.org/10.1016/j.physrep.2012.07.001}
  {\path{doi:10.1016/j.physrep.2012.07.001}}.

\bibitem{Tamir2013}
B.~Tamir, E.~Cohen, {Introduction to weak measurements and weak values}, Quanta
  2~(1) (2013) 7--17.
\newblock \href {http://dx.doi.org/10.12743/quanta.v2i1.14}
  {\path{doi:10.12743/quanta.v2i1.14}}.

\bibitem{Aharonov2014-2}
Y.~Aharonov, E.~Cohen, A.~C. Elitzur, {Foundations and applications of weak
  quantum measurements}, Physical Review A 89~(5) (2014) 052105.
\newblock \href {http://dx.doi.org/10.1103/PhysRevA.89.052105}
  {\path{doi:10.1103/PhysRevA.89.052105}}.

\bibitem{Dressel2014}
J.~Dressel, M.~Malik, F.~M. Miatto, A.~N. Jordan, R.~W. Boyd, {{\em
  Colloquium}: Understanding quantum weak values: Basics and applications},
  Reviews of Modern Physics 86~(1) (2014) 307.
\newblock \href {http://dx.doi.org/10.1103/RevModPhys.86.307}
  {\path{doi:10.1103/RevModPhys.86.307}}.

\bibitem{Hosten2008}
O.~Hosten, P.~Kwiat, {Observation of the spin hall effect of light via weak
  measurements}, Science 319~(5864) (2008) 787--790.
\newblock \href {http://dx.doi.org/10.1126/science.1152697}
  {\path{doi:10.1126/science.1152697}}.

\bibitem{Starling2009}
D.~J. Starling, P.~B. Dixon, A.~N. Jordan, J.~C. Howell, {Optimizing the
  signal-to-noise ratio of a beam-deflection measurement with interferometric
  weak values}, Physical Review A 80~(4) (2009) 041803.
\newblock \href {http://dx.doi.org/10.1103/PhysRevA.80.041803}
  {\path{doi:10.1103/PhysRevA.80.041803}}.

\bibitem{Brunner2010}
N.~Brunner, C.~Simon, {Measuring small longitudinal phase shifts: Weak
  measurements or standard interferometry?}, Physical Review Letters 105~(1)
  (2010) 010405.
\newblock \href {http://dx.doi.org/10.1103/PhysRevLett.105.010405}
  {\path{doi:10.1103/PhysRevLett.105.010405}}.

\bibitem{Hofmann2011a}
H.~F. Hofmann, {Uncertainty limits for quantum metrology obtained from the
  statistics of weak measurements}, Physical Review A 83~(2) (2011) 022106.
\newblock \href {http://dx.doi.org/10.1103/PhysRevA.83.022106}
  {\path{doi:10.1103/PhysRevA.83.022106}}.

\bibitem{Lundeen11}
J.~S. Lundeen, B.~Sutherland, A.~Patel, C.~Stewart, C.~Bamber, {Direct
  measurement of the quantum wavefunction}, Nature 474~(7350) (2011) 188--191.
\newblock \href {http://dx.doi.org/10.1038/nature10120}
  {\path{doi:10.1038/nature10120}}.

\bibitem{Kedem2012}
Y.~Kedem, {Using technical noise to increase the signal-to-noise ratio of
  measurements via imaginary weak values}, Physical Review A 85~(6) (2012)
  060102.
\newblock \href {http://dx.doi.org/10.1103/PhysRevA.85.060102}
  {\path{doi:10.1103/PhysRevA.85.060102}}.

\bibitem{Viza2013}
G.~I. Viza, J.~Mart{\'{i}}nez-Rinc{\'{o}}n, G.~A. Howland, H.~Frostig,
  I.~Shomroni, B.~Dayan, J.~C. Howell, {Weak-values technique for velocity
  measurements}, Optics Letters 38~(16) (2013) 2949--2952.
\newblock \href {http://dx.doi.org/10.1364/ol.38.002949}
  {\path{doi:10.1364/ol.38.002949}}.

\bibitem{Jordan2014}
A.~N. Jordan, J.~Mart{\'{i}}nez-Rinc{\'{o}}n, J.~C. Howell, {Technical
  advantages for weak-value amplification: When less is more}, Physical Review
  X 4~(1) (2014) 011031.
\newblock \href {http://dx.doi.org/10.1103/PhysRevX.4.011031}
  {\path{doi:10.1103/PhysRevX.4.011031}}.

\bibitem{Knee2014}
G.~C. Knee, E.~M. Gauger, {When amplification with weak values fails to
  suppress technical noise}, Physical Review X 4~(1) (2014) 011032.
\newblock \href {http://dx.doi.org/10.1103/PhysRevX.4.011032}
  {\path{doi:10.1103/PhysRevX.4.011032}}.

\bibitem{Salazar2014}
L.~J. Salazar-Serrano, D.~Janner, N.~Brunner, V.~Pruneri, J.~P. Torres,
  {Measurement of sub-pulse-width temporal delays via spectral interference
  induced by weak value amplification}, Physical Review A 89~(1) (2014) 012126.
\newblock \href {http://dx.doi.org/10.1103/PhysRevA.89.012126}
  {\path{doi:10.1103/PhysRevA.89.012126}}.

\bibitem{Gross2015}
J.~A. Gross, N.~Dangniam, C.~Ferrie, C.~M. Caves, {Novelty, efficacy, and
  significance of weak measurements for quantum tomography}, Physical Review A
  92~(6) (2015) 062133.
\newblock \href {http://dx.doi.org/10.1103/PhysRevA.92.062133}
  {\path{doi:10.1103/PhysRevA.92.062133}}.

\bibitem{VMH15}
G.~I. Viza, J.~Mart{\'{i}}nez-Rinc{\'{o}}n, G.~B. Alves, A.~N. Jordan, J.~C.
  Howell, {Experimentally quantifying the advantages of weak-value-based
  metrology}, Physical Review A 92~(3) (2015) 032127.
\newblock \href {http://dx.doi.org/10.1103/PhysRevA.92.032127}
  {\path{doi:10.1103/PhysRevA.92.032127}}.

\bibitem{Zhang2015}
L.~Zhang, A.~Datta, I.~A. Walmsley, {Precision metrology using weak
  measurements}, Physical Review Letters 114~(21) (2015) 210801.
\newblock \href {http://dx.doi.org/10.1103/PhysRevLett.114.210801}
  {\path{doi:10.1103/PhysRevLett.114.210801}}.

\bibitem{Knee2016}
G.~C. Knee, J.~Combes, C.~Ferrie, E.~M. Gauger, {Weak-value amplification:
  state of play}, Quantum Measurements and Quantum Metrology 1~(open-issue).
\newblock \href {http://dx.doi.org/10.1515/qmetro-2016-0006}
  {\path{doi:10.1515/qmetro-2016-0006}}.

\bibitem{Steinberg2017}
M.~Hallaji, A.~Feizpour, G.~Dmochowski, J.~Sinclair, A.~M. Steinberg,
  {Weak-value amplification of the nonlinear effect of a single photon}, Nature
  Physics 13~(6) (2017) 540--544.
\newblock \href {http://dx.doi.org/10.1038/nphys4040}
  {\path{doi:10.1038/nphys4040}}.

\bibitem{Ren2020}
J.~Ren, L.~Qin, W.~Feng, X.-Q. Li, {Weak-value-amplification analysis beyond
  the Aharonov-Albert-Vaidman limit}, Physical Review A 102~(4) (2020) 042601.
\newblock \href {http://dx.doi.org/10.1103/PhysRevA.102.042601}
  {\path{doi:10.1103/PhysRevA.102.042601}}.

\bibitem{RohAha02}
D.~Rohrlich, Y.~Aharonov, {Cherenkov radiation of superluminal particles},
  Physical Review A 66~(4) (2002) 042102.
\newblock \href {http://dx.doi.org/10.1103/PhysRevA.66.042102}
  {\path{doi:10.1103/PhysRevA.66.042102}}.

\bibitem{Wiseman2007-2}
H.~M. Wiseman, {Grounding Bohmian mechanics in weak values and bayesianism},
  New Journal of Physics 9~(6) (2007) 165.
\newblock \href {http://dx.doi.org/10.1088/1367-2630/9/6/165}
  {\path{doi:10.1088/1367-2630/9/6/165}}.

\bibitem{Brunner2004}
N.~Brunner, V.~Scarani, M.~Wegm{\"{u}}ller, M.~Legr{\'{e}}, N.~Gisin, {Direct
  measurement of superluminal group velocity and signal velocity in an optical
  fiber}, Physical Review Letters 93~(20) (2004) 203902.
\newblock \href {http://dx.doi.org/10.1103/PhysRevLett.93.203902}
  {\path{doi:10.1103/PhysRevLett.93.203902}}.

\bibitem{Mir07}
R.~Mir, J.~Lundeen, M.~Mitchell, A.~Steinberg, J.~Garretson, H.~Wiseman, {A
  double-slit ``which-way" experiment on the complementarity--uncertainty
  debate}, New Journal of Physics 9~(8) (2007) 287.
\newblock \href {http://dx.doi.org/10.1088/1367-2630/9/8/287}
  {\path{doi:10.1088/1367-2630/9/8/287}}.

\bibitem{Lundeen09}
J.~S. Lundeen, A.~M. Steinberg, {Experimental joint weak measurement on a
  photon pair as a probe of Hardy's paradox}, Physical Review Letters 102~(2)
  (2009) 020404.
\newblock \href {http://dx.doi.org/10.1103/PhysRevLett.102.020404}
  {\path{doi:10.1103/PhysRevLett.102.020404}}.

\bibitem{Yokota09}
K.~Yokota, T.~Yamamoto, M.~Koashi, N.~Imoto, {Direct observation of Hardy's
  paradox by joint weak measurement with an entangled photon pair}, New Journal
  of Physics 11~(3) (2009) 033011.
\newblock \href {http://dx.doi.org/10.1088/1367-2630/11/3/033011}
  {\path{doi:10.1088/1367-2630/11/3/033011}}.

\bibitem{Dressel2011}
J.~Dressel, C.~Broadbent, J.~Howell, A.~N. Jordan, {Experimental violation of
  two-party Leggett-Garg inequalities with semiweak measurements}, Physical
  Review Letters 106~(4) (2011) 040402.
\newblock \href {http://dx.doi.org/10.1103/PhysRevLett.106.040402}
  {\path{doi:10.1103/PhysRevLett.106.040402}}.

\bibitem{KBR2011}
S.~Kocsis, B.~Braverman, S.~Ravets, M.~J. Stevens, R.~P. Mirin, L.~K. Shalm,
  A.~M. Steinberg, {Observing the average trajectories of single photons in a
  two-slit interferometer}, Science 332~(6034) (2011) 1170--1173.
\newblock \href {http://dx.doi.org/10.1126/science.1202218}
  {\path{doi:10.1126/science.1202218}}.

\bibitem{Pryde2011}
M.~E. Goggin, M.~P. Almeida, M.~Barbieri, B.~P. Lanyon, J.~L. O'Brien, A.~G.
  White, G.~J. Pryde, {Violation of the {L}eggett-{G}arg inequality with weak
  measurements of photons}, Proceedings of the National Academy of Sciences of
  the United States of America 108~(4) (2011) 1256--1261.
\newblock \href {http://dx.doi.org/10.1073/pnas.1005774108}
  {\path{doi:10.1073/pnas.1005774108}}.

\bibitem{RozSte12}
L.~A. Rozema, A.~Darabi, D.~H. Mahler, A.~Hayat, Y.~Soudagar, A.~M. Steinberg,
  {Violation of Heisenberg's measurement-disturbance relationship by weak
  measurements}, Physical Review Letters 109~(10) (2012) 100404.
\newblock \href {http://dx.doi.org/10.1103/PhysRevLett.109.100404}
  {\path{doi:10.1103/PhysRevLett.109.100404}}.

\bibitem{HWP13}
M.~M. Weston, M.~J. Hall, M.~S. Palsson, H.~M. Wiseman, G.~J. Pryde,
  {Experimental test of universal complementarity relations}, Physical Review
  Letters 110~(22) (2013) 220402.
\newblock \href {http://dx.doi.org/10.1103/PhysRevLett.110.220402}
  {\path{doi:10.1103/PhysRevLett.110.220402}}.

\bibitem{Kaneda14}
F.~Kaneda, S.-Y. Baek, M.~Ozawa, K.~Edamatsu, {Experimental test of
  error-disturbance uncertainty relations by weak measurement}, Physical Review
  Letters 112~(2) (2014) 020402.
\newblock \href {http://dx.doi.org/10.1103/PhysRevLett.112.020402}
  {\path{doi:10.1103/PhysRevLett.112.020402}}.

\bibitem{HWP15}
B.~L. Higgins, M.~S. Palsson, G.~Y. Xiang, H.~M. Wiseman, G.~J. Pryde, {Using
  weak values to experimentally determine ``negative probabilities" in a
  two-photon state with Bell correlations}, Physical Review A 91~(1) (2015)
  012113.
\newblock \href {http://dx.doi.org/10.1103/PhysRevA.91.012113}
  {\path{doi:10.1103/PhysRevA.91.012113}}.

\bibitem{MRF2016}
D.~H. Mahler, L.~Rozema, K.~Fisher, L.~Vermeyden, K.~J. Resch, H.~M. Wiseman,
  A.~Steinberg, {Experimental nonlocal and surreal Bohmian trajectories},
  Science advances 2~(2) (2016) e1501466.
\newblock \href {http://dx.doi.org/10.1126/sciadv.1501466}
  {\path{doi:10.1126/sciadv.1501466}}.

\bibitem{YX2017}
Y.~Xiao, Y.~Kedem, J.-S. Xu, C.-F. Li, G.-C. Guo, {Experimental nonlocal
  steering of Bohmian trajectories}, Optics Express 25~(13) (2017)
  14463--14472.
\newblock \href {http://dx.doi.org/10.1364/OE.25.014463}
  {\path{doi:10.1364/OE.25.014463}}.

\bibitem{XW2019}
Y.~Xiao, H.~M. Wiseman, J.-S. Xu, Y.~Kedem, C.-F. Li, G.-C. Guo, {Observing
  momentum disturbance in double-slit ``which-way'' measurements}, Science
  Advances 5~(6) (2019) eaav9547.
\newblock \href {http://dx.doi.org/10.1126/sciadv.aav9547}
  {\path{doi:10.1126/sciadv.aav9547}}.

\bibitem{Ramos2020}
R.~Ramos, D.~Spierings, I.~Racicot, A.~M. Steinberg, {Measurement of the time
  spent by a tunnelling atom within the barrier region}, Nature 583~(7817)
  (2020) 529--532.
\newblock \href {http://dx.doi.org/10.1038/s41586-020-2490-7}
  {\path{doi:10.1038/s41586-020-2490-7}}.

\bibitem{Hulet1997}
N.~Ritchie, J.~G. Story, R.~G. Hulet, {Realization of a measurement of a ``weak
  value''}, Physical Review Letters 66~(9) (1991) 1107.
\newblock \href {http://dx.doi.org/10.1103/PhysRevLett.66.1107}
  {\path{doi:10.1103/PhysRevLett.66.1107}}.

\bibitem{PryWis2005}
G.~J. Pryde, J.~L. O'Brien, A.~G. White, T.~C. Ralph, H.~M. Wiseman,
  {Measurement of quantum weak values of photon polarization}, Physical Review
  Letters 94~(22) (2005) 220405.
\newblock \href {http://dx.doi.org/10.1103/PhysRevLett.94.220405}
  {\path{doi:10.1103/PhysRevLett.94.220405}}.

\bibitem{Hofmann2011c}
M.~Iinuma, Y.~Suzuki, G.~Taguchi, Y.~Kadoya, H.~F. Hofmann, {Weak measurement
  of photon polarization by back-action-induced path interference}, New Journal
  of Physics 13 (2011) 033041.
\newblock \href {http://dx.doi.org/10.1088/1367-2630/13/3/033041}
  {\path{doi:10.1088/1367-2630/13/3/033041}}.

\bibitem{SDG2015}
S.~Sponar, T.~Denkmayr, H.~Geppert, H.~Lemmel, A.~Matzkin, J.~Tollaksen,
  Y.~Hasegawa, {Weak values obtained in matter-wave interferometry}, Physical
  Review A 92~(6) (2015) 062121.
\newblock \href {http://dx.doi.org/10.1103/PhysRevA.92.062121}
  {\path{doi:10.1103/PhysRevA.92.062121}}.

\bibitem{Vaidman2017-2}
L.~Vaidman, A.~Ben-Israel, J.~Dziewior, L.~Knips, M.~Wei{\ss}l, J.~Meinecke,
  C.~Schwemmer, R.~Ber, H.~Weinfurter, {Weak value beyond conditional
  expectation value of the pointer readings}, Physical Review A 96~(3) (2017)
  032114.
\newblock \href {http://dx.doi.org/10.1103/PhysRevA.96.032114}
  {\path{doi:10.1103/PhysRevA.96.032114}}.

\bibitem{Johansen2004}
L.~M. Johansen, {What is the value of an observable between pre- and
  postselection?}, Physics Letters, Section A: General, Atomic and Solid State
  Physics 322~(5-6) (2004) 298--300.
\newblock \href {http://dx.doi.org/10.1016/j.physleta.2004.01.041}
  {\path{doi:10.1016/j.physleta.2004.01.041}}.

\bibitem{Jozsa2007}
R.~Jozsa, {Complex weak values in quantum measurement}, Physical Review A
  76~(4) (2007) 044103.
\newblock \href {http://dx.doi.org/10.1103/PhysRevA.76.044103}
  {\path{doi:10.1103/PhysRevA.76.044103}}.

\bibitem{Hosoya2010}
A.~Hosoya, Y.~Shikano, {Strange weak values}, Journal of Physics A:
  Mathematical and Theoretical 43~(38) (2010) 385307.
\newblock \href {http://dx.doi.org/10.1088/1751-8113/43/38/385307}
  {\path{doi:10.1088/1751-8113/43/38/385307}}.

\bibitem{DreAga2010}
J.~Dressel, S.~Agarwal, A.~N. Jordan, {Contextual values of observables in
  quantum measurements}, Physical Review Letters 104~(24) (2010) 240401.
\newblock \href {http://dx.doi.org/10.1103/PhysRevLett.104.240401}
  {\path{doi:10.1103/PhysRevLett.104.240401}}.

\bibitem{Dressel2012}
J.~Dressel, A.~N. Jordan, {Significance of the imaginary part of the weak
  value}, Physical Review A 85~(1) (2012) 012107.
\newblock \href {http://dx.doi.org/10.1103/PhysRevA.85.012107}
  {\path{doi:10.1103/PhysRevA.85.012107}}.

\bibitem{Dressel2015}
J.~Dressel, {Weak values as interference phenomena}, Physical Review A 91~(3)
  (2015) 032116.
\newblock \href {http://dx.doi.org/10.1103/PhysRevA.91.032116}
  {\path{doi:10.1103/PhysRevA.91.032116}}.

\bibitem{Hall2016}
M.~J. Hall, A.~K. Pati, J.~Wu, {Products of weak values: Uncertainty relations,
  complementarity, and incompatibility}, Physical Review A 93~(5) (2016)
  052118.
\newblock \href {http://dx.doi.org/10.1103/PhysRevA.93.052118}
  {\path{doi:10.1103/PhysRevA.93.052118}}.

\bibitem{nielsen2010quantum}
M.~A. Nielsen, I.~L. Chuang, Quantum computation and quantum information,
  Cambridge university press, 2010.

\bibitem{BookJacobs}
K.~Jacobs, Quantum Measurement Theory and its Applications, Cambridge
  University Press, 2014.

\bibitem{BookKraus}
K.~Kraus, A.~B{\"o}hm, J.~D. Dollard, W.~H. Wootters, {States, effects, and
  operations fundamental notions of quantum theory}, Springer, Berlin, 1983.

\bibitem{Vai14}
L.~Vaidman, {Tracing the past of a quantum particle}, Physical Review A 89~(2)
  (2014) 024102.
\newblock \href {http://dx.doi.org/10.1103/PhysRevA.89.024102}
  {\path{doi:10.1103/PhysRevA.89.024102}}.

\bibitem{BookDavies}
E.~B. Davies, {Quantum Theory of Open Systems}, Academic Press, London, 1976.

\bibitem{BookCarmichael}
H.~J. Carmichael, {An open systems approach to quantum optics}, Springer,
  Berlin, 1993.

\bibitem{ZhaMol17}
J.~Zhang, K.~M{\o}lmer, {Prediction and retrodiction with continuously
  monitored Gaussian states}, Physical Review A 96~(6) (2017) 062131.
\newblock \href {http://dx.doi.org/10.1103/PhysRevA.96.062131}
  {\path{doi:10.1103/PhysRevA.96.062131}}.

\bibitem{Budini2018a}
A.~A. Budini, {Quantum non-Markovian processes break conditional past-future
  independence}, Physical Review Letters 121~(24) (2018) 240401.
\newblock \href {http://dx.doi.org/10.1103/PhysRevLett.121.240401}
  {\path{doi:10.1103/PhysRevLett.121.240401}}.

\bibitem{Budini2019}
A.~A. Budini, {Conditional past-future correlation induced by non-Markovian
  dephasing reservoirs}, Physical Review A 99~(5) (2019) 052125.
\newblock \href {http://dx.doi.org/10.1103/PhysRevA.99.052125}
  {\path{doi:10.1103/PhysRevA.99.052125}}.

\bibitem{GreMol15}
Q.~Xu, E.~Greplova, B.~Julsgaard, K.~M{\o}lmer, {Correlation functions and
  conditioned quantum dynamics in photodetection theory}, Physica Scripta
  90~(12) (2015) 128004.
\newblock \href {http://dx.doi.org/10.1088/0031-8949/90/12/128004}
  {\path{doi:10.1088/0031-8949/90/12/128004}}.

\bibitem{GreMol16}
E.~Greplova, K.~M{\o}lmer, C.~K. Andersen, {Quantum teleportation with
  continuous measurements}, Physical Review A 94~(4) (2016) 042334.
\newblock \href {http://dx.doi.org/10.1103/PhysRevA.94.042334}
  {\path{doi:10.1103/PhysRevA.94.042334}}.

\bibitem{GreMol17}
E.~Greplova, E.~A. Laird, G.~A.~D. Briggs, K.~M{\o}lmer, {Conditioned spin and
  charge dynamics of a single-electron quantum dot}, Physical Review A 96~(5)
  (2017) 052104.
\newblock \href {http://dx.doi.org/10.1103/PhysRevA.96.052104}
  {\path{doi:10.1103/PhysRevA.96.052104}}.

\bibitem{Gough2020}
J.~Gough, {How to estimate past quantum measurement interventions after
  continuous monitoring}, Russian Journal of Mathematical Physics 27~(2) (2020)
  218--227.
\newblock \href {http://dx.doi.org/10.1134/S1061920820020089}
  {\path{doi:10.1134/S1061920820020089}}.

\bibitem{RybMol15}
T.~Rybarczyk, B.~Peaudecerf, M.~Penasa, S.~Gerlich, B.~Julsgaard, K.~M{\o}lmer,
  S.~Gleyzes, M.~Brune, J.~M. Raimond, S.~Haroche, I.~Dotsenko,
  {Forward-backward analysis of the photon-number evolution in a cavity},
  Physical Review A 91~(6) (2015) 062116.
\newblock \href {http://dx.doi.org/10.1103/PhysRevA.91.062116}
  {\path{doi:10.1103/PhysRevA.91.062116}}.

\bibitem{TanMol15}
D.~Tan, S.~J. Weber, I.~Siddiqi, K.~M{\o}lmer, K.~W. Murch, {Prediction and
  retrodiction for a continuously monitored superconducting qubit}, Physical
  Review Letters 114~(9) (2015) 090403.
\newblock \href {http://dx.doi.org/10.1103/PhysRevLett.114.090403}
  {\path{doi:10.1103/PhysRevLett.114.090403}}.

\bibitem{TanMol16}
D.~Tan, M.~Naghiloo, K.~M{\o}lmer, K.~W. Murch, {Quantum smoothing for
  classical mixtures}, Physical Review A 94~(5) (2016) 050102.
\newblock \href {http://dx.doi.org/10.1103/PhysRevA.94.050102}
  {\path{doi:10.1103/PhysRevA.94.050102}}.

\bibitem{FTM16}
N.~Foroozani, M.~Naghiloo, D.~Tan, K.~M{\o}lmer, K.~W. Murch, {Correlations of
  the time dependent signal and the state of a continuously monitored quantum
  system}, Physical Review Letters 116~(11) (2016) 110401.
\newblock \href {http://dx.doi.org/10.1103/PhysRevLett.116.110401}
  {\path{doi:10.1103/PhysRevLett.116.110401}}.

\bibitem{TanMol17}
D.~Tan, N.~Foroozani, M.~Naghiloo, A.~H. Kiilerich, K.~M{\o}lmer, K.~W. Murch,
  {Homodyne monitoring of postselected decay}, Physical Review A 96~(2) (2017)
  022104.
\newblock \href {http://dx.doi.org/10.1103/PhysRevA.96.022104}
  {\path{doi:10.1103/PhysRevA.96.022104}}.

\bibitem{BaoMol20a}
H.~Bao, J.~Duan, S.~Jin, X.~Lu, P.~Li, W.~Qu, M.~Wang, I.~Novikova, E.~E.
  Mikhailov, K.~F. Zhao, K.~M{\o}lmer, H.~Shen, Y.~Xiao, {Spin squeezing of
  $10^{11}$ atoms by prediction and retrodiction measurements}, Nature
  581~(7807) (2020) 159--163.
\newblock \href {http://dx.doi.org/10.1038/s41586-020-2243-7}
  {\path{doi:10.1038/s41586-020-2243-7}}.

\bibitem{BaoMol20b}
H.~Bao, S.~Jin, J.~Duan, S.~Jia, K.~M{\o}lmer, H.~Shen, Y.~Xiao, Retrodiction
  beyond the heisenberg uncertainty relation, Nature communications 11~(1)
  (2020) 1--7.
\newblock \href {http://dx.doi.org/10.1038/s41467-020-19495-1}
  {\path{doi:10.1038/s41467-020-19495-1}}.

\bibitem{BookGardiner1}
C.~W. Gardiner, P.~Zoller, {Quantum noise: A handbook of Markovian and
  non-Markovian quantum stochastic methods with applications to quantum
  optics}, Springer, Berlin, 2004.

\bibitem{Weber2014}
S.~J. Weber, A.~Chantasri, J.~Dressel, A.~N. Jordan, K.~W. Murch, I.~Siddiqi,
  {Mapping the optimal route between two quantum states}, Nature 511~(7511)
  (2014) 570--573.
\newblock \href {http://dx.doi.org/10.1038/nature13559}
  {\path{doi:10.1038/nature13559}}.

\bibitem{ChaKim2016}
A.~Chantasri, M.~E. Kimchi-Schwartz, N.~Roch, I.~Siddiqi, A.~N. Jordan,
  {Quantum trajectories and their statistics for remotely entangled quantum
  bits}, Physical Review X 6~(4) (2016) 041052.
\newblock \href {http://dx.doi.org/10.1103/PhysRevX.6.041052}
  {\path{doi:10.1103/PhysRevX.6.041052}}.

\bibitem{JorMur2017}
M.~Naghiloo, D.~Tan, P.~M. Harrington, P.~Lewalle, A.~N. Jordan, K.~W. Murch,
  {Quantum caustics in resonance-fluorescence trajectories}, Physical Review A
  96~(5) (2017) 053807.
\newblock \href {http://dx.doi.org/10.1103/PhysRevA.96.053807}
  {\path{doi:10.1103/PhysRevA.96.053807}}.

\bibitem{Silveri2016}
M.~Silveri, E.~Zalys-Geller, M.~Hatridge, Z.~Leghtas, M.~H. Devoret, S.~M.
  Girvin, {Theory of remote entanglement via quantum-limited phase-preserving
  amplification}, Physical Review A 93~(6) (2016) 062310.
\newblock \href {http://dx.doi.org/10.1103/PhysRevA.93.062310}
  {\path{doi:10.1103/PhysRevA.93.062310}}.

\bibitem{Lewalle2018chaos}
P.~Lewalle, J.~Steinmetz, A.~N. Jordan, {Chaos in continuously monitored
  quantum systems: An optimal-path approach}, Physical Review A 98~(1) (2018)
  012141.
\newblock \href {http://dx.doi.org/10.1103/PhysRevA.98.012141}
  {\path{doi:10.1103/PhysRevA.98.012141}}.

\bibitem{JorCha2016}
A.~N. Jordan, A.~Chantasri, P.~Rouchon, B.~Huard, {Anatomy of fluorescence:
  quantum trajectory statistics from continuously measuring spontaneous
  emission}, Quantum Studies: Mathematics and Foundations 3~(3) (2016)
  237--263.
\newblock \href {http://dx.doi.org/10.1007/s40509-016-0075-9}
  {\path{doi:10.1007/s40509-016-0075-9}}.

\bibitem{ChaAta2018}
A.~Chantasri, J.~Atalaya, S.~Hacohen-Gourgy, L.~S. Martin, I.~Siddiqi, A.~N.
  Jordan, {Simultaneous continuous measurement of noncommuting observables:
  Quantum state correlations}, Physical Review A 97~(1) (2018) 012118.
\newblock \href {http://dx.doi.org/10.1103/PhysRevA.97.012118}
  {\path{doi:10.1103/PhysRevA.97.012118}}.

\bibitem{Amini2011}
H.~Amini, M.~Mirrahimi, P.~Rouchon, {On stability of continuous-time quantum
  filters}, in: {50th IEEE Conference on Decision and Control and European
  Control Conference (CDC-ECC) 2011}, Orlando, United States, 2011, pp. 6242 --
  6247.
\newblock \href {http://dx.doi.org/10.1109/CDC.2011.6160631}
  {\path{doi:10.1109/CDC.2011.6160631}}.

\bibitem{Rouchon2015}
P.~Rouchon, J.~F. Ralph, Efficient quantum filtering for quantum feedback
  control, Phys. Rev. A 91 (2015) 012118.
\newblock \href {http://dx.doi.org/10.1103/PhysRevA.91.012118}
  {\path{doi:10.1103/PhysRevA.91.012118}}.

\bibitem{LewCha17}
P.~Lewalle, A.~Chantasri, A.~N. Jordan, {Prediction and characterization of
  multiple extremal paths in continuously monitored qubits}, Physical Review A
  95~(4) (2017) 042126.
\newblock \href {http://dx.doi.org/10.1103/PhysRevA.95.042126}
  {\path{doi:10.1103/PhysRevA.95.042126}}.

\bibitem{Armen2009}
M.~A. Armen, A.~E. Miller, H.~Mabuchi, {Spontaneous Dressed-State Polarization
  in the Strong Driving Regime of Cavity QED}, Physical Review Letters 103
  (2009) 173601.
\newblock \href {http://dx.doi.org/10.1103/PhysRevLett.103.173601}
  {\path{doi:10.1103/PhysRevLett.103.173601}}.

\bibitem{Sarkka13}
S.~S{\"a}rkk{\"a}, {Bayesian filtering and smoothing}, Vol.~3, Cambridge
  University Press, 2013.

\bibitem{LavGueWis21}
K.~T. Laverick, I.~Guevara, H.~M. Wiseman, Quantum state smoothing as an
  optimal estimation problem with three different cost functions\href
  {http://arxiv.org/abs/2106.02354} {\path{arXiv:2106.02354}}.

\bibitem{Laverick21}
K.~T. Laverick, The quantum {Rauch-Tung-Striebel} smoothed state\href
  {http://arxiv.org/abs/2010.11027} {\path{arXiv:2010.11027}}.

\bibitem{LavWarWis21}
K.~T. Laverick, P.~Warszawski, H.~M. Wiseman, {Quantum state smoothing can be
  non-classical even when the filtering and retrofiltering are classical}, in
  Preparation.

\bibitem{GamWis2005}
J.~Gambetta, H.~M. Wiseman, {Stochastic simulations of conditional states of
  partially observed systems, quantum and classical}, Journal of Optics B:
  Quantum and Semiclassical Optics 7~(10) (2005) S250.
\newblock \href {http://dx.doi.org/10.1088/1464-4266/7/10/008}
  {\path{doi:10.1088/1464-4266/7/10/008}}.

\bibitem{Jozsa1994}
R.~Jozsa, {Fidelity for mixed quantum states}, Journal of Modern Optics 41~(12)
  (1994) 2315--2323.
\newblock \href {http://dx.doi.org/10.1080/09500349414552171}
  {\path{doi:10.1080/09500349414552171}}.

\bibitem{NLevelBloch}
G.~Kimura, {The Bloch vector for N-level systems}, Physics Letters, Section A:
  General, Atomic and Solid State Physics 314~(5-6) (2003) 339--349.
\newblock \href {http://dx.doi.org/10.1016/S0375-9601(03)00941-1}
  {\path{doi:10.1016/S0375-9601(03)00941-1}}.

\bibitem{Wiseman1993-2}
H.~M. Wiseman, G.~J. Milburn, {Interpretation of quantum jump and diffusion
  processes illustrated on the Bloch sphere}, Physical Review A 47~(3) (1993)
  1652.
\newblock \href {http://dx.doi.org/10.1103/PhysRevA.47.1652}
  {\path{doi:10.1103/PhysRevA.47.1652}}.

\bibitem{LiLi2018}
L.~Li, M.~J. Hall, H.~M. Wiseman, {Concepts of quantum non-Markovianity: A
  hierarchy}, Physics Reports 759 (2018) 1--51.
\newblock \href {http://dx.doi.org/10.1016/j.physrep.2018.07.001}
  {\path{doi:10.1016/j.physrep.2018.07.001}}.

\bibitem{Plenio1998}
M.~B. Plenio, P.~L. Knight, The quantum-jump approach to dissipative dynamics
  in quantum optics, Reviews of Modern Physics 70 (1998) 101--144.

\bibitem{MinMun2019}
Z.~K. Minev, S.~O. Mundhada, S.~Shankar, P.~Reinhold,
  R.~Guti{\'{e}}rrez-J{\'{a}}uregui, R.~Schoelkopf, M.~Mirrahimi,
  H.~Carmichael, M.~Devoret, {To catch and reverse a quantum jump mid-flight},
  Nature 570~(7760) (2019) 200--204.
\newblock \href {http://dx.doi.org/10.1038/s41586-019-1287-z}
  {\path{doi:10.1038/s41586-019-1287-z}}.

\bibitem{smiRei2002}
W.~P. Smith, J.~E. Reiner, L.~A. Orozco, S.~Kuhr, H.~M. Wiseman, Capture and
  release of a conditional state of a cavity qed system by quantum feedback,
  Physical Review Letters 89 (2002) 133601.
\newblock \href {http://dx.doi.org/10.1103/PhysRevLett.89.133601}
  {\path{doi:10.1103/PhysRevLett.89.133601}}.

\bibitem{BookGardiner2}
C.~W. Gardiner, {Handbook of stochastic methods for physics}, Springer, Berlin,
  2004.

\bibitem{BookJacobsSto}
K.~Jacobs, {Stochastic processes for physicists: Understanding noisy systems},
  Cambridge University Press, New York, 2010.
\newblock \href {http://dx.doi.org/10.1017/CBO9780511815980}
  {\path{doi:10.1017/CBO9780511815980}}.

\end{thebibliography}


%\begin{thebibliography}{00}


%\end{thebibliography}

\end{document}